\documentclass[12pt]{elsart}
\usepackage{epsfig}
\usepackage{colordvi}
\begin{document}

\begin{frontmatter}

\title{
Systematics of central heavy ion collisions in the $1A$ GeV regime}

\author[GSI]{W.~Reisdorf, \thanksref{info}},
\author[GSI]{A.~Andronic},
\author[GSI]{R.~Averbeck},
\author[HEID]{M.L.~Benabderrahmane},
\author[GSI]{O.N.~Hartmann},
\author[HEID]{N.~Herrmann},
\author[GSI]{K.D.~Hildenbrand},
\author[GSI,KOR]{T.I.~Kang},
\author[GSI]{Y.J.~Kim},
\author[GSI,ZAG]{M.~Ki\v{s}},
\author[GSI]{P.~Koczo\'{n}},
\author[GSI]{T.~Kress},
\author[GSI]{Y.~Leifels},
\author[HEID]{M.~Merschmeyer},
\author[HEID,WAR]{K.~Piasecki},
\author[GSI]{A.~Sch\"{u}ttauf},
\author[HEID]{M.~Stockmeier},
\author[CLER]{V.~Barret},
\author[ZAG]{Z.~Basrak},
\author[CLER]{N.~Bastid},
\author[ZAG]{R.~\v{C}aplar},
\author[CLER]{P.~Crochet},
\author[CLER]{P.~Dupieux},
\author[ZAG]{M.~D\v{z}elalija},
\author[BUD]{Z.~Fodor},
\author[WAR]{P.~Gasik},
\author[ITEP]{Y.~Grishkin},
\author[KOR]{B.~Hong},
\author[BUD]{J.~Kecskemeti},
\author[WAR]{M.~Kirejczyk},
\author[ZAG]{M.~Korolija},
\author[ROSS]{R.~Kotte},
\author[ITEP]{A.~Lebedev},
\author[CLER]{X.~Lopez},
\author[WAR]{T.~Matulewicz},
\author[ROSS]{W.~Neubert},
\author[BUC]{M.~Petrovici},
\author[STRAS]{F.~Rami},
\author[KOR]{M.S.~Ryu},
\author[BUD]{Z.~Seres},
\author[WAR]{B.~Sikora},
\author[KOR]{K.S.~Sim},
\author[BUC]{V.~Simion},
\author[WAR]{K.~Siwek-Wilczy\'nska},
\author[ITEP]{V.~Smolyankin},
\author[BUC]{G.~Stoicea},
\author[WAR]{Z.~Tymi\'{n}ski},
\author[WAR]{K.~Wi\'{s}niewski},
\author[ROSS]{D.~Wohlfarth},
\author[GSI,IMP]{Z.G.~Xiao},
\author[IMP]{H.S.~Xu},
\author[KUR]{I.~Yushmanov},
\author[ITEP]{A.~Zhilin}

(FOPI Collaboration)
\address[GSI]{GSI Helmholtzzentrum f\"ur Schwerionenforschung GmbH,
 Darmstadt, Germany}
\address[BUC]{National Institute for Nuclear Physics and Engineering,
Bucharest,Romania}
\address[BUD]{Central Research Institute for Physics, Budapest, Hungary}
\address[CLER]{Clermont Universit\'e, Universit\'e Blaise Pascal, CNRS/IN2P3,
 Laboratoire de Physique Corpusculaire, Clermont-Ferrand, France}
\address[ROSS]{ Institut f\"ur Strahlenphysik, Forschungszentrum Rossendorf,
Dresden, Germany}
\address[HEID]{Physikalisches Institut der Universit\"at Heidelberg,Heidelberg,
Germany}
\address[ITEP]{Institute for Theoretical and Experimental Physics,
Moscow,Russia}
\address[KUR]{Kurchatov Institute, Moscow, Russia}
\address[IMP]{Institute of Modern Physics, Chinese Academy of Sciences,
Lanzhou, China}
\address[KOR]{Korea University, Seoul, South Korea}
\address[STRAS] {Institut Pluridisciplinaire Hubert Curien, IN2P3-CNRS,
Universit\'e de Strasbourg, Strasbourg, France}
\address[WAR]{Institute of Experimental Physics, University of Warsaw, Poland}
\address[ZAG]{Rudjer Boskovic Institute, Zagreb, Croatia}

\thanks[info]{Email:~W.Reisdorf@gsi.de}

\begin{abstract}
Using the large acceptance apparatus FOPI, we study central collisions in
the reactions (energies in $A$ GeV are given in parentheses):
$^{40}$Ca+$^{40}$Ca (0.4, 0.6, 0.8, 1.0, 1.5, 1.93),
$^{58}$Ni+$^{58}$Ni (0.15, 0.25, 0.4),
$^{96}$Ru+$^{96}$Ru (0.4, 1.0, 1.5),
$^{96}$Zr+$^{96}$Zr (0.4, 1.0, 1.5),
$^{129}$Xe+CsI (0.15, 0.25, 0.4),
$^{197}$Au+$^{197}$Au (0.09, 0.12, 0.15, 0.25, 0.4, 0.6, 0.8, 1.0, 1.2, 1.5).
The observables include
cluster multiplicities,
longitudinal and transverse rapidity distributions and stopping,
and radial flow.                
The data are compared to earlier data where possible and to transport model
simulations.

\end{abstract}

\begin{keyword}
 heavy ions,  rapidity, stopping, viscosity,
 radial flow, cluster production, nucleosynthesis,
 nuclear equation of state, isospin 

 \PACS 25.75.-q, 25.75.Dw, 25.75.Ld
\end{keyword}
\end{frontmatter}

\section{Introduction}\label{intro}
In the past twenty five years studies of nucleus-nucleus collisions over a
very broad range of energies have been performed.
The goal was, and still is, to learn more about the properties of hot
and dense nuclear matter, namely its equation of state (EOS) and transport
coefficients such as viscosity and thermal conductivity.
In the energy range (0.1 to $2A$ GeV) of the Schwerionen Synchrotron (SIS)
 accelerator located at GSI, Darmstadt, one expects from
simple-minded hydrodynamic considerations that densities
twice to three times saturation density are reached.
Another important characteristic of this energy range is that the
compressed overlap zone created in the collision expands at a rate
comparable to the rate at which the accelerated nuclei interpenetrate,
a phenomenon that provides a very useful clock: the passage time
$2R/(p_{cm}/m_N)$ where R is the nuclear radius, $p_{cm}$ and $m_N$ are
the incoming nucleon's c.m. momentum and mass.
In a historic paper~\cite{scheid74} a strong ejection of matter in the
{\it transverse} direction to the beam was predicted.
In the present work on very central collisions, this transverse direction, 
in particular in comparison to the longitudinal (beam) direction, will play
a crucial role. 

The discovery \cite{gustafsson84,renfordt84} of collective flow seemed
 to confirm
that fluid dynamics was the proper language to theoretically accompany
the experimental efforts.
However, soon after it became clear that a quantitative extraction of
fundamental properties of nuclear matter required the development
of microscopic transport theories \cite{bertsch88} that did not  have to rely on
the local equilibrium postulate needed to apply the (one-fluid) 
hydrodynamic approach \cite{clare86,stoecker86}.
Immediately it also became then clear that the static EOS was not the only
 unknown as momentum dependent potentials \cite{aichelin87,gale87}
 and in-medium modifications
of hadron masses \cite{brown91} and cross sections \cite{fuchs01} 
came into play. 
Despite of these increased difficulties the past decade has witnessed
reasonably successful attempts to come to a first-generation of nuclear
EOS using experimental observations in heavy ion collisions as constraints
\cite{danielewicz02,fuchs06,fuchs06a}.
One approach, \cite{danielewicz02}, choose to base its conclusions on
 azimuthally asymmetric flow measurements for the Au+Au system 
(reviewed in \cite{reisdorf97r,herrmann99}) done at
the BEVALAC and AGS accelerators (USA), while the other approach
\cite{fuchs06,hartnack06} was based on a comparison of kaon production in
C+C and Au+Au collisions using primarily data obtained at SIS by the KaoS
Collaboration (\cite{sturm01}, for a review see \cite{senger99,foerster07}).

Despite this encouraging progress a critical look at the present situation
reveals that more work is necessary.
 Fundamental conclusions based on one particular observable, such as
the $K^+$ production varying the system size \cite{sturm01} or
centrality \cite{hartnack06}, are not
sufficiently convincing to settle the problem.
At SIS energies kaons are a rare probe: besides the condition that their
production in a dense medium must be under sufficient theoretical control,
one must also assume that the change of environment, i.e. the bulk matter
produced, is correctly described. This involves a complete mastery of
the degree of stopping reached in the collisions, since the idea
\cite{aichelin85}
behind the kaon observable is that it is (in the subthreshold regime) 
sensitive to the achieved density.
Further, the possible influence of isospin dependences has to be accounted
for: the switch from the iso-symmetric system $^{12}$C+$^{12}$C to the 
asymmetric $^{197}$Au+$^{197}$Au is expected to produce a shift of the
ratio of the isospin pair $K^+/K^0$, but the $K^0$ production was not
measured.

On the other hand,
EOS determinations based on reproducing flow data are also not straightforward
as has become clear from a number of earlier works by our Collaboration
\cite{ramillien95,bastid97,crochet97,andronic01,andronic01b,andronic03,bastid04,stoicea04,andronic05} and the EoS Collaboration \cite{swang96}.
Estimates using Rankine-Hugoniot-Taub shock equations
\cite{reisdorf97r,stoecker81},
(that ignore geometrical constraints and non-equilibrium, as well as
 momentum-dependent phenomena),
show that the pressures generated in the collisions, which are at the
origin of the measured flow observables, differ only relatively modestly
when the assumed stiffness of the 'cold' (zero temperature) EOS is varied
\footnote{it is primarily the density that is modified},
 leading already in 1981
the authors of ref. \cite{stoecker81} to conclude
 {\it that high accuracy of flow and
entropy measurements would be needed to deduce information on the cold EOS}.
In this connection it is interesting to note that
 a promising new observable has been proposed  
\cite{stoicea04}
 suggesting to use azimuthal dependences of the kinetic energies,
rather than the yields of emitted particles. 
A disadvantage of the azimuthally asymmetric flow signal,
is that it converges to zero for the most central collisions where compression
is maximal.

It is also
 a somewhat unsatisfactory situation,  that in the past no clean
isolation of the isospin dependence of the EOS has been made: most
of the conclusions based on heavy ion data involve the Au+Au system, which is
neither isospin symmetric nor close to the composition of neutron matter
(or neutron stars).

In the present work, devoted to {\it bulk} observations in very {\it central}
collisions, we set up a large data base of additional challenging constraints
for microscopic transport simulations that claim to extract basic nuclear
matter properties from heavy ion reactions.
We shall show that sensitivity to the EOS is not limited to rare probes
or non-central collisions.

One observable, the ratio of variances of transverse and longitudinal
rapidities \cite{reisdorf04a}, that we shall loosely speaking call 'stopping',
 will catch our particular attention.
Developing further on the original idea \cite{scheid74}, 
we shall show evidence that
this ratio is influenced by the ratio of the speed of expansion and
the speed at which the incoming material is progressing and in this sense
the 'clock' is not limited to non-central collisions. 
A first attempt to use simultaneously the 'stopping' observable
 and the directed flow
to constrain the EOS was published in \cite{andronic06}.
Another signal of the EOS that we see is the degree of cooling following
the decompression of the system: correlations between stopping, radial flow
and clusterization serve as evidence that at least a partial memory of the
 original compression exists \cite{reisdorf04b}. 
Cooling (reabsorption for produced particles, clusterization for nucleonic
matter) can be more efficient if it was preceded by more compression (softer
EOS). 
This has  different, subtle, consequences for pion, kaon
and nucleonic cluster yields  at freeze out.

It is well known by now that this kind of physics has overlap with some areas
of astrophysics. 
The stiffness of the EOS limits the density in the center of neutron stars as
it does limit the maximum densities achievable in heavy ion reactions.
The synthesis of nuclei during the expansion-cooling processes in the Universe
and more locally of supernovae is partially imitated in the expansion-cooling
of the heavy-ion systems we observe. 

With this exciting context in mind
 we have studied twenty five system-energies varying
the incident energy by a factor twenty and choosing 
three, and for some incident energies more, different system sizes and
compositions
in an effort to elucidate the finite size and isospin effects
 on the observables:
it is well known that nuclear ground state masses are heavily affected by
direct (surface) and indirect (Coulomb and neutron-proton asymmetry) 
contributions due to finite size: the 'correction' to the bulk energy is by
a factor two. 
Nuclei are therefore truly 'small' objects.
 In the case of nuclear reactions
with incident energies not larger than the rest masses an added finite dimension
comes to the nuclear 'surface': the finite time duration. 
There is no reason to expect then, that finite size effects are smaller
in reaction physics than for ground state masses.
Data analyses, in particular thermal model analyses, that ignore this fact
are likely to be misleading.

Relative to other earlier experimental studies we shall make a substantial
effort to eliminate, or at least clarify, apparatus specific effects on
 the observables.
First, the choice of the centrality selection is done in such a way that
a precise matching of centralities between experiment and simulation is
possible: this is important, as some bulk observables (yields, stopping)
 turn out to be very
sensitive to centrality for  good physical reasons.
Second, we try to make superfluous the use of apparatus filters
when comparing data from experiment and simulation.
A large effort is devoted to reconstruct the full $4\pi$ data wherever
this is possible with a reasonably low uncertainty. The large
acceptance of our setup, the FOPI apparatus \cite{gobbi93,ritman95}, makes this
possible.
   
 We start our paper with a more technical part, 
a brief description of the apparatus, section \ref{apparatus},
followed by some frequently used definitions and the centrality selection,
section \ref{centrality}, and a detailed account of our $4\pi$ reconstruction
method, section \ref{reconstruction}.
The description of results will then start with a presentation in section
\ref{rapidity} of a systematics of longitudinal and transverse
rapidity distributions for various identified ejectiles. 
We  proceed with deduced information on the three stages of
 the reaction:
compression, or stopping, section \ref{stopping}, expansion, or radial flow,
section \ref{radial} and freeze-out, or chemistry, section \ref{chemistry}.
We will use the transport code IQMD \cite{hartnack98} both for technical checks and
for an assessment of the basic information that can be extracted from our
data. 
References to some of the copious earlier literature will be presented
wherever relevant in the various sections.
We provide a summary (besides the abstract and the definition section) 
for the quick reader.

The spirit of the present work is similar to the one on pion emission at
SIS energies \cite{reisdorf07} and another paper in preparation on
azimuthally asymmetric flow: provide a large systematic data base 
(for the SIS energy range) that
hopefully can be used for a reassessment of a second generation of 
nuclear EOS and nuclear transport properties, accompanied by the hope
that the various existing or still developed transport codes will
 eventually converge to common
conclusions.

\section{Apparatus and data analysis}\label{apparatus}
The experiments were performed at the heavy ion accelerator SIS of
GSI/Darmstadt using the large acceptance FOPI detector \cite{gobbi93,ritman95}.
A total of 25 system-energies are analysed for this work (energies 
per nucleon, $E/A$, in  GeV
are given in parentheses):
$^{40}$Ca+$^{40}$Ca (0.4, 0.,6 0.8, 1.0, 1.5, 1.93),
$^{58}$Ni+$^{58}$Ni (0.15, 0.25),
$^{129}$Xe+CsI (0.15, 0.25),
$^{96}$Ru+$^{96}$Ru (0.4, 1.0, 1.5),
$^{96}$Zr+$^{96}$Zr (0.4, 1.5),
$^{197}$Au+$^{197}$Au (0.09, 0.12, 0.15, 0.25, 0.4, 0.6, 0.8, 1.0, 1.2, 1.5).

Two somewhat different setups were used for the low energy data
($E/A < 0.4$ GeV) and the high energy data.
For the latter,
particle tracking  and energy loss determination were
done using two drift chambers, the CDC (covering polar angles
between $35^\circ$ and $135^\circ$) and
the Helitron ($9^\circ-26^\circ$), both
located inside a superconducting solenoid operated at a magnetic field of
0.6\,T.
A set of scintillator arrays, Plastic Wall $(7^\circ-30^\circ)$,
Zero Degree Detector
$(1.2^\circ-7^\circ)$, and Barrel $(42^\circ-120^\circ)$,
 allowed us to measure the time of flight
and, below $30^\circ$, also the energy loss.
The velocity resolution below $30^\circ$ was $(0.5-1.5)\%$,
the momentum resolution in the
CDC was $(4-12)\%$ for momenta of 0.5 to 2 GeV/c, respectively.
Use of CDC and Helitron allowed the identification of pions, as well as
good isotope separation for  hydrogen and
helium clusters in a large part of momentum space.
Heavier clusters are separated  by nuclear charge.
More features of the experimental method, some of them specific to the CDC,
have been described in ref. \cite{pelte97au}.

In the low energy run the Helitron drift chamber was not yet available
and the Barrel surrounding the CDC covered only $1/8$ of the full $2\pi$
azimuth.
On the other hand a set of gas ionization chambers (Parabola) installed between 
the solenoid magnet (enclosing the CDC) and the Plastic Wall was operated
that extended to lower velocities and higher charges the range of 
charge identified  highly ionizing fragments and covered approximately
the same angular range as the Plastic Wall.
The Parabola together with the Plastic Wall represent what we used
to call the early PHASE I of FOPI the technical details of which can be
found in ref. \cite{gobbi93}.
Also in the low energy run, the CDC was operated in a 'split mode'
in an effort to extend the dynamical range of measured energy losses along
the particle tracks:
of the 60 potential wires the inner 30 wires were held under a lower
voltage, see ref. \cite{andronic01} for more details.
We note that the combination of the two setups allowed us to cover
a rather large range of incident $E/A$: there is approximately a factor of
twenty between the lowest and the highest energy. 

\section{Some definitions, centrality selection}\label{centrality}

Choosing the $c.m.$ as reference frame, orienting the z-axis in the beam
direction, and ignoring  deviations from axial symmetry 
the two remaining dimensions are characterized by the longitudinal
rapidity $y\equiv y_z$,
given by $exp(2y)=(1+\beta_z)/(1-\beta_z)$ and the transverse
(spatial) component $t$ of the four-velocity $u$, given by $u_t=\beta_t\gamma$.
The 3-vector $\vec{\beta}$ is the velocity in units of the light
velocity and $\gamma=1/\sqrt{1-\beta^2}$.
In order to be able to compare longitudinal and transversal degrees of
freedom on a common basis,
we shall also use the {\em transverse} rapidity, $y_x$,
which is defined by
replacing $\beta_z$ by $\beta_x$ in the expression for the longitudinal
rapidity.
The $x$-axis is laboratory fixed and hence randomly oriented relative to the
reaction plane, i.e. we average over deviations from axial symmetry.
The transverse rapidities $y_x$ (or $y_y$) should not be confused with
$y_t$ which is defined by replacing $\beta_z$ by
$\beta_t\equiv\sqrt{\beta_x^2+\beta_y^2}$.

For thermally equilibrated systems $\beta_t=\sqrt{2}\beta_x$ and the local
rapidity distributions $dN/dy_x$ and $dN/dy_y$ (rather than $dN/dy_t$)
should have the same shape and height than the usual longitudinal
rapidity distribution $dN/dy_z$, where we will omit the subscript $z$ when no 
confusion is likely.
Throughout we use scaled units $y_0=y/y_p$ and $u_{t0}=u_t/u_p$,
with $u_p=\beta_p \gamma_p$, the index p referring to the incident projectile
in the $c.m.$.
In these units the initial target-projectile rapidity gap always extends from
$y_0=-1$ to $y_0=1$.
Besides the 'scaled transverse rapidity', $y_{x0}$, distributions
$dN/dy_{x0}$ we
will also present data on 'constrained scaled transverse rapidity' distributions
$dN/dy_{xm0}$ which are obtained using a midrapidity-cut $|y_{z0}|<0.5$ on the
longitudinal rapidities $y_{z0}$.
In this work the shapes of the various rapidity distributions will often
be characterized by their variances which we term {\it varz, varx, and varxm}
for the longitudinal, transverse and constrained transverse distributions,
respectively, adding a suffix zero if scaled units are used.
In section \ref{varxz} we will introduce the ratio $varxz \equiv varx/varz$
which we propose as a measure of {\it stopping} to be discussed in detail in
section \ref{stopping}.
This ratio, naturally, is scale invariant.

Collision centrality selection was obtained by binning 
distributions of either the detected
charge particle multiplicity, MUL, or the ratio of total transverse to
longitudinal kinetic energies in the center-of-mass (c.m.) system, ERAT.
To the degree where ERAT is  evidently less influenced by clusterization than
MUL, there is some advantage of using ERAT when trying to simulate the
selections with transport codes that do not reproduce well the measured
multiplicities.
As will be clear from our data, a careful matching of centralities between
data and simulations is important for correct quantitative conclusions.
ERAT is related to the stopping observable {\it varxz}  defined earlier.
 To avoid autocorrelations we have always excluded the particle
of interest, for which we build up spectra, from the definition of
ERAT.

We estimate  the impact parameter $b$
from the measured differential cross sections for the ERAT or the
multiplicity distributions, using a geometrical sharp-cut approximation.
The ERAT selections show better impact parameter resolution for the most
central collisions than the multiplicity selections.
In the present work we will limit ourselves to ERAT selected data.
More detailed discussions of the centrality selection methods used here
can be found in refs. \cite{andronic06,reisdorf97}.
In the sequel, rather than using cryptic names for the various centrality
intervals, we shall characterize the 
centrality by the interval of
the scaled impact parameter $b_0$ defined by $b_0=b/b_{max}$.
We take $b_{max} = 1.15 (A_{P}^{1/3} + A_{T}^{1/3})$~fm 
giving values known \cite{myers73} to describe well effective sharp radii
of nuclei with mass higher than 40u.
This scaling is useful when comparing systems of different size.
In this work we limit ourselves to four centrality bins:
$b_0<0.15$, $b_0<0.25$, $0.25<b_0<0.45$ and $0.45<b_0<0.55$.
Most of the data presented here concern the most central bin.
In this most central bin typical accepted event numbers vary between
$(45-90)\times 10^3$ for all but the Ca+Ca data where we registered 
$(20-35)\times 10^3$
most central events. 
The total number of registered events was approximately ten times higher
and was limited by a multiplicity filter.
Some minimum bias events were registered at a lower rate.

When comparing to IQMD simulations we also take the ERAT observable
for binning and carefully match the so defined centralities to those of
the experiment.
Apparatus effects tend to modify the observed shape of the ERAT distributions.
However, the influence of a realistic apparatus filter on the
ERAT binning quality was found to be very small and was therefore not critical
as long as the corresponding cross sections remained matched.

The IQMD version we use in the present work largely corresponds to the
description given in Ref.~\cite{hartnack98}. 
More specifically, the following 'standard' parameters were used throughout:
$L=8.66 fm^2$ (wave packet width parameter),
 $t=200fm$ (total integration time), $K=200, 380$ MeV (compressibility of
the momentum dependent soft, resp. stiff EoS, $E_{sym}=25 \rho/\rho_0$ MeV
(symmetry energy, with $\rho_0$ the saturation value of the nuclear density
$\rho$).
The versions with $K=200$, resp. 380 MeV are called IQMD-SM, resp. IQMD-HM.
The clusterization was determined from a separate routine using the minimum
spanning tree method in configuration space with a clusterization radius
$R_c=3$ fm.

\section{$4\pi$ reconstruction}\label{reconstruction}
It is of high interest to know the event topology in full $4\pi$ coverage.
Thus, $4\pi$ yields are needed for convincing estimates of chemical freeze-out
characteristics, chemical temperatures, entropy generation etc.
Also stopping characteristics to be extensively discussed later, require
a large coverage of phase space.
Since measured 
momentum space distributions  do not
cover the complete $4\pi$ phase space they must be complemented by
interpolations and extrapolations.
As this is a rather important feature of our data analysis, we devote
a whole section to this undertaking.

Since this study is limited 
to symmetric systems, we require
reflection symmetry in the center of momentum ($c.m.$).

To start the data treatment,
we  filter the data to eliminate regions of
distorted measurements, such as edge effects.
One advantage of filtering {\it experimental} data, rather than just the
simulated events, is that well defined sharp limits
are introduced which can then be applied in exactly the same form to
simulations (see later). 
Examples of these sharp filtered data are given in Figs.~1,2,4,5 and 6.
A second advantage of these sharp filters, that are more restrictive than
just the nominal geometries of the various sub-detectors, is that within
the remaining acceptances detector efficiencies are close to $100\%$.
For example, studies of measured correlated adjacent detector hit
probabilities in the Plastic Wall showed a worst case (Au+Au at $1.5A$ GeV)
double hit probability of $8\%$, lowering the apparent total multiplicity,
which was nearly compensated by a cross-talk between detector units of
$6\%$, raising the apparent multiplicity. This cross talk was mainly due
to imperfect alignment of the beam axis with the apparatus.
The rationale for dropping further going detailed efficiency considerations
{\it within} the filtered acceptances was that the charge balances, after
$4\pi$ reconstruction, were found to be accurate to typically $5\%$ (see
the tables in the appendix) for all the 25 system-energies studied,
i.e. for a statistically significant sampling encompassing a large
variation in energy and multiplicity.
In our analysis the Plastic Wall and the CDC are treated as 'master'
detectors required for the registration of an ejectile.
The Helitron efficiency was taken care of by matching its signals
for H isotopes to Z=1 hits in the Plastic Wall after subtracting from
the Plastic Wall hits the estimated pion contributions known from our
earlier study \cite{reisdorf07}.
At incident energies beyond $1A$ GeV, where the Plastic Wall showed a
large background for $Z\geq 2$, a matching of Z with the Helitron
was required (in addition to track matching) using the energy loss signals
 and assuming for the Helitron
the same efficiency as for Z=1. Other assumptions were found to be in conflict
with the global charge balances.
The Barrel, as a slave detector to the CDC, was used to improve isotope
identification (for H and He) whenever its signal matched the CDC tracking.
The mixing of adjacent particles is
estimated to be less or equal to $10\%$, except for tritons and $^3$He in the
low energy run where it could be up to $20\%$. 

In the following step each measured phase-space cell
$dy_0\times du_{t0}$ and its local surrounding $N_{yu}-1$ cells, with
$N_{yu} = (2n_y+1)(2n_u+1)$, is  least squares fitted using the ansatz
$$  \frac{1}{u_{t0}} \frac{d^2 N} {du_{t0} dy_0 }
 = \exp \, [f(y_0,u_{t0})] $$
where 
$f(y_0,u_{t0})=a_2 y_0^2 + b_2 u_{t0}^2 + d |y_0| u_{t0} + a_1 |y_0| +b_1
 u_{t0} +c_0$
is a five parameter function.

This procedure smoothens out statistical errors and allows
subsequently a well defined  extension to gaps in the data.
Within errors the smoothened representation of the data follows  the
topology of the original data:
 deviations of $5\%$ are caused by local distortions of the apparatus
response, revealing typical systematic uncertainties.

The technical parameters of the procedure were chosen to be
$dy_0=0.1$, $du_{t0}=0.04$, $n_y=2$, $n_u=5$, i.e. the local fitting domain
consisted of a maximum of 55 cells. 
These choices were governed by the available statistics and the need to follow
the measured topology  within statistical and systematic errors.
Variations of these parameters were investigated to determine the 
systematic errors of the procedure.

The smoothened data 
are well suited for the final step: the extrapolation
to zero transverse momenta.
The low $p_t$ extrapolation  was extensively checked by  event
simulations using both thermal models or transport codes.
For fragments separated only by charge the phase space coverage of FOPI
is rather good and the extension to $4\pi$ is unproblematic.
When reconstructing the isotope separated distributions we have used
the constraint that the sum of isotopes of Z=1 and 2 should be consistent 
{\it in every phase space cell} $dy_0\times du_{t0}$ 
with the total value for Z=1, resp. Z=2
fragments (henceforth dubbed 'sum-of-isotopes check'). 
This ensures at low $p_t$ that Coulomb effects are approximately respected.

Finally, a word on systematic errors:
 Systematic errors are dominated by forward-backward inconsistencies
and extrapolation uncertainties.
The estimates for the latter were  guided by the accuracy of the total
 charge balances
and the isotope balances.
The global systematic errors on particle yields are given in the 
appendix. The small straggling of differential data due to generally 
good statistics and the smoothening fits suggest point to point errors
smaller than the global errors.

In the following we show a few illustrative examples of the $4\pi$
reconstructions.


\subsection{Thermal model reconstruction}
\begin{figure}
\epsfig{file=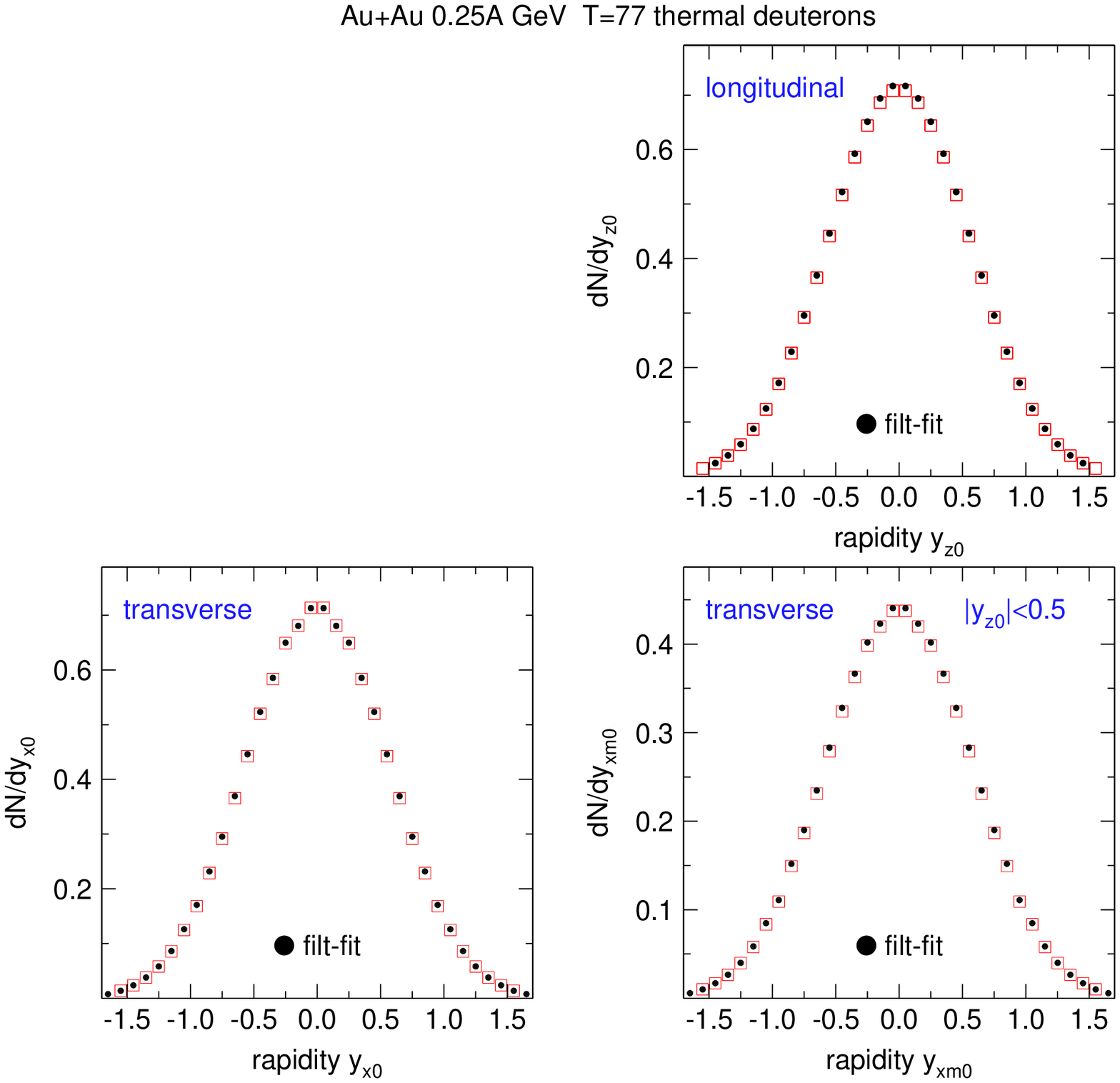,width=140mm}

\vspace{-140mm}
\hspace{-5mm}
\epsfig{file=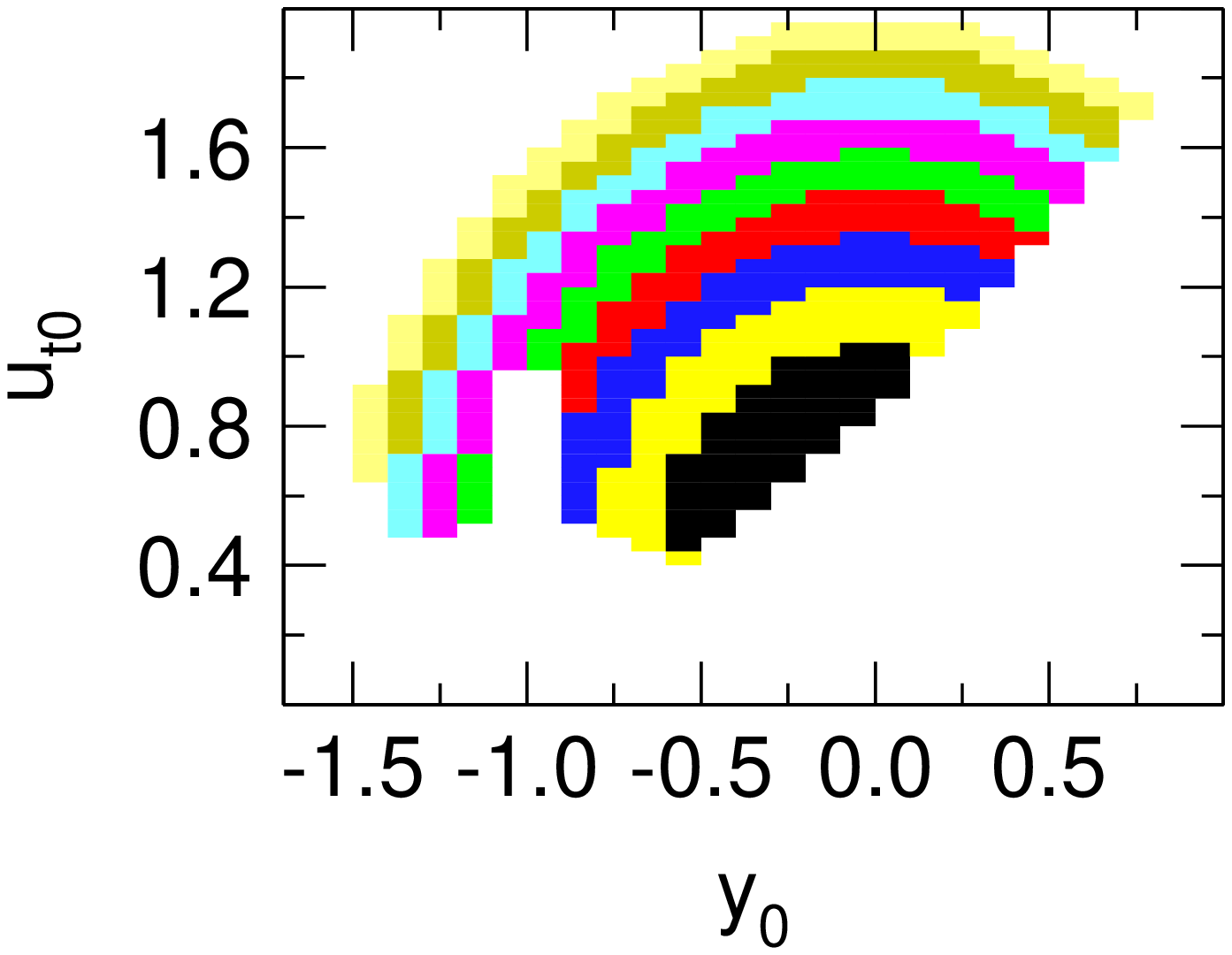,width=80mm}

\vspace{70mm}
\caption{
Rapidity distributions of deuterons in Au+Au reactions at
$0.25A$ GeV obtained from a thermal model with $T=77$ MeV.
The original predictions are given by (red) open squares, the
reconstructed distributions after filtering and then extrapolating
are given by (black) filled dots.
The upper left panel shows the filtered distribution in scaled
momentum versus rapidity space, the various color tones corresponding 
to yields differing by a factor 1.5.
The lower panels show the integrated (left) and the constrained transverse
rapidity distributions, while the longitudinal distribution is plotted
in the upper right panel. 
}
\label{au250c1Z1A2}
\end{figure}

The above ansatz in terms of exponential functions to limited two-dimensional
phase space cells 
is inspired by the
thermal model, but does allow for deviations from it.
Of course in case the thermal model would represent the data well, we
would recover the thermal model parameterization (i.e. the 'temperature').
This is shown in our first example below (Fig.~\ref{au250c1Z1A2}).
The sharp filter used in this case is representative of the actual
acceptance for isotope separated fragments in the low energy runs.
As can be seen the reconstruction in this case is close to perfect and
purely statistical errors can be neglected.

\subsection{Tests of passage to $4\pi$ using IQMD}
\begin{figure}[h]
%
\begin{minipage}{80mm}
\epsfig{file=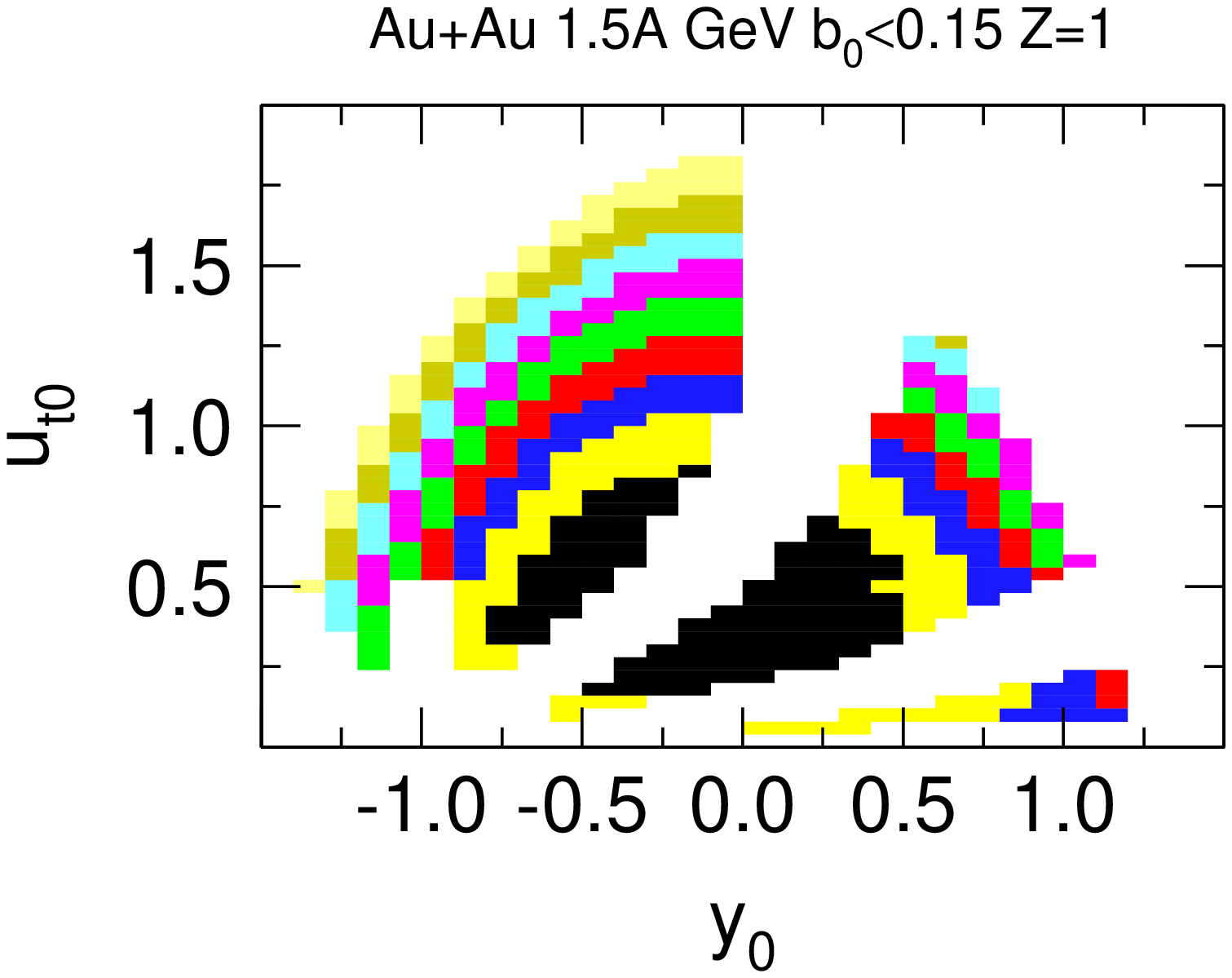,width=80mm}
\end{minipage}
\begin{minipage}{80mm}
\epsfig{file=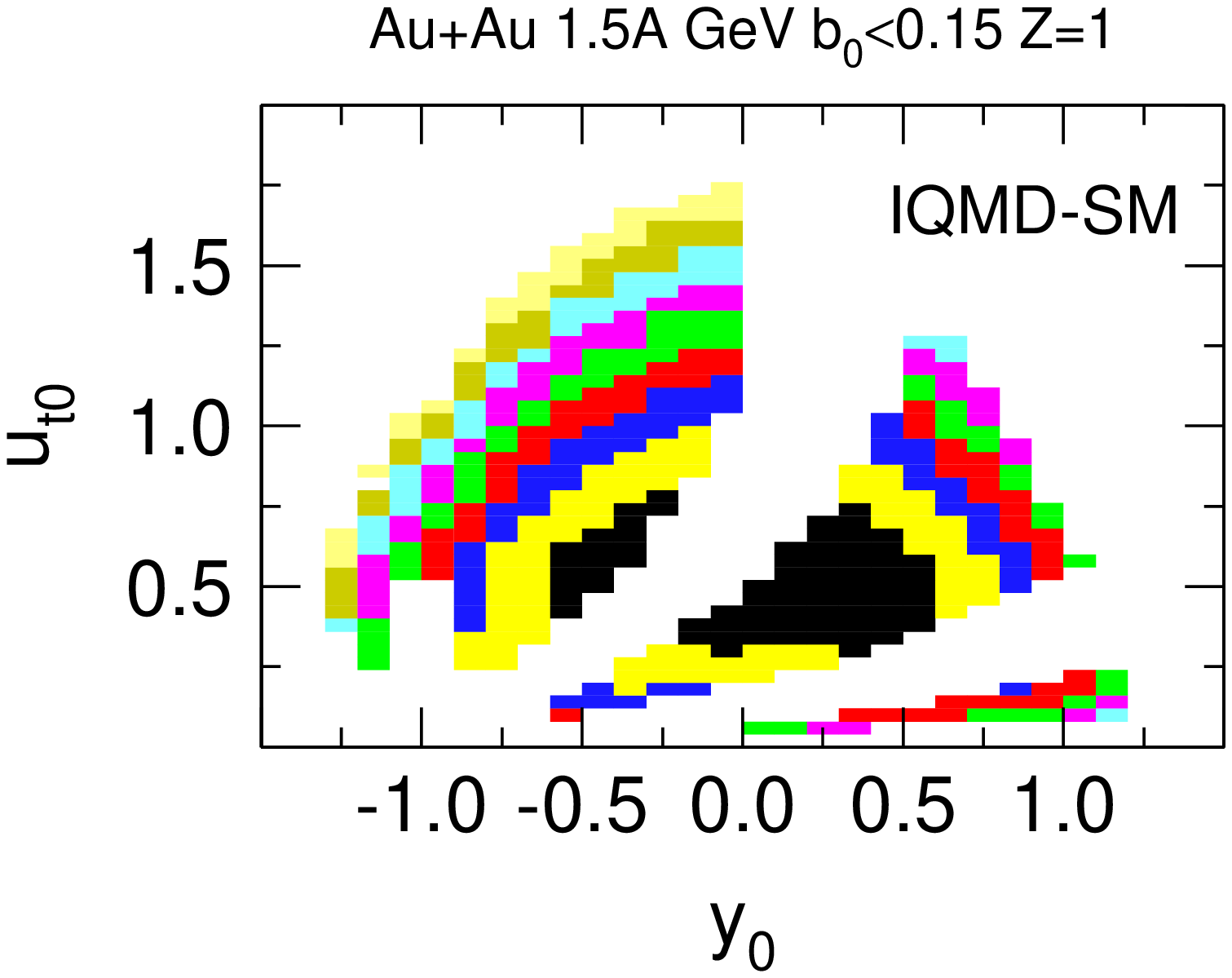,width=80mm}
\end{minipage}
\caption{
Distributions $dN/du_{t0}dy_0$ of hydrogen fragments emitted in central
collisions of Au+Au at $1.5A$ GeV. 
The various color tones correspond to cuts differing by factors 1.5.
The distributions are filtered by a sharp-cut filter the left panel
showing the experimental data, while the right panel shows the
simulation (applying the same centrality  and filter cuts) with IQMD.
}
\label{uty-au1500c1Z1}
\end{figure}

The following example shows the application to a case which cannot be
a thermal distribution since it is the superposition of the three isotopes
(p, d, t) of hydrogen.
As can be seen from Fig. \ref{uty-au1500c1Z1}, there are three separate
 contributions to the
phase space distribution after applying the filter: they are from the CDC,
the Helitron  and the Zero Degree detectors. 
We show both the experimental data and the simulation with IQMD. 
Pion contributions are excluded or subtracted using the information obtained
in \cite{reisdorf07}.

\begin{figure}
\epsfig{file=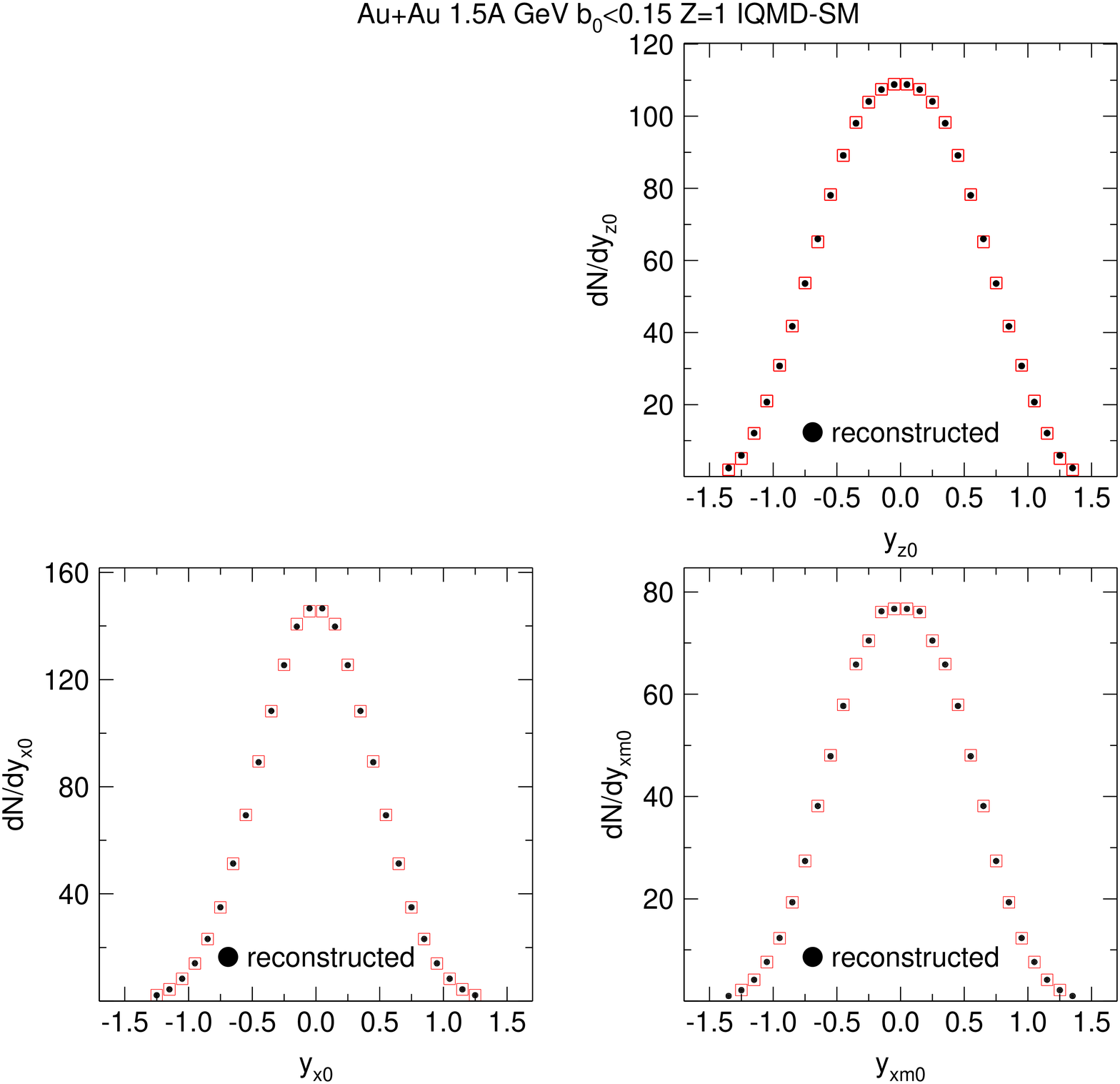,width=140mm}
\caption{
IQMD simulations:
Longitudinal (upper panel), transverse and constrained transverse rapidity
distributions of hydrogen fragments emitted in central
collisions of Au+Au at $1.5A$ GeV. 
The original predictions are given by (red) open squares, the
reconstructed distributions after filtering (see Fig.~\ref{uty-au1500c1Z1})
and then extrapolating
are given by (black) filled dots.
}
\label{dndy-au1500c1Z1}
\end{figure}

The reconstruction of the full acceptance rapidity distribution of the
simulation causes no difficulty as demonstrated in Fig.~\ref{dndy-au1500c1Z1}.
The existence of low $p_t$ data due to the Zero Degree proves to be a
 very useful constraint on the reconstruction.

\begin{figure}

\hspace{-6mm}
\epsfig{file=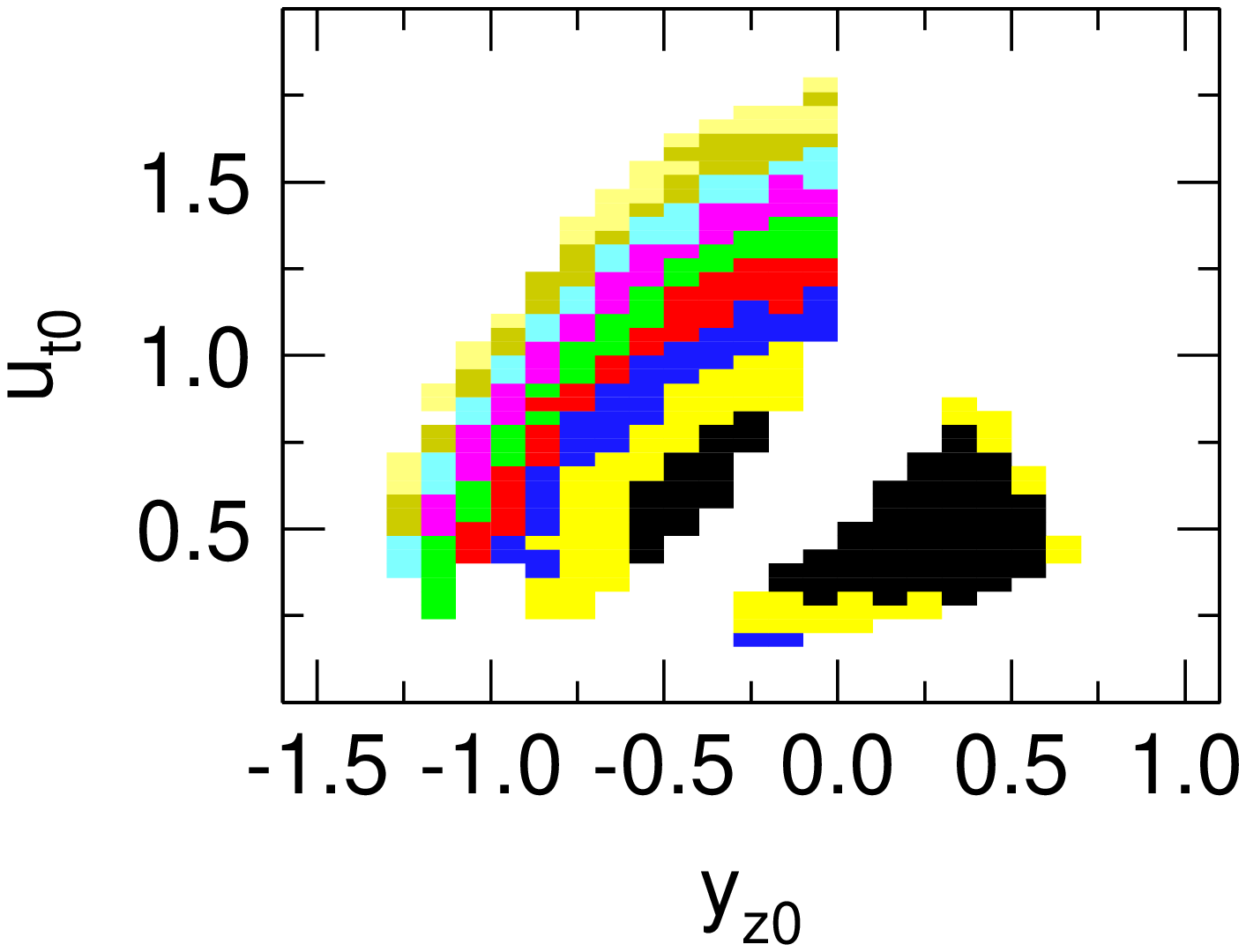,width=80mm}

\vspace{-66mm}

\epsfig{file=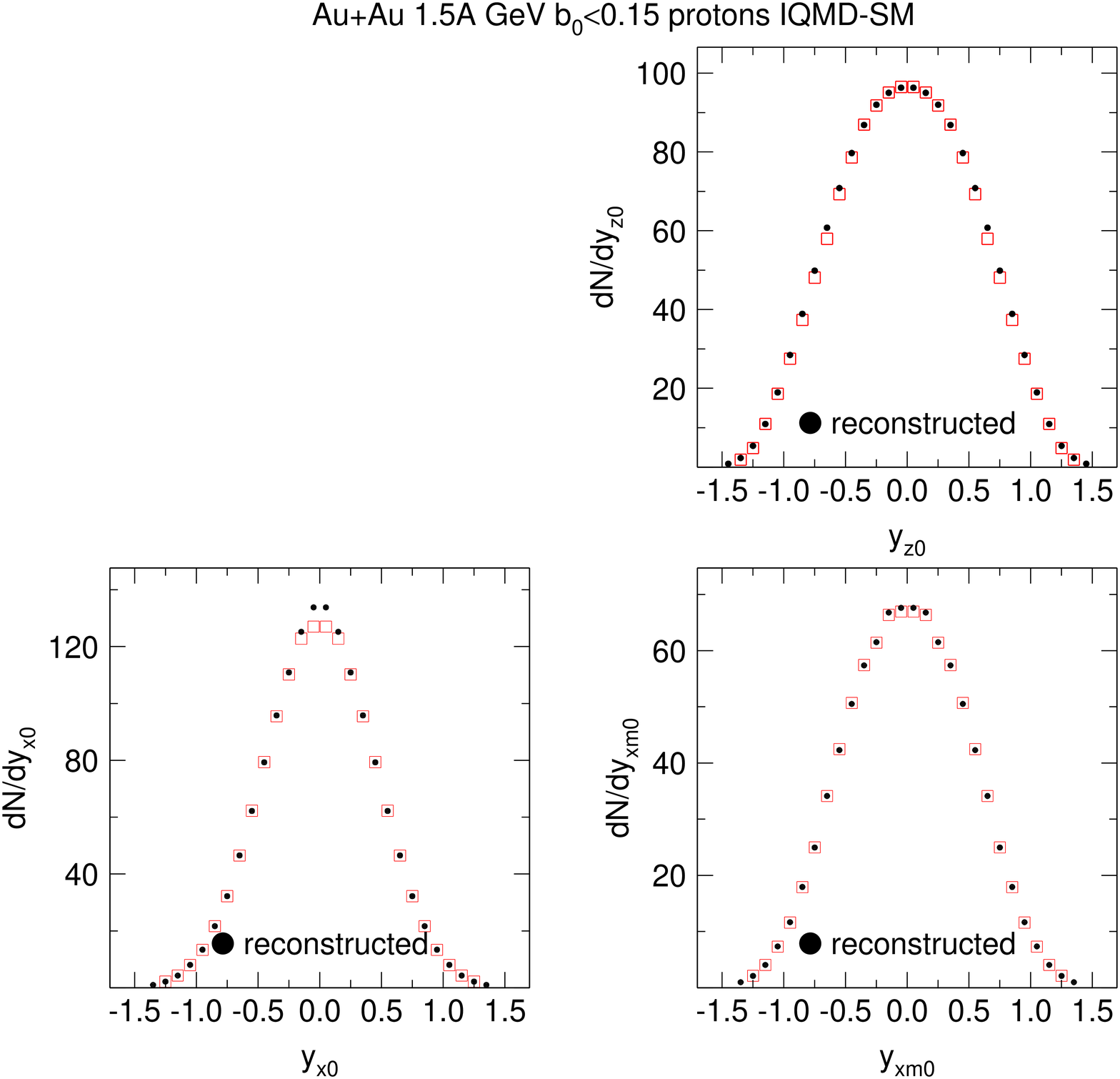,width=140mm}
\caption{
Rapidity distributions of protons in Au+Au reactions at
$1.5A$ GeV obtained from IQMD-SM.
The original predictions are given by (red) open squares, the
reconstructed distributions after filtering and then extrapolating
are given by (black) filled dots.
The upper left panel shows the filtered distribution in scaled
momentum versus rapidity space.
}
\label{au1500c1Z1A1}
\end{figure}

The acceptance for identified protons in the higher energy runs was
more restricted since the Zero Degree detector did not allow isotope
separation and the need to use the Helitron in addition to the
Plastic Wall restricted the forward part of the detector somewhat.
This is shown in Fig.~\ref{au1500c1Z1A1}.
Still, we find that the reconstruction, for {\it central} collisions
(which are of prime interest here) is very satisfactory.

\begin{figure}

\hspace{-5mm}
\epsfig{file=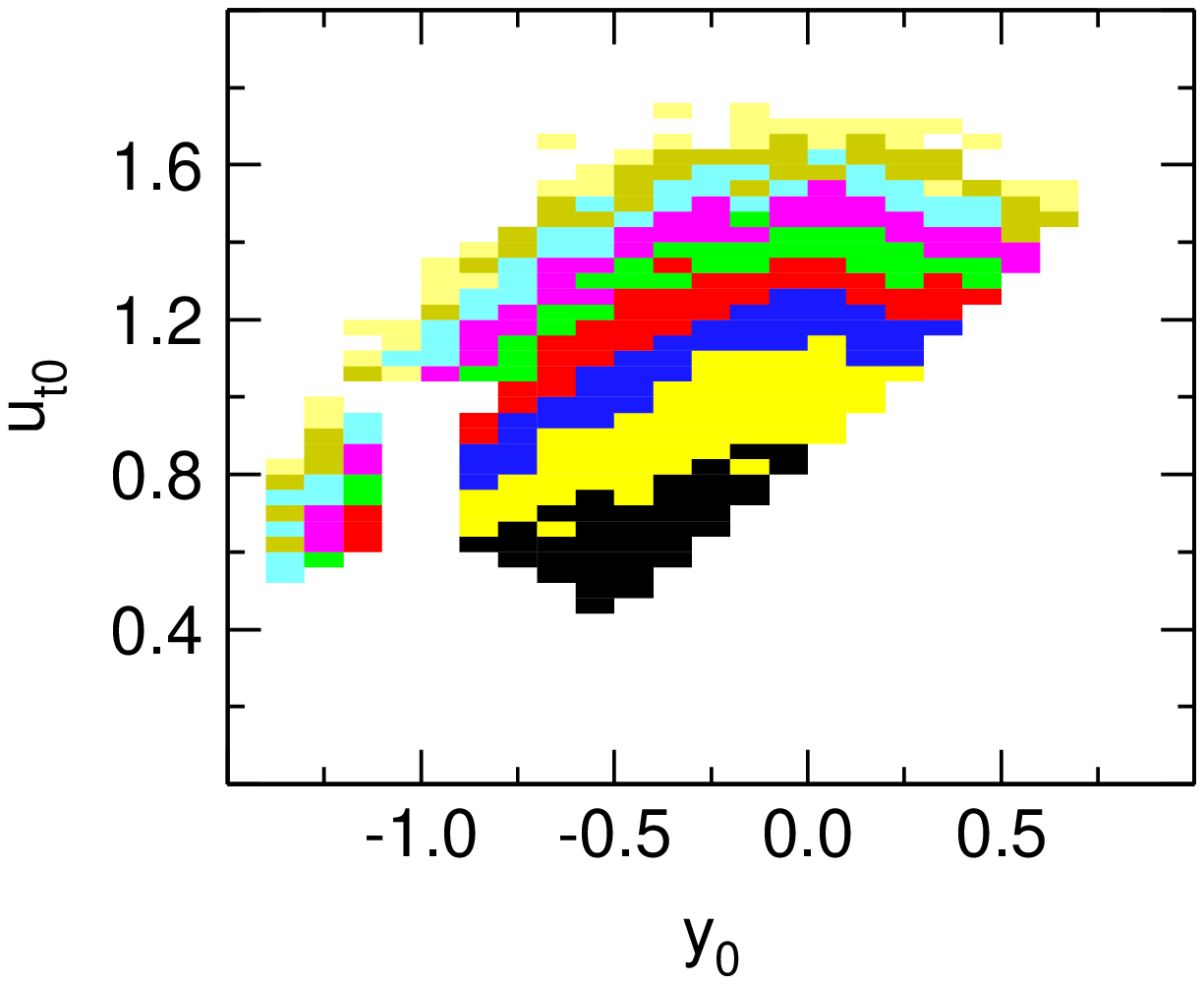,width=80mm}

\vspace{-70mm}

\epsfig{file=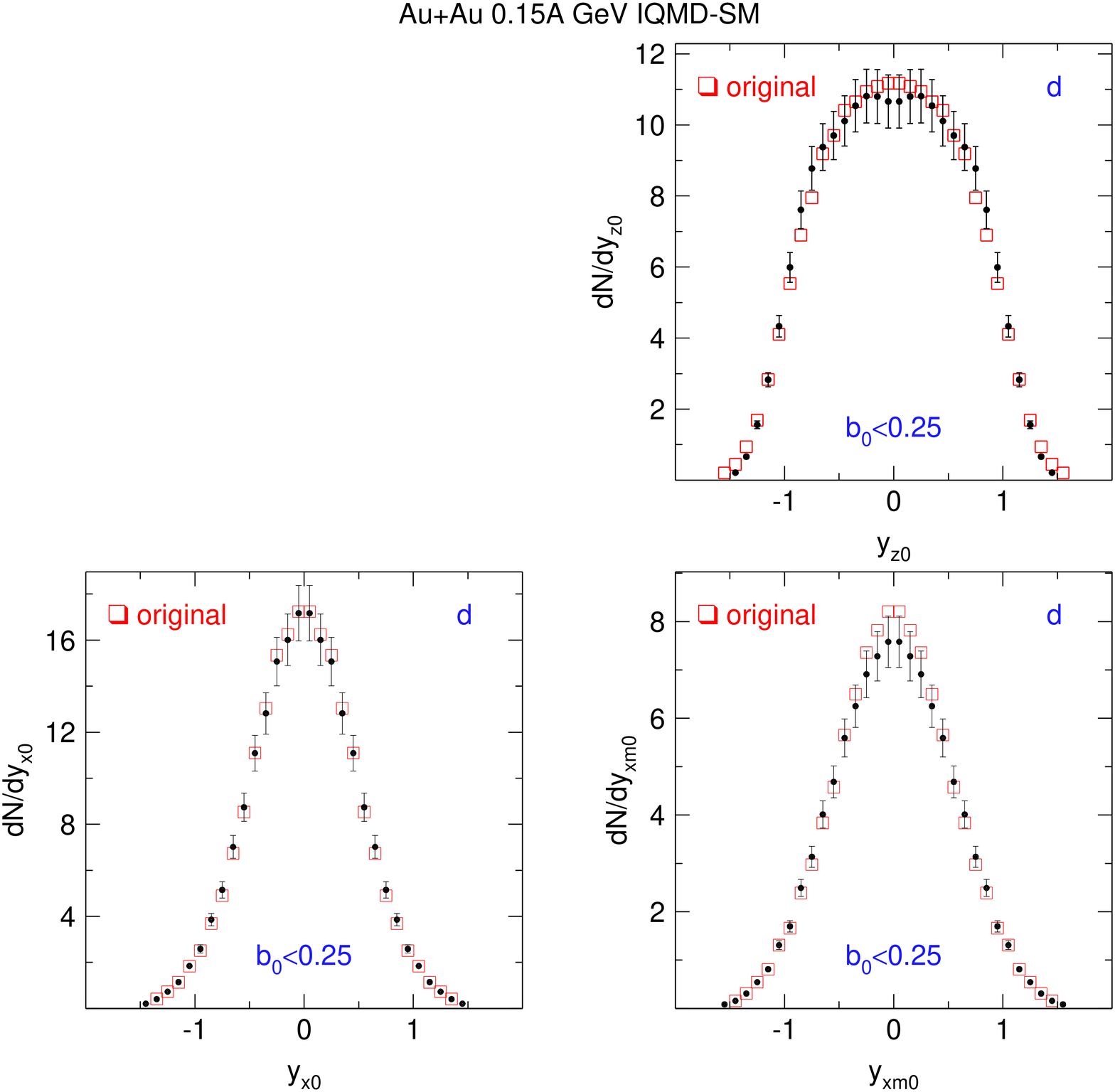,width=140mm}
\caption{
Rapidity distributions of deuterons in Au+Au reactions at
$0.15A$ GeV obtained from IQMD-SM.
The original predictions are given by (red) open squares, the
reconstructed distributions after filtering and then extrapolating
are given by (black) filled dots.
The upper left panel shows the filtered distribution in scaled
momentum versus rapidity space.
}
\label{au150c4Z1A2}
\end{figure}

A case with a typical 'isotope acceptance' (for deuterons here) in the low
 energy runs, where only the CDC was available for isotope separation,
 is shown in Fig.~\ref{au150c4Z1A2}. 
With the sharp filter distorted phase space cells near $y_0=-1$ due
to multiple scattering effects in the target are also cut out.
Due to missing low $p_t$ information and despite the mentioned constraints
from the more complete data on $Z=1$ fragments, the reconstruction has to
allow for about $7\%$ errors indicated by the plotted systematic error bars.
The accumulated counts
 in the experiment were actually 10 times higher than in
the shown IQMD simulation, so that uncertainties originating here also from 
finite statistics were smaller in the experiment.
On the other hand, again, the relatively faithful $4\pi$ reconstruction
is limited to {\it central} collisions, where low $p_t$ spectator material
is not so copious. 
Here the sum-of-isotopes check proved to be mandatory to make sure there
was no 'extrapolation catastrophe'.
Constraints from charge separated data obtained with the ZERO Degree
Detector, covering low transverse momenta, were used for these checks. 

\subsection{Heavy clusters}
\begin{figure}
\epsfig{file=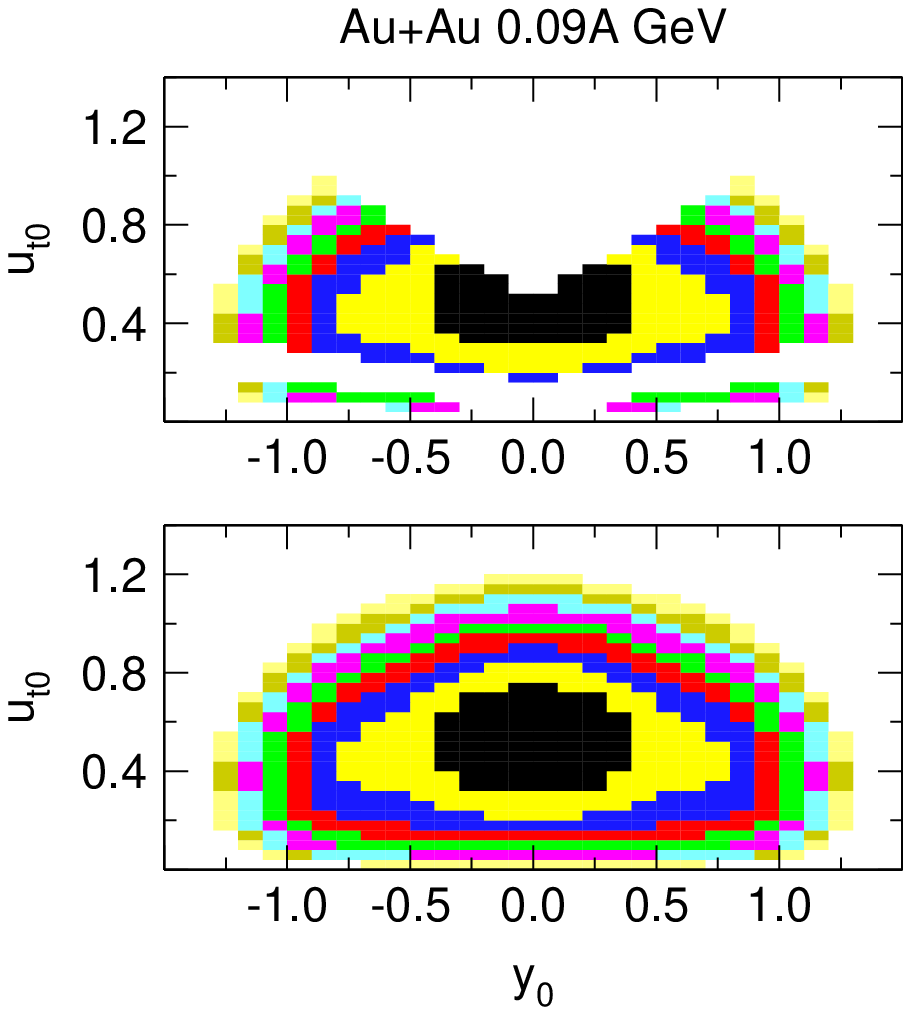,width=90mm}
\hspace{-10mm}
\epsfig{file=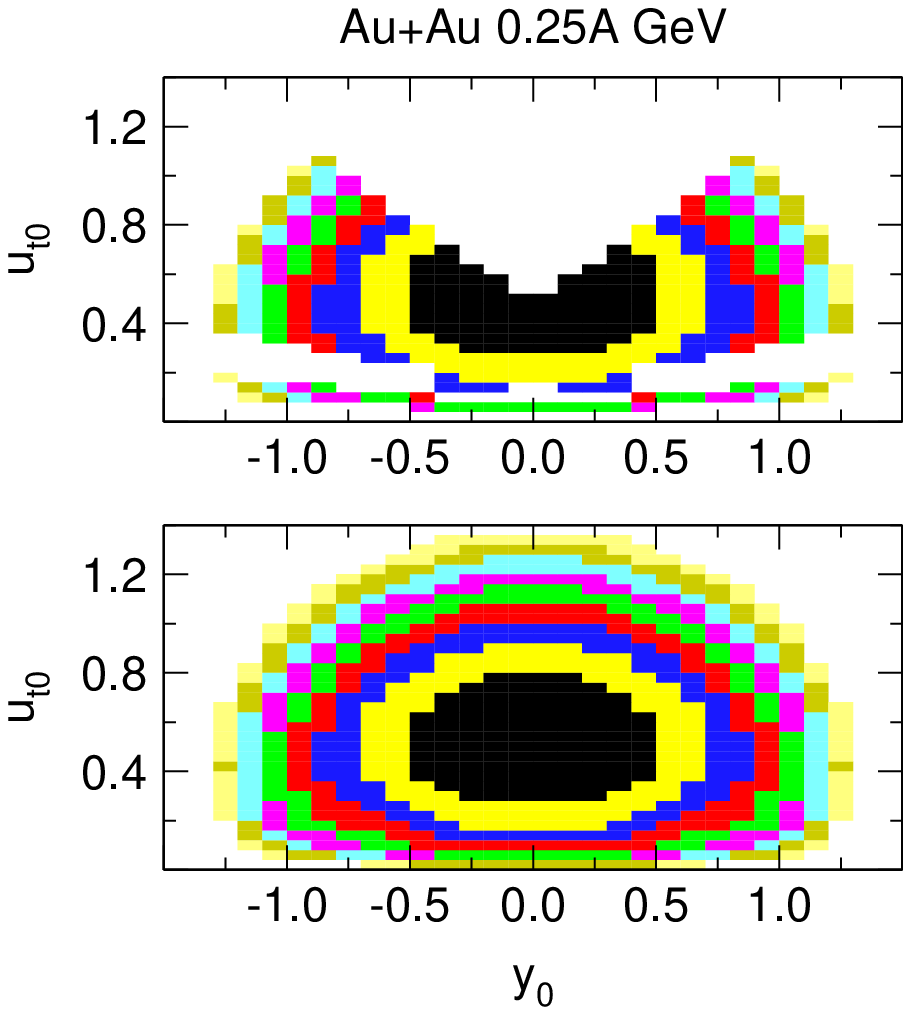,width=90mm}
\caption{
Example of  $4\pi$ reconstruction of the phase space population for
$Z=3$ fragments in central ($b_0<0.15)$ Au+Au collisions 
at two indicated incident energies.
The upper two panels show the measured part after applying some smoothing,
a sharp filter cut and reflection symmetry, the lower two panels show
the result of a two-dimensional extrapolation to $4\pi$ geometry.
}
\label{ecmtcm-au90c1}
\end{figure}

The $4\pi$ reconstruction of fragments with $Z > 2$, that we call here
'heavy clusters' ({\it hc}) is somewhat problematic due to the more
limited acceptance, although the two-dimensional extrapolation procedure is
an improvement over the conventional one-dimensional extrapolation with
separated rapidity bins. The results presented here for {\it hc},
see also ref. \cite{reisdorf04b}, should
be considered to be  estimates  when
applied for example to $90^{\circ}$ spectra. 
In Fig.~\ref{ecmtcm-au90c1} we show an illustrative picture of what it meant
to 'reconstruct to $4\pi$'.
A stringent test using IQMD was not possible here due to limited
statistics available for {\it hc} in the simulation.
However, observables, such as stopping, relying on $4\pi$ coverage,
were found to be in very reasonable agreement with INDRA data
\cite{andronic06}.
As can be seen by inspecting Fig.~\ref{ecmtcm-au90c1},
the increased 'stopping' at an incident energy  of $0.25A$ GeV
as compared to $0.09A$ GeV (larger transverse
momenta relative to the longitudinal ones) is already suggested
{\it before} the extrapolation.

\newpage



\subsection{Adjusting the low-energy runs to high energy runs} \label{adjusting}
\begin{figure}
\epsfig{file=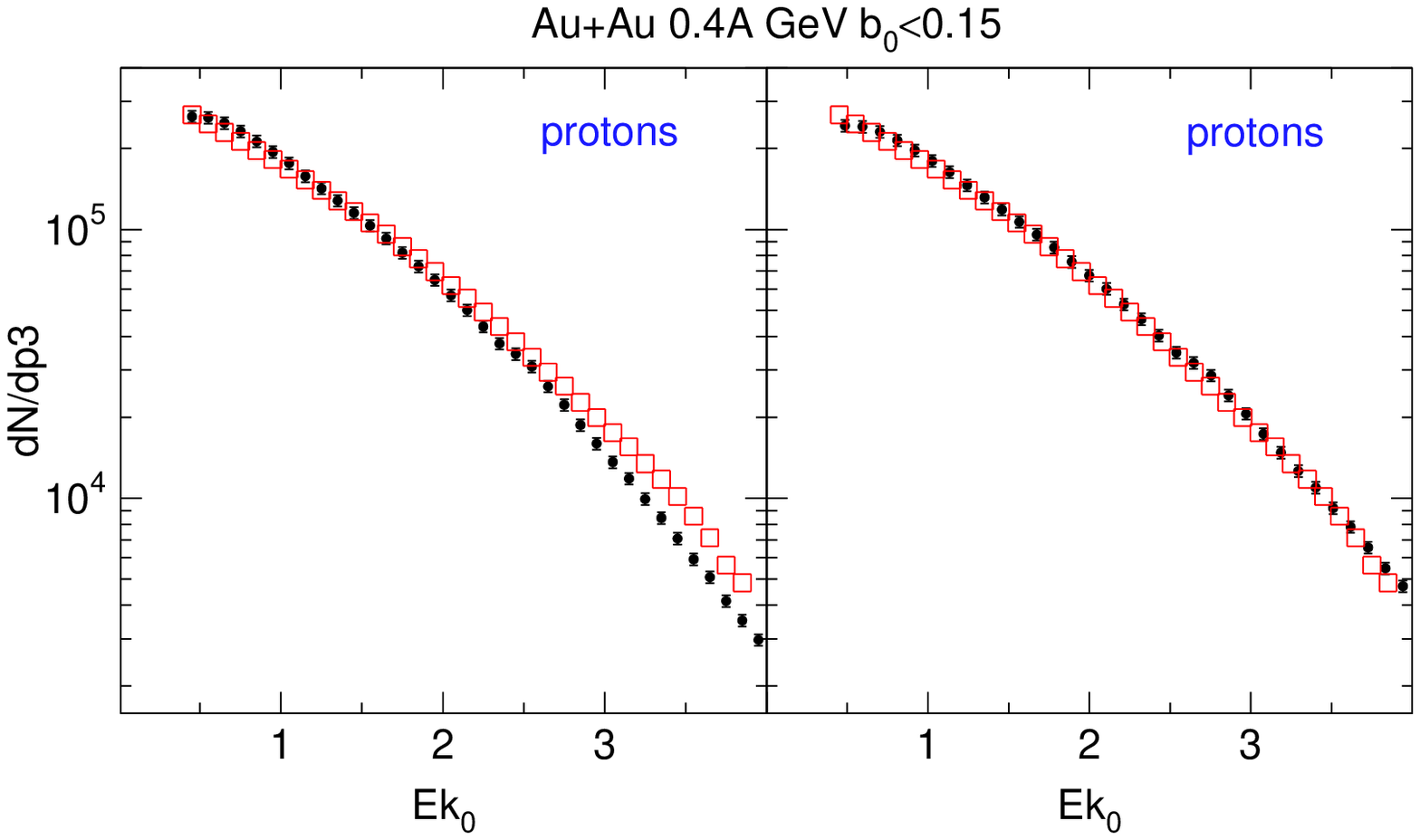,width=140mm}
\caption{
Left panel:
kinetic energies (scaled units) of protons emitted at $90^{\circ}$
in central collisions of Au+Au. Comparison of two
experiments performed in the low energy (full black circles) and the high
energy campaigns (red squares).
In the right panel
the low energy experiment is rescaled along the abscissa by a factor 1.1
 to achieve agreement with the high
energy  experiment.
}
\label{ekin-au400c1Z1A1}
\end{figure}

\begin{figure}
\begin{center}
\epsfig{file=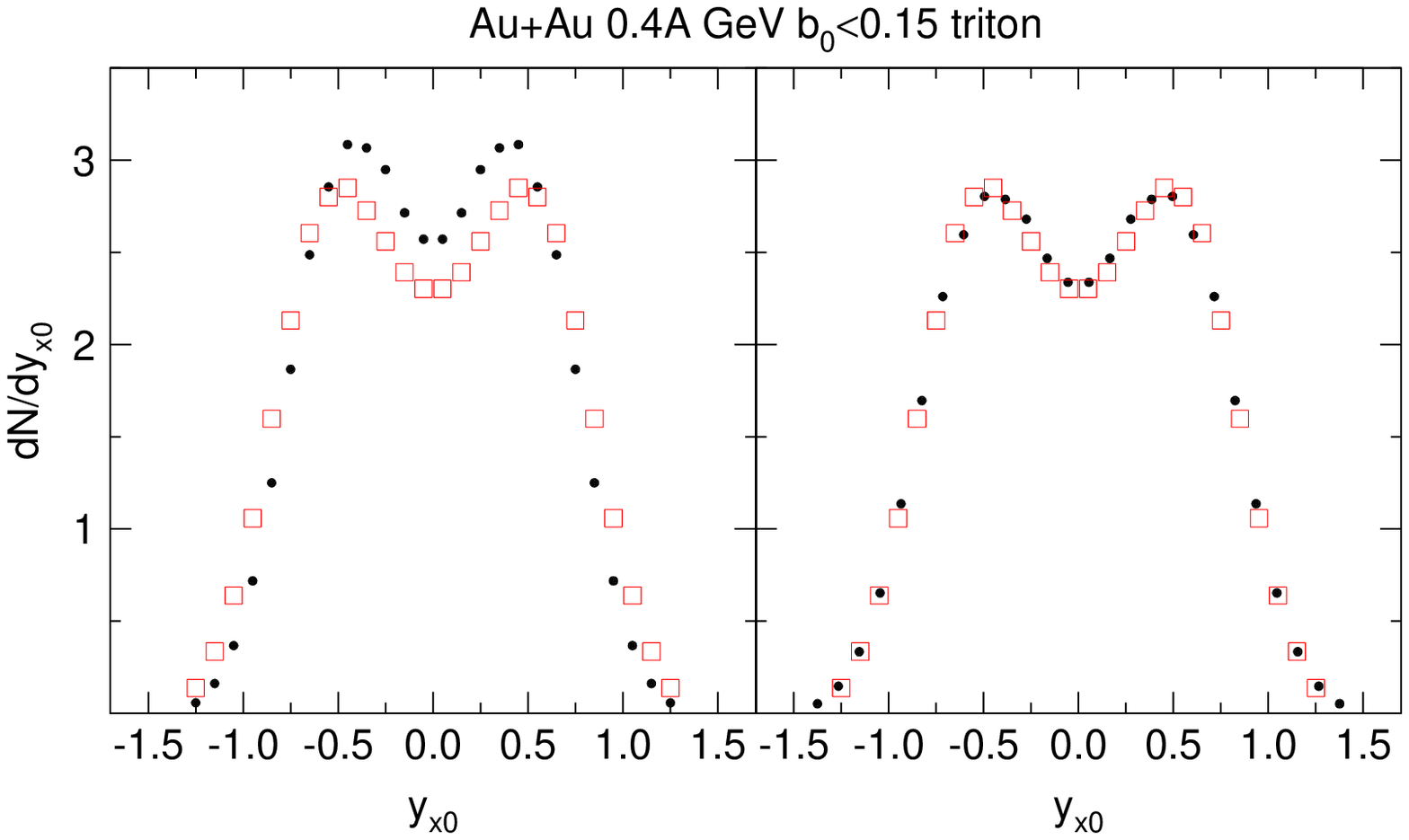,width=105mm}
\end{center}
\caption{
Sharp filtered transverse rapidity distributions for tritons
in central Au+Au collisions at $E/A=0.4A$ GeV.
 Comparison of two
experiments performed in the high energy (open red squares) and the low
energy campaigns (full black circles).
Left panel: original data. Right panel with the low energy data rescaled
 by factor 1.1 along the abscissa.
}
\label{dndy-au400c1Z1A3f-resc}
\end{figure}

As mentioned in section \ref{apparatus}, the CDC was operated in split mode in
 the low energy runs, i.e. with a different voltage, \cite{andronic01}, on the
inner wires.
Despite careful calibration to take this into account, it turned out
by  comparison with later runs made without the split voltage feature,
that an additional correction was needed for the low energy runs.
Fortunately this correction could be shown to involve just a simple rescaling
factor. 
This is illustrated in
 Figs.~\ref{ekin-au400c1Z1A1} and \ref{dndy-au400c1Z1A3f-resc}
 which consist both of two
panels, the left one showing for central Au+Au collisions at $0.4A$ GeV
differences of the two operating modes
 (the earlier one with split voltage) before,
and, in the right panel, after correction with a constant rescaling factor.
Fig.~\ref{ekin-au400c1Z1A1} shows kinetic energy spectra for protons, 
while Fig.~\ref{dndy-au400c1Z1A3f-resc} shows
 transverse rapidity spectra for tritons.
A common sharp filter is applied to the data from both experiments.
The two peak-structure in Fig.~\ref{dndy-au400c1Z1A3f-resc}
is a consequence of the limited acceptance, see
the example in Fig.~\ref{au150c4Z1A2}.
The rescalings in both figures \ref{ekin-au400c1Z1A1}
 and \ref{dndy-au400c1Z1A3f-resc} 
was primarily along the abscissa, the rescaling of the ordinates then
followed from the condition that the integrated yields were not affected.

\begin{figure}
\begin{minipage}{80mm}
\epsfig{file=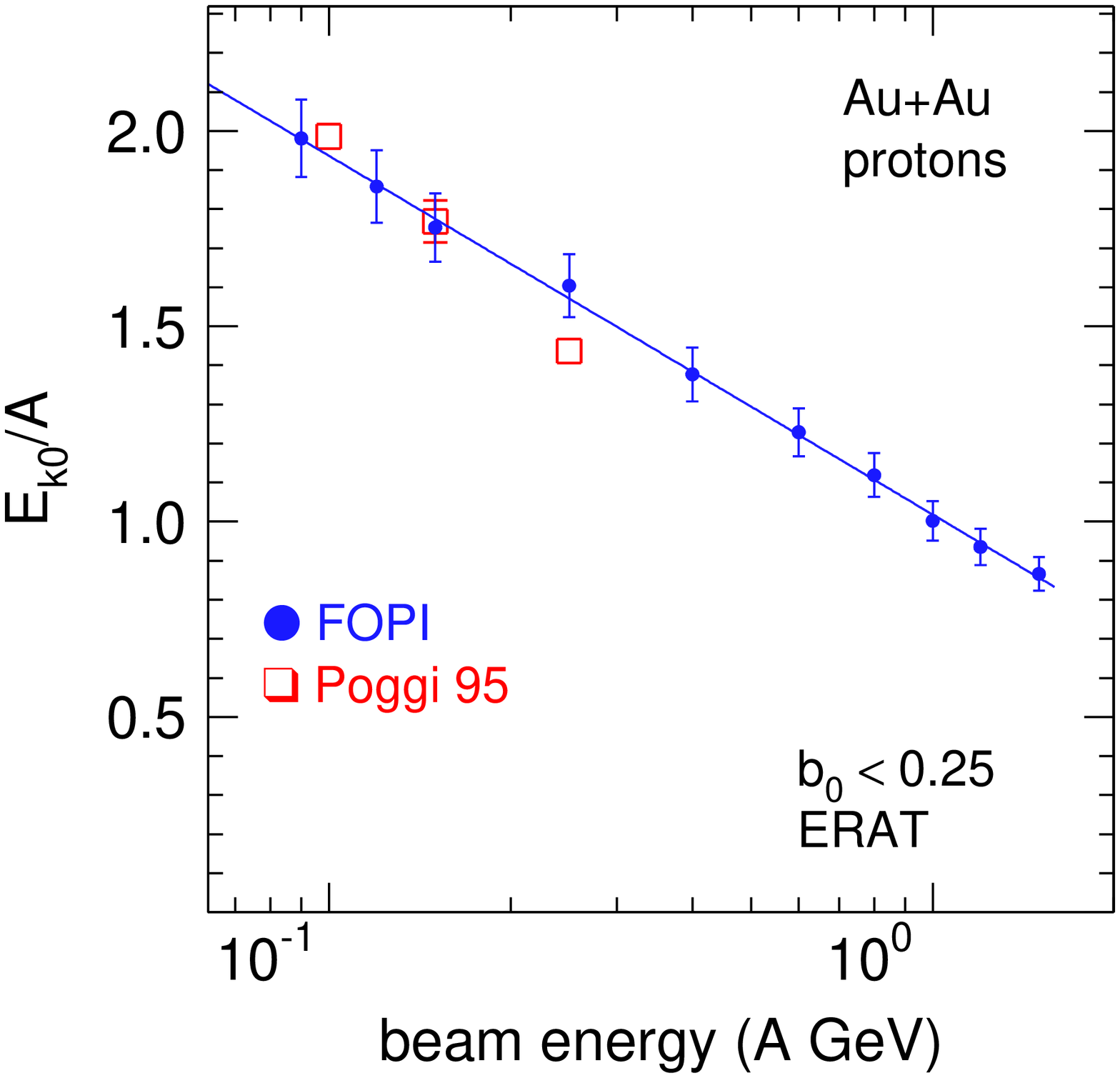,width=76mm}
\end{minipage}
\begin{minipage}{80mm}
\caption{
Average kinetic energy (scaled units) of protons emitted
 around $90^\circ$ (c.m.) in
central collisions of Au+Au as a function of incident beam energy.
The straight line fitted to the combination of the low/high energy data
 (logarithmic abscissa) reproduces, after correction of the low energy data,
the individual points with an average accuracy of $1.2\%$.
The data of \cite{poggi95} (red squares) are also shown.
}
\label{ekinavc4-r-p}
\end{minipage}
\end{figure}

To avoid too many arbitrary parameters, the rescaling factors were fixed
for all the low energy data down to $0.09A$ GeV.
An example of the resulting extremely smooth behaviour of excitation
functions for average kinetic energies 
covering the complete data is shown in Fig.~\ref{ekinavc4-r-p}.
As will be shown later, this surprisingly regular trend is also
predicted by IQMD simulations.
We use scaled energy units here: a value equal to one indicates an
energy equal to the incoming c.m. kinetic energy per nucleon. 
The figure also shows the reasonable agreement with data from ref. 
\cite{poggi95}
which were obtained by our Collaboration using a different method.
While we show here only a comparison for protons for reasons of space economy,
similar conclusions hold for all other isotopes of hydrogen and helium.
 
\begin{figure}
\epsfig{file=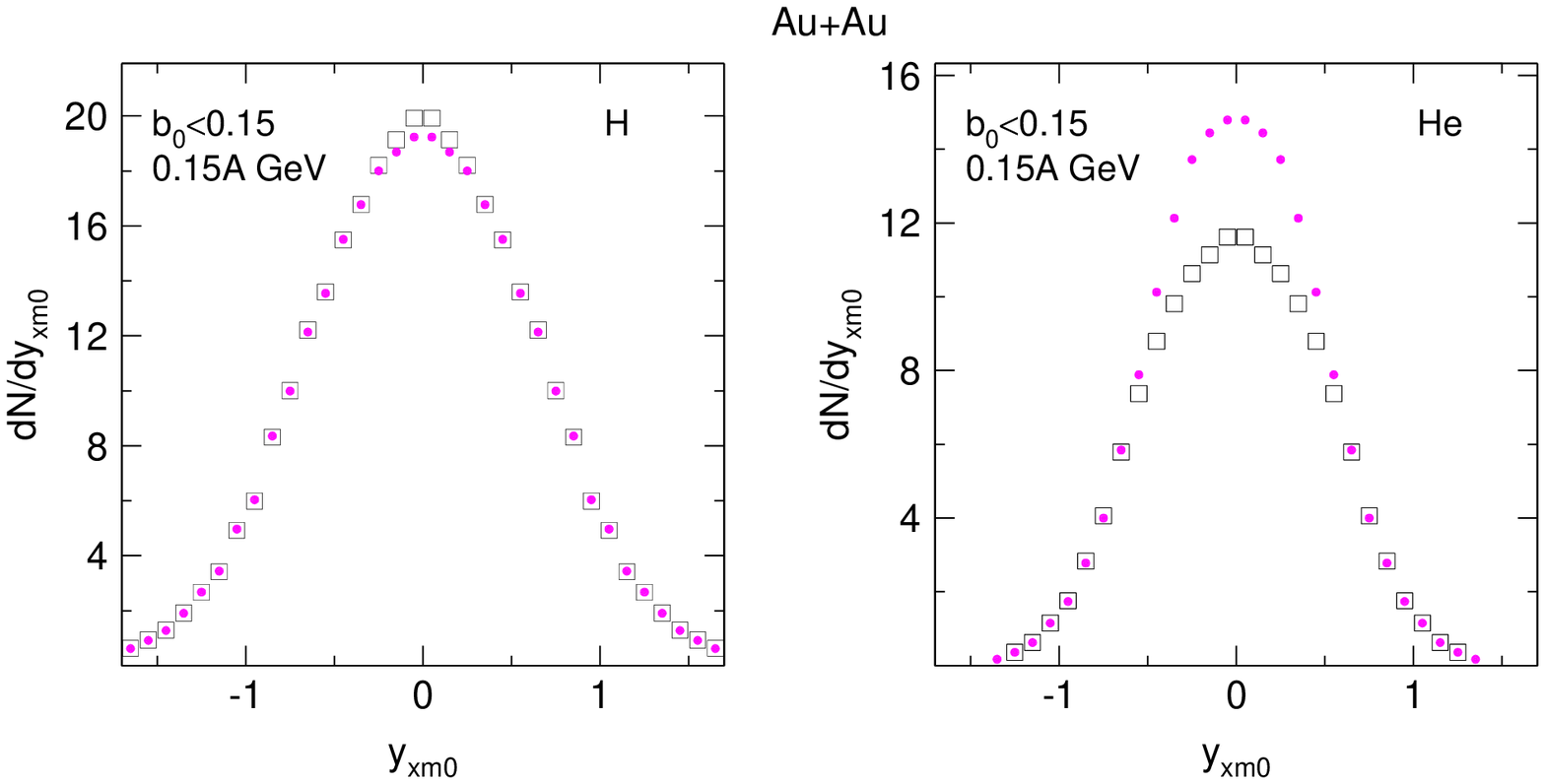,width=140mm}
\caption{
Left panel:
Constrained transverse rapidity distribution of H fragments in
central Au+Au collisions.
(Black) open squares: demanding Z identification only, (pink) closed circles:
sum of identified isotopes (p, d, t).
Right panel: same for He. 
}
\label{dndym-au150c1pdt-r}
\end{figure}



\subsection{Z-renormalization}
By Z-renormalization we address the sum-of-isotopes check and correction after
$4\pi$ reconstruction, which uses the fact, shown earlier, that we had
a more complete acceptance for fragments identified by charge only.
This was most important for the low energy runs.
We illustrate the necessary 'renormalization' in Fig.~\ref{dndym-au150c1pdt-r}
for hydrogen and helium isotopes in central collisions of Au+Au
at $0.15A$ GeV. 
While renormalization is seen to be only a small effect for hydrogen,
the worst case scenario shown for helium requires a significant correction
for low transverse velocities, which can be traced back to errors due
to not accounting properly for Coulomb repulsion effects on low momenta not
covered by our data.


\section{Rapidity distributions}\label{rapidity}
 We present the phase space distributions in a way which
deviates somewhat from the traditional way:
instead of showing standard (longitudinal) rapidity distributions on
a linear ordinate scale and then switching for the transverse direction
to transverse momentum distributions on a logarithmic scale, we show the
transverse degree of freedom in the same representation as the longitudinal
degree of freedom, namely in terms of transverse rapidities plotted on a
linear scale as well (see definitions in section \ref{centrality}).
This allows to compare more directly longitudinal and transverse directions,
a recurrent theme in the present work which is much concerned with
assessing the degree of stopping and equilibration in these complex
 central collisions.
We shall also be interested in the {\it constrained} 
transverse rapidity distributions (defined in section \ref{centrality}) 
which are likely to be closest to thermalized distributions.

 $4\pi$ phase space distributions of five  'light
charged particles' (LCP: p, d, t, $^3$He,$^4$He) have been 
reconstructed using four
different centralities in 25 system energies, i.e. 500 2d-spectra are
available.
This number could be doubled if we consider that two different
selection criteria (ERAT and MUL, see section \ref{centrality})
 were applied throughout.
However, unless otherwise stated, we shall show generally only the ERAT
selected data.
Further, for $E/A \leq 0.4$ GeV we have the data for heavier charge separated
clusters (typically up to Z=8) and for $E/A \geq 0.4$ GeV charged 
pion data of both
polarities (see \cite{reisdorf07}).
It is out of question to present here all this rich information.
Instead we shall present  samples which serve to
illustrate some of the typical aspects of the LCP data and add a few
remarks on heavier clusters and pion data \cite{reisdorf07}.
In later sections some of these aspects will be summarized in terms
of simple concepts:  'stopping', 'radial flow' and 'chemistry'.
These concepts reflect the time sequence of the central reactions.
Earlier work on these aspects will be referred to.

Some of the rapidity distributions will be compared with distributions
expected from a Boltzmann thermal model.
 These thermal distributions are generated 
 varying T
until the variance of the constrained transverse rapidity distribution
(i.e. with the scaled longitudinal distribution constrained by $|y_{z0}| < 0.5$
and within the $y_{xm0}$ range shown in the respective panels)
is reproduced (the same constraints are used on the thermal simulation).
The obtained T will be dubbed 'equivalent' temperatures, $T_{eq}$. 
The longitudinal constraint leads  to a cutoff of
part of the yields, a cutoff which is particle-mass and T dependent.
This cut is corrected for when extracting so called effective
'mid-rapidity yields' from the experimental data by comparison to the
thermal model (see section \ref{chemistry}).


\subsection{Light charged particles (LCP)} \label{LCP}
\begin{figure}
\epsfig{file=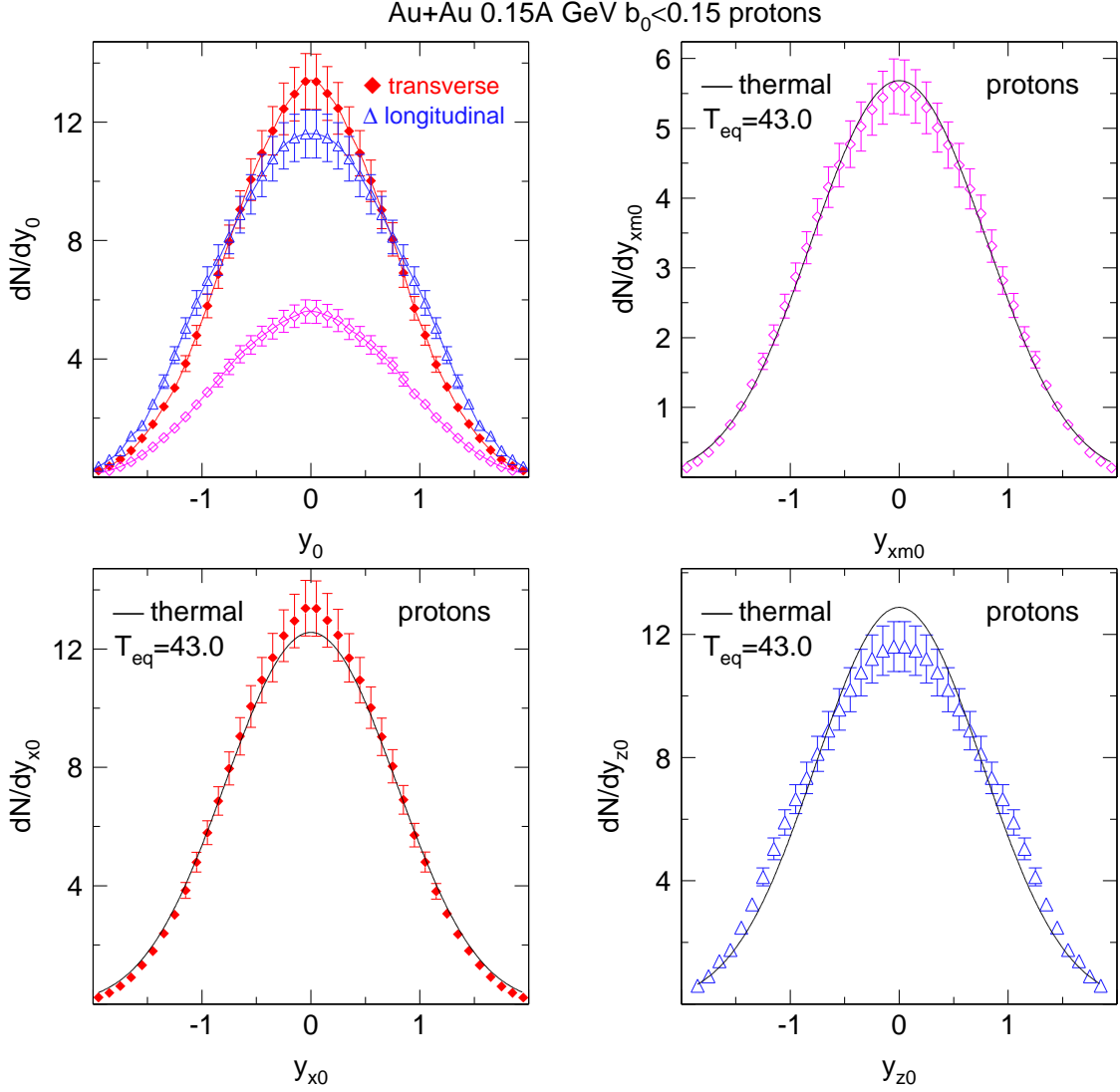,width=160mm}
%
\caption{
Upper left panel: Longitudinal (blue open triangles) and transverse
rapidity distributions (red full diamonds) of protons in central
collisions of Au+Au at $0.15A$ GeV beam energy.
The third, lower yield curve (magenta open diamonds) in the panel represents
the 'constrained' transverse rapidity distribution: here
a cut $|y_{z0}|<0.5$ on the longitudinal rapidity is applied.
The smooth curves are just guides for the eye here.
In the other three panels the three kinds of distributions
 are compared to a thermal distribution (corresponding to an 
 equivalent temperature of 43 MeV)
having the same area as the respective data, but a variance as implied by
 the constrained transverse distribution (seen in upper right panel).
}
\label{dndy-au150c1Z1A1th}
\end{figure}

\begin{figure}
\hspace*{\fill}
\epsfig{file=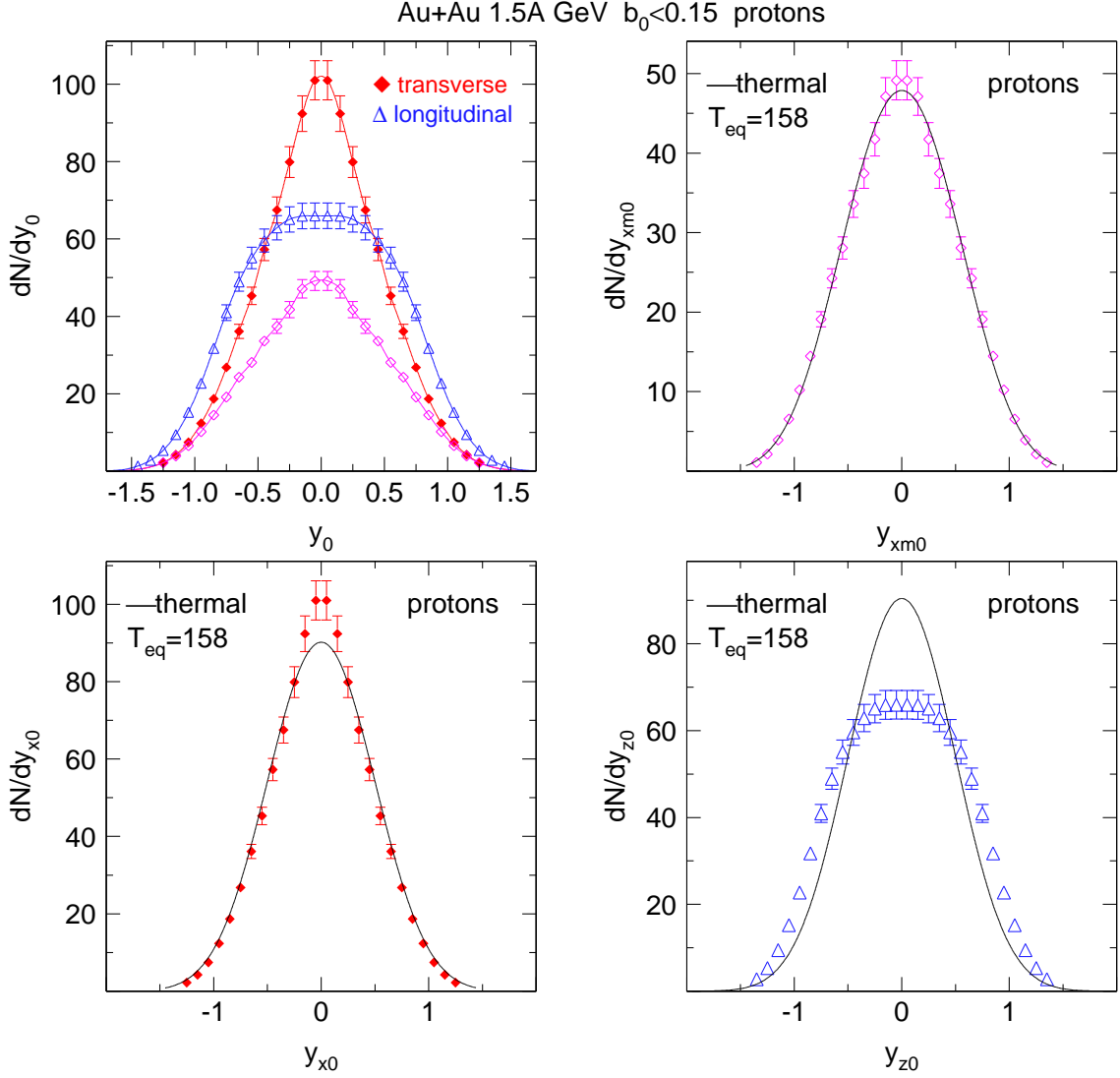,width=160mm}
%
\caption{
Upper left panel: Longitudinal (blue open triangles) and transverse
rapidity distributions (red full diamonds) of protons in central
collisions of Au+Au at $1.5A$ GeV beam energy.
The 'constrained' transverse rapidity distribution is also shown: here
a cut $|y_{z0}|<0.5$ on the longitudinal rapidity is applied.
The data points are joined by smooth curves to guide the eye.
In the other three panels the three kinds of experimental distributions
 are compared to a thermal distribution (smooth black curves)
having the same area as the respective data and a variance as implied by the
 constrained transverse distribution (seen in upper right panel).
}
\label{dndy-au1500c1Z1A1}
\end{figure}

The next two 4-panel figures, Figs.~\ref{dndy-au150c1Z1A1th}, 
\ref{dndy-au1500c1Z1A1},
 show the three kinds of rapidity distributions
for protons emitted in the most central Au+Au collisions at two incident
energies differing by one order of magnitude: $0.15A$ GeV and $1.5A$ GeV.

The upper-left panels in the two figures show that the longitudinal
rapidity distributions are broader than the transverse distributions, but while
the effect is relatively modest at the lower energy, it is more conspicuous
at the higher energy. 
The constrained transverse distributions are generally somewhat broader than
the integrated transverse distributions and can be  reproduced rather well
 by a
thermal model simulation having an adjusted 'equivalent' temperature,
$T_{eq}$, and the same total area. 
We find $T_{eq}=43$ MeV, and $158$ MeV, respectively,
 for protons at the two energies.
This $T_{eq}$ comes closest to the 'inverse slopes' usually fitted to
$p_t$ spectra in the literature, but is different in the sense that it is
obtained by just demanding a reproduction of the bulk experimental variances
taken in the shown rapidity ranges and hence excluding far out tails of
the distributions.
On the {\it linear} scales applied in the figures one finds a very satisfactory
reproduction of the shape of the $dN/dy_{xm0}$ distributions (upper right
panels).
A comparison to the equivalent thermal representations fixed from
the constrained data, to the full transverse and longitudinal distributions
 in the lower panels
stresses the differences to the naive thermal expectations.
The full transverse distributions are modestly narrower only, but the
longitudinal distributions clearly deviate.

\begin{figure}
\hspace{\fill}
\epsfig{file=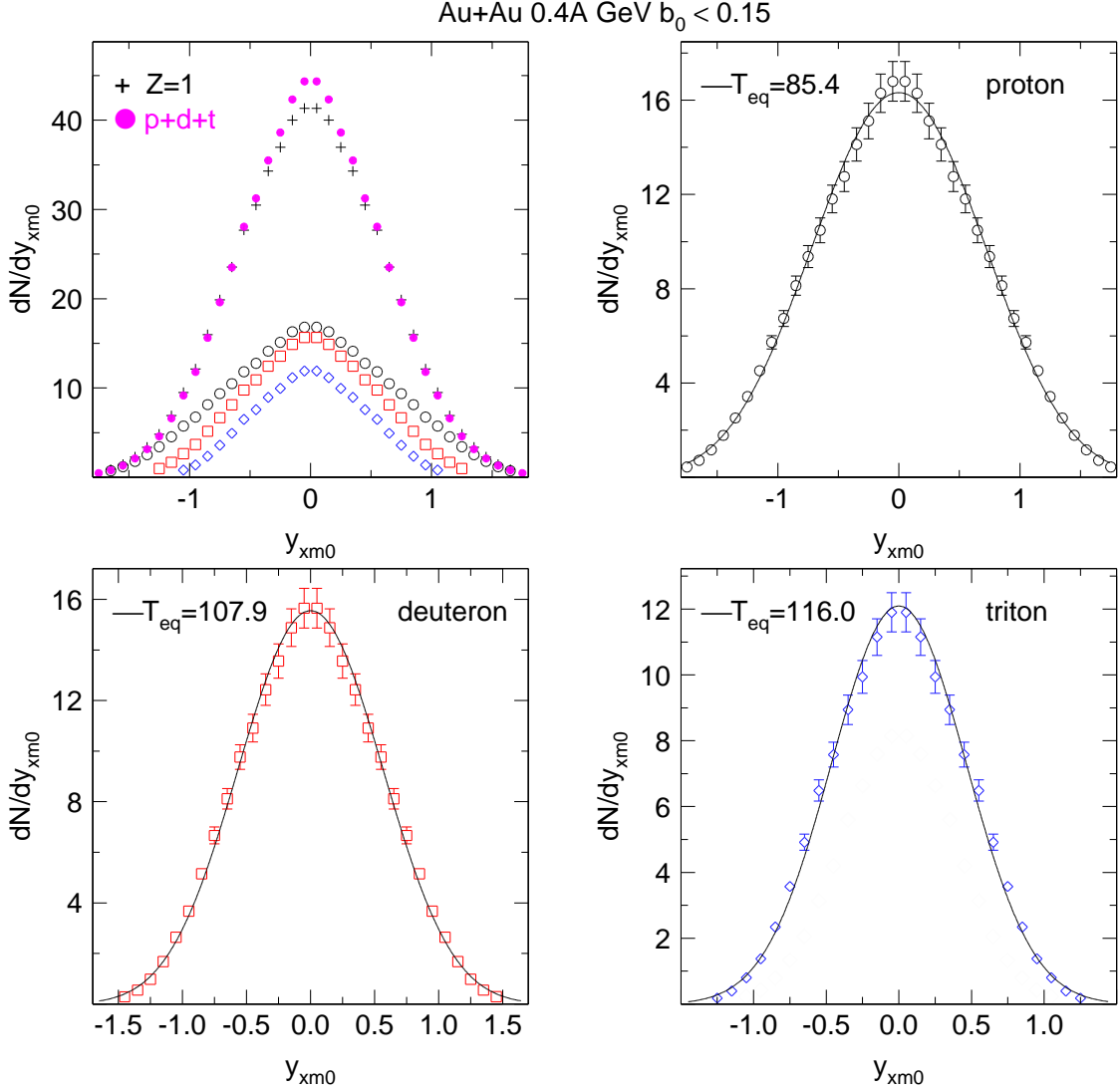,width=160mm}
\hspace{\fill}
%
\caption{
Central ($b_0 < 0.15$) Au+Au collisions at 0.4A GeV.
Comparison of experimental scaled and constrained transverse rapidity
 distributions
 $dN/dy_{xm0}$ of protons (upper right), deuterons (lower left)
and tritons (lower right)
with thermal distributions (smooth black curves) assuming the
indicated equivalent temperatures $T_{eq}$.
The upper left panel shows the distributions of the three H isotopes,
their sum (pink full dots) and the distribution for Z=1 (excluding
created particles) obtained requiring only charge identification.
}
\label{dndym-au400c1pdt}
\end{figure}

\begin{figure}
\hspace{\fill}
\epsfig{file=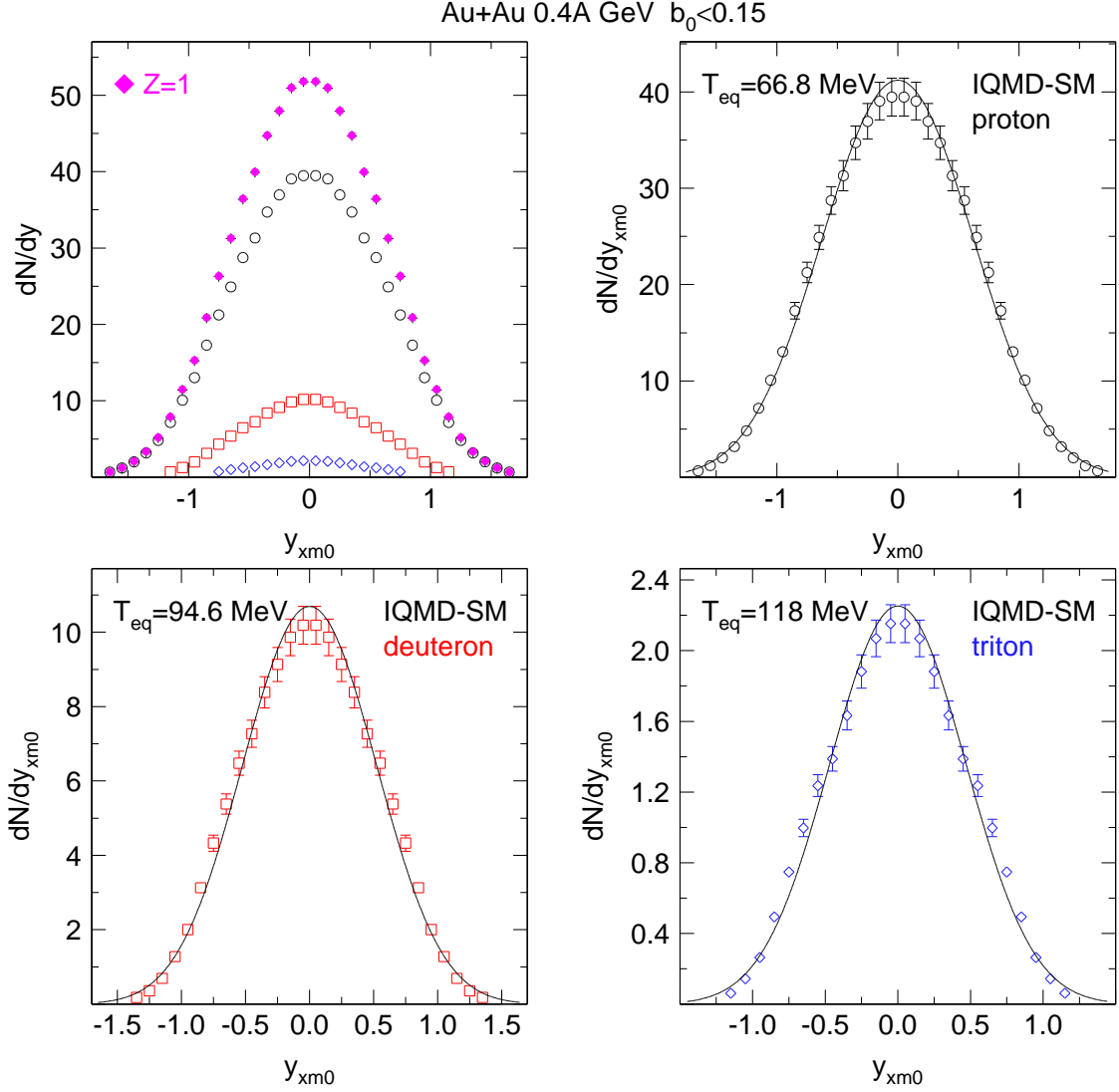,width=160mm}
\hspace{\fill}
%
\caption{
Central ($b_0 < 0.15$) Au+Au collisions at 0.4A GeV.
Comparison of simulated (IQMD-SM) scaled and constrained transverse rapidity
 distributions
 $dN/dy_{xm0}$ of protons (upper right), deuterons (lower left)
and tritons (lower right)
with thermal distributions (smooth black curves) assuming the
indicated equivalent temperatures $T_{eq}$.
}
\label{dndym-au400c1smpdt-a}
\end{figure}

The next two 4-panel figures,
Figs.~\ref{dndym-au400c1pdt}, \ref{dndym-au400c1smpdt-a},
 take a closer look at constrained transverse
rapidities varying the (hydrogen) isotope mass. 
The system is Au+Au at $0.4A$ GeV with $b_0<0.15$. 
First we show the data, then a simulation with IQMD. 
In each case we make in the upper left panel the sum-of-isotopes check
 (which is fulfilled by definition in the simulation)
 and we compare in the other panels  with a thermal model calculation
adjusting $T_{eq}$.
These one-shape parameter fits are close to perfect, but require a rising
equivalent temperature with the isotope mass, a well known phenomenon
often interpreted as 'radial flow' in the literature that we shall come back
to in section \ref{radial}. 
Clearly there is no global equilibrium, but at very best a 'local' equilibrium
(in the hydrodynamic sense).

A closer look at the simulated data 
reveals small but systematic shape differences that exceed
those found when using experimental data instead of the IQMD events.
One could argue that the experimental data look more 'thermal'.
Note the rather strong decrease of yields with isotope mass in contrast
to the more comparable yields in the experiment shown in the previous figure.
Also, if one were to associate, naively, the strength of radial flow with the
mass dependence of $T_{eq}$, then one would conclude that the simulation
overestimates the flow somewhat, since, as indicated in the various panels
the $T_{eq}$ rise from 85 MeV for protons to 116 MeV for tritons in the data,
while for the simulation we find 67 MeV and 118 MeV, respectively. 
For further discussions of the $T_{eq}$ values, see later 
section \ref{radial}.

\begin{figure}
\hspace{\fill}
\epsfig{file=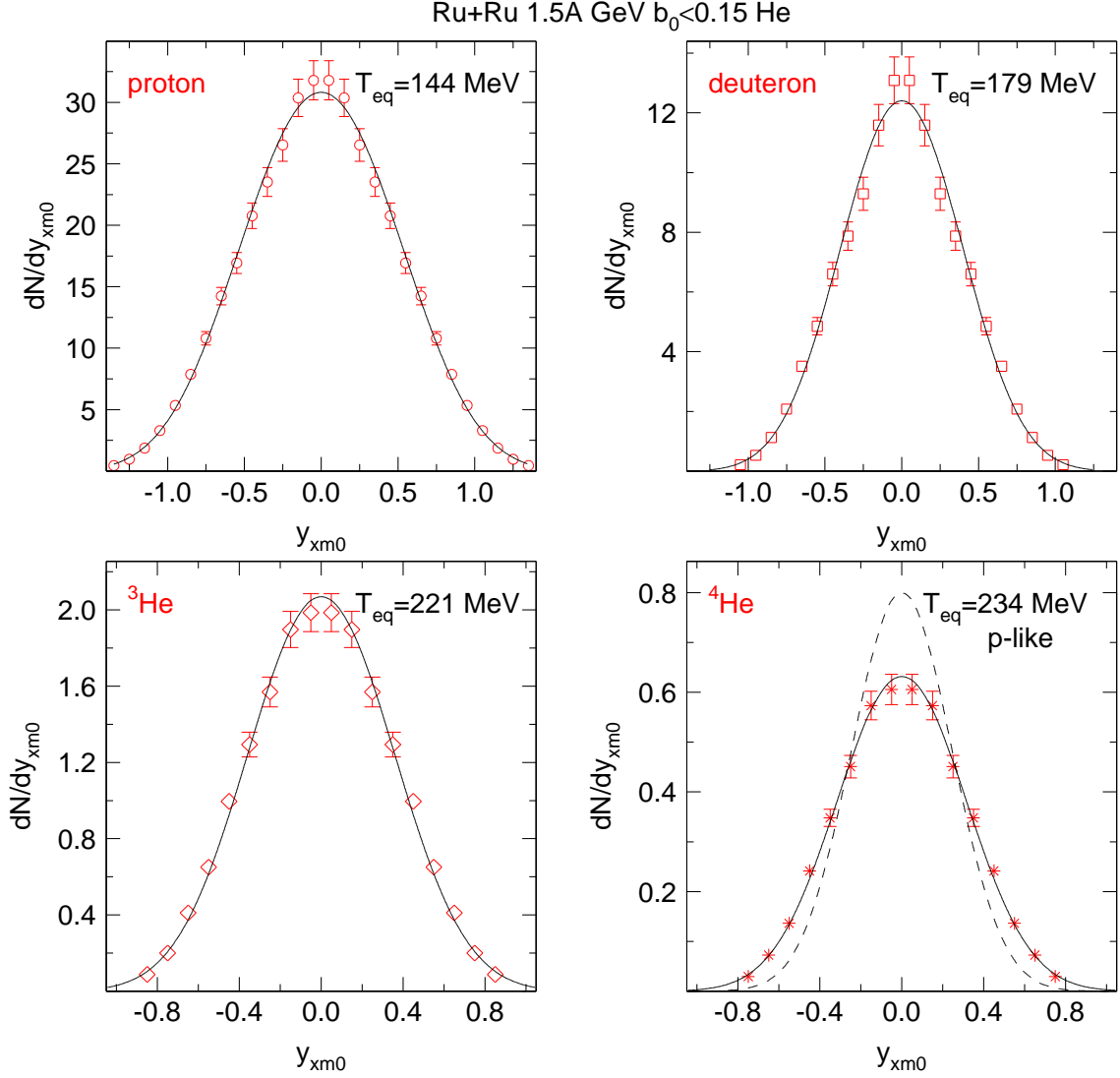,width=160mm}
\hspace{\fill}
%
\caption{
Comparison of experimental constrained transverse rapidity distributions
$dN/dy_{xm0}$ with thermal distributions (black smooth curves) 
having the same first and second moments in central collisions of
Ru+Ru at $1.5A$ GeV.
The various ejectiles and the equivalent temperatures $T_{eq}$ are indicated.
In the lower right panel we also plotted the $^4$He distribution
(dashed smooth curve, 'p-like') expected if the equivalent temperature was
the same as for protons, but with the integrated yields unchanged.
}
\label{dndym-ruru1500c1he-b}
\end{figure}

The Fig.~\ref{dndym-ruru1500c1he-b} showing constrained transverse rapidity
distributions for central collisions in the system Ru+Ru at 1.5A GeV
serves two purposes:
 1) besides hydrogen isotopes (p and d) it also shows data
for $^3$He and $^4He$ in the lower panels.
Again thermal model calculations with the indicated $T_{eq}$ are shown along
with the data;
 2) We show that the $^4$He data are not well represented by a
coalescence model using the measured proton data as generating
spectrum: see the dashed curve.
Indeed the observed rising equivalent temperatures with mass contradict
 the naive model.
This is at variance with ref. \cite{swang95} where the validity of the power law
for a wide range of incident energies and centralities in Au+Au systems
was stressed.  The authors mentioned that a cutoff of  $p_t/A$ smaller than
0.2 GeV/c was required and that the single ('free') proton spectra had
 to be used, rather than the total spectra including those bound in clusters
as one would naively expect. 
 Non-perturbative features of clusterization are suggested
strongly by our data, as we shall show later in this section
and in section \ref{chemistry}.

\begin{figure}
\hspace{\fill}
\epsfig{file=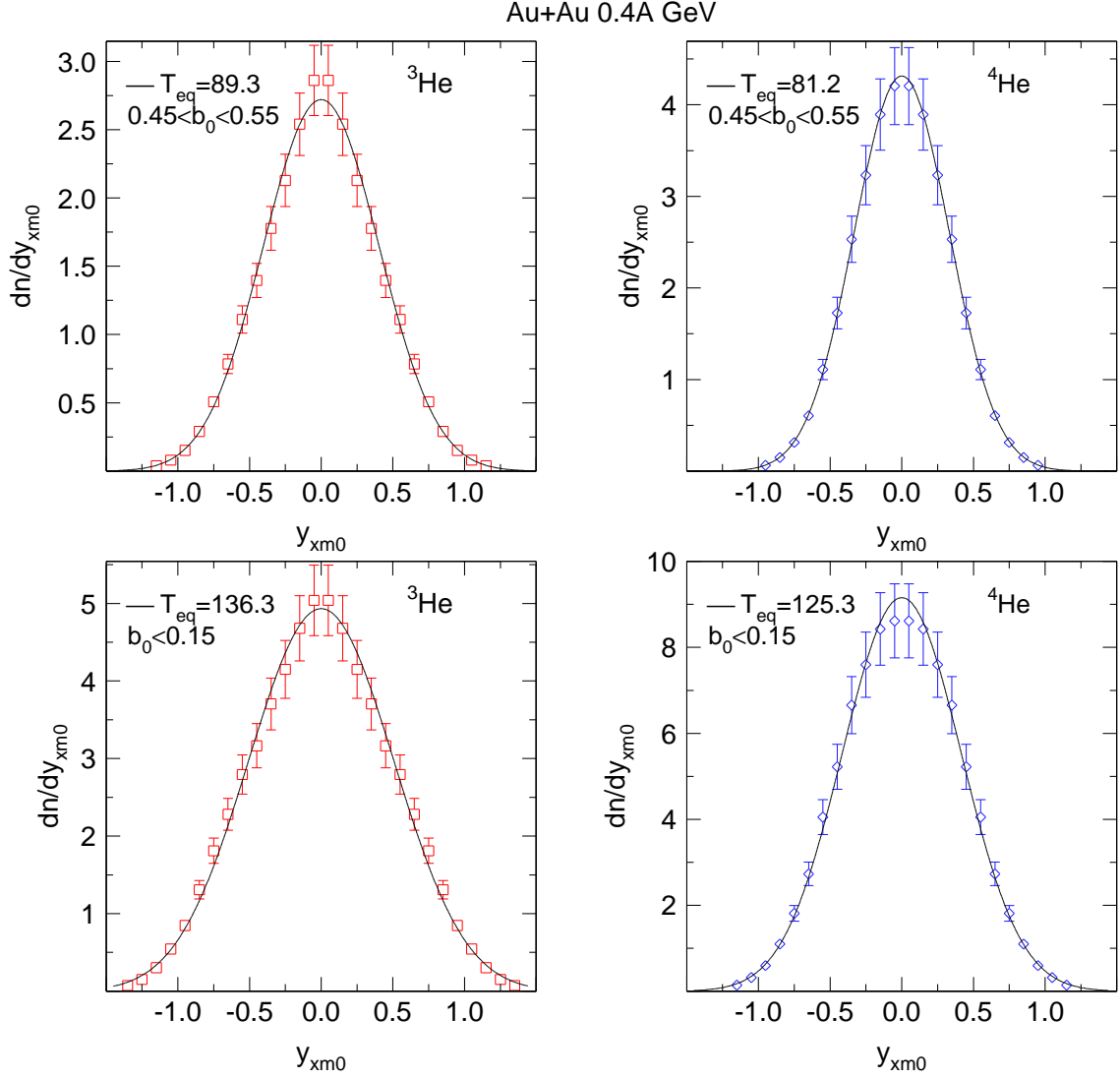,width=160mm}
\hspace{\fill}
%
\caption{
Central ($b_0 < 0.15$) (bottom) and semi-central (top)
 Au+Au collisions at 0.4A GeV.
Comparison of experimental scaled and constrained transverse rapidity
 distributions
 $dN/dy_{xm0}$ of $^3$He  (left) and $^4$He (right) fragments
with thermal distributions (smooth black curves) assuming the
indicated equivalent temperatures $T_{eq}$.
}
\label{dndym-au400c1he}
\end{figure}

In the cases shown so far we observe a regular rise of $T_{eq}$ with the
mass of the ejectile.
However, frequently in the literature, the so-called 'helium anomaly' is
mentioned \cite{neubert00}, namely the observation that $^3$He kinetic energies
 are
not lower, but higher than the $^4$He energies, a phenomenon that to
our knowledge has not been reproduced by {\it microscopic} dynamic
 reaction models.  
In Fig.~\ref{dndym-au400c1he} we show for Au+Au that even at the
relatively high incident energy of $E/A=0.4$ GeV this phenomenon still
subsists to some degree as the $T_{eq}$ for $^4$He are {\it not} found to
exceed those of $^3$He.
Further we show in the panels the rather strong drop of $T_{eq}$ with decreasing
centrality, which is varied from $b_0<0.15$ (lower panels) to $0.45<b_0<0.55$
(upper panels).
This is a general feature also observed at other energies and was reported
earlier \cite{lisa95}.

\subsection{The influence of clusterization} \label{clusterization}
The failure of the naive coalescence approach consisting in trying to
reproduce the spectra of heavier clusters from the {\it measured}
nucleon (proton) spectra (Fig.~\ref{dndym-ruru1500c1he-b}) and
some of the 'He-anomaly' discussed before may be connected with
our current failure to understand clusterization quantitatively on a
 microscopic level (compare also the upper right panels of 
Figs.~\ref{dndym-au400c1pdt} and
\ref{dndym-au400c1smpdt-a}).
Concerning the probability of nucleons to cluster,
 we are at SIS in a non-perturbative
regime (see section \ref{chemistry}): heavier cluster formation has a 
back-influence on the lighter generating transverse rapidity spectra.

\begin{figure}
\begin{minipage}{80mm}
\begin{center}
\epsfig{file=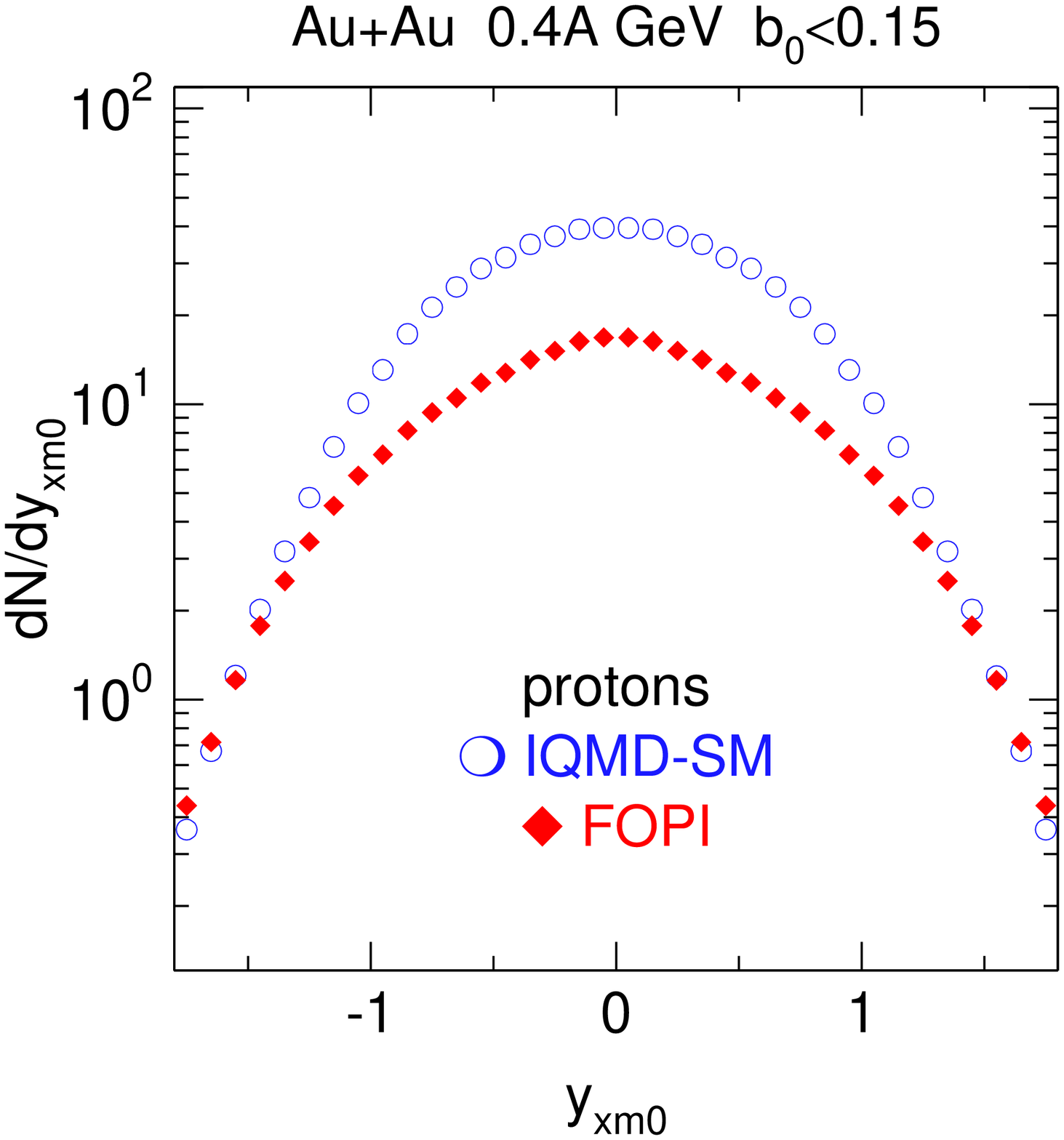,width=77mm}
\end{center}
\end{minipage}
\begin{minipage}{80mm}
\begin{center}
\epsfig{file=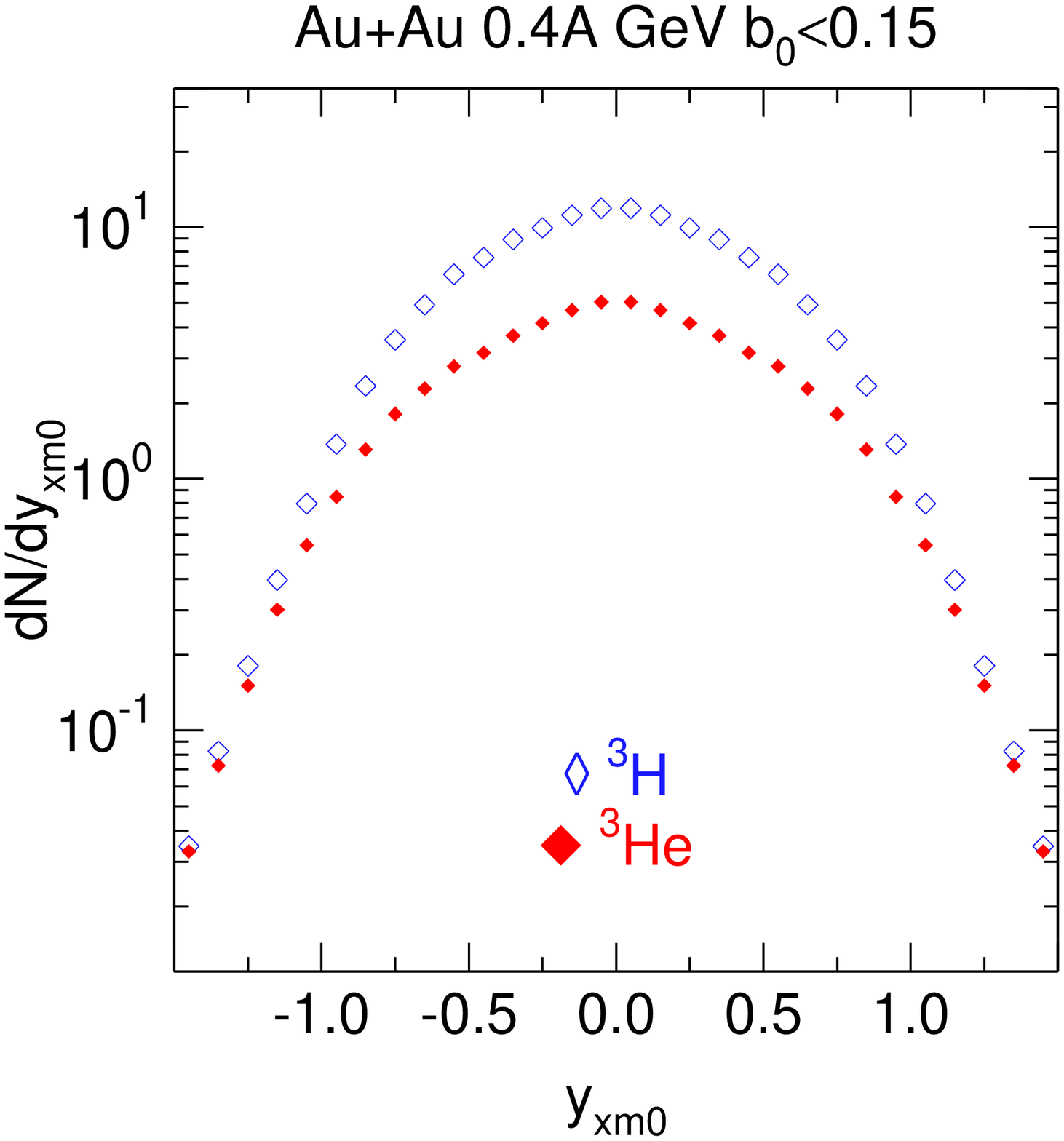,width=77mm}
\end{center}
\end{minipage}
\caption{
Left panel: Constrained transverse rapidity distributions of protons in
central collisions of Au+Au. Comparison of data (red closed diamonds) and
IQMD simulation (blue open circles).
Right panel: comparison of experimental data for $^3$He (red closed diamonds)
 and $^3$H (blue open diamonds) in the  same reaction.
}
\label{dndym-au400c1p}
\end{figure}

The  two-panel Fig.~\ref{dndym-au400c1p} illustrates the non-perturbative 
features at SIS.
In the left panel we show  the remarkable difference
between FOPI and IQMD for the transverse rapidity distributions of protons.
The surplus of IQMD for lower transverse velocities or momenta is due to
a lack of sufficient clusterization: in the experiment more copious
cluster formation massively depletes the {\it low} momenta.
The right panel compares two experimental distributions:
now it is the $^3$He that appears to have its low momenta depleted
relative to $^3$H
(the naive perturbative coalescence model does not predict any difference here).
In view of the finding of the left panel, it is tempting to
associate the effect to the formation of heavier clusters from the
  'primeval' $^3$He (created in earlier expansion stages).

\begin{figure}
\begin{minipage}{80mm}
\begin{center}
\epsfig{file=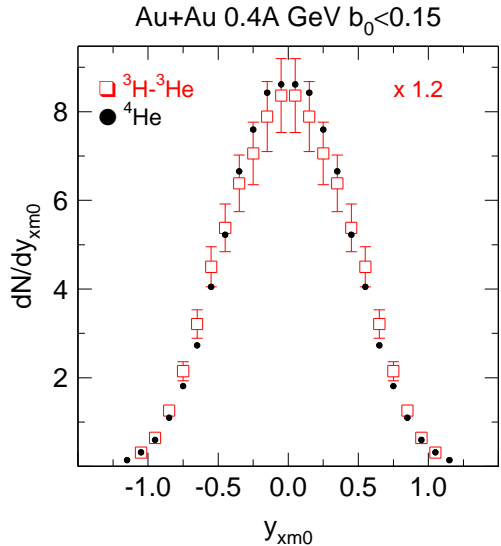,width=77mm}
\end{center}
\end{minipage}
\begin{minipage}{80mm}
\caption{
Comparison of the difference spectrum (see right panel previous figure)
$^3$H-$^3$He (rescaled by a factor 1.2) to the $^4$He (black dots) 
transverse rapidity spectrum.
}
\label{dndym-au400c1A3-a}
\end{minipage}
\end{figure}

\vspace{4mm}

This conjecture is supported by Fig.~\ref{dndym-au400c1A3-a} which shows the
$^3$H-$^3$He difference spectrum together with data for $^4$He.
The $^3$H and the $^3$He compete to be a condensation nucleus to a possible
$^4$He.
If both mass 3 isotopes are in a neutron-rich environment, the $^3$He will
'win' for two reasons:\\ 
a) it is easier to 'find' a single neutron to attach to $^3$He than a
single proton to attach to $^3$H;\\
b) in contrast to $^3$H, the $^3$He nucleus does not Coulomb-repulse
its needed  partner.\\
A quantitative transport model theory must include the formation of $\alpha$
clusters if these conjectures are correct.

\subsection{How to define stopping: {\it varxz}} \label{varxz}

\begin{figure}
\epsfig{file=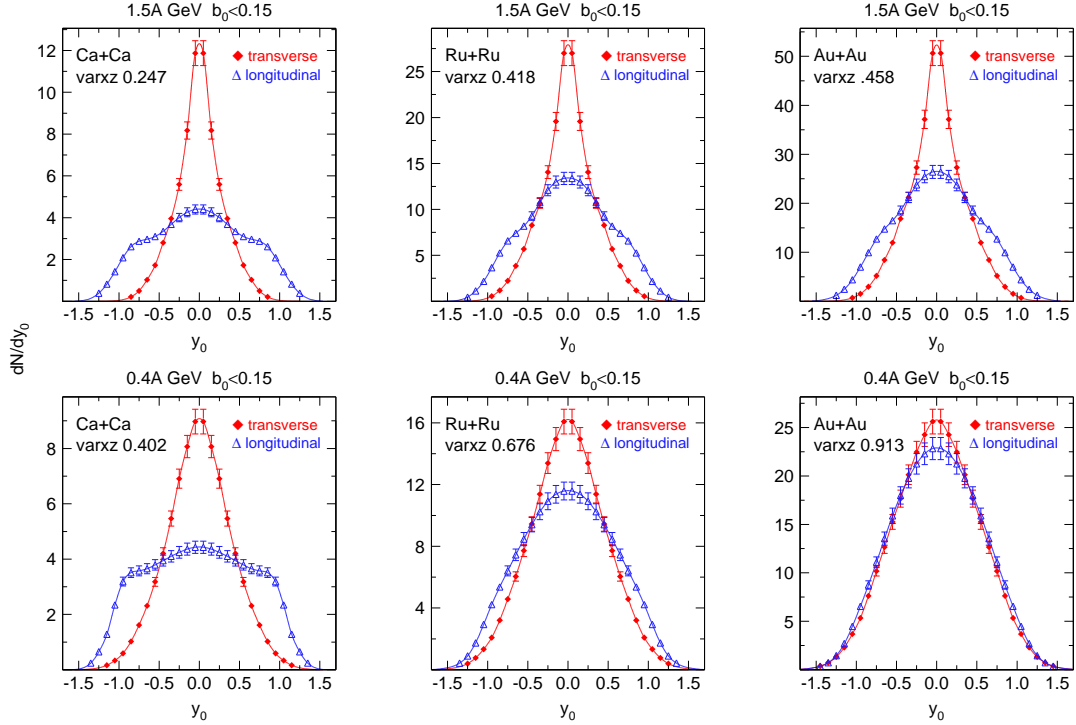,width=150mm}
\caption{
Comparison of scaled rapidity distributions for deuterons
in central collisions at
$E/A=1.5$ GeV (top panels) and 0.4 GeV (bottom panels):
shown are data for
Ca+Ca, Ru+Ru and Au+Au (from left to right).
In each panel,
the scaled longitudinal rapidity $y_{z0}$ distributions (blue open triangles)
are shown together
with the scaled transverse rapidity $y_{x0}$ distributions (red).
The derived value of the stopping observable {\it varxz} is also
indicated.
}
\label{dndy-400c1Z1A2}
\end{figure}

The final figure in this section on rapidity distributions,
Fig.~\ref{dndy-400c1Z1A2}, serves to
illustrate what we shall call 'partial transparency' or 'incomplete stopping'.
Transparency increases as either the studied
system mass is lowered, or the incident energy is raised by going from
$E/A=0.4$ GeV (lower panels) to 1.5 GeV (upper panels).
With this figure, which shows data for the most central collisions
($b_0<0.15$ or $2.2\%$ of the equivalent sharp cross section),
 we argue that a good measure of the degree of stopping
consists in evaluating the ratio {\it varxz} of variances {\it varx} and
{\it varz} of the
transverse relative to the longitudinal rapidity distribution, respectively,
 both
of which are shown in the figure for deuterons.
Note also the subtle evolution of shapes, especially at the
higher energy: there is no trivial subdivision into 
'participants' or 'spectators'  possible.
The interpretation in terms of two remnant counterflowing
 (not completely stopped) 'fluids' is strongly supported  by isospin tracer
methods (see section \ref{stopping}) and suggests that not only global,
but also {\it local} equilibrium is not achieved.

\section{Stopping}\label{stopping}
As introduced in section \ref{varxz},
we quantify the degree of stopping by comparing the variances of 
 transverse rapidity distributions (defined in section \ref{centrality})
 with that in the
longitudinal $(z)$ direction.
Throughout we use scaled rapidities (i.e. the rapidity gap is
projected unto the fixed interval $(-1,+1)$) and typical samples have
been shown and discussed already (section \ref{rapidity}).
All rapidities  are evaluated in the c.m..
However, for the ratios we discuss the scaling is immaterial.
We shall call
{\it varxz}  the ratio of the variances in the $x$-direction and the $z$
direction.
The choice of x is arbitrary, i.e. it is {\it not} connected with the
azimuth of the reaction plane.
 {\it varxz(1)} is defined like {\it varxz} except that the integrations
for calculating the variance are limited to the range $|y_{z0}|<1$;
in \cite{reisdorf04a} this was called {\it vartl} and $|y_{x0}|$ was also
constrained to be $<1$.
Such a  restricted measure of stopping can be more adequate if
measured data outside the range $|y_{z0}|<1$ are missing or less reliable.
For {\it varxz(0.5)} the transverse rapidity distribution is taken under the
constraint $|y_{z0}|<0.5$ on the scaled {\it longitudinal} rapidity.
This measure of stopping is clearly biased and has systematically higher
values, but is of interest when comparing with many data in the 
(higher energy) literature, where often attention is focused on 
transverse momentum spectra
 around (longitudinal) midrapidity.
When no number in parenthesis is given, the intervals in rapidity are
sufficiently broad to make the value of {\it varxz} asymptotically stable
 within the indicated, mostly systematic,  errors.
 
There are other ways of characterizing stopping. 
Videbaek and Hansen \cite{videbaek95} have introduced the
{\it mean rapidity shift} $<\delta y>$ which is defined by
%
\[ <\delta y> \equiv \frac{\int_{-\infty}^{y_{cm}} |y-y_{tgt}| (dN_p/dy) dy}
   {\int_{-\infty}^{y_{cm}}(dN_p/dy) dy}    \]
where $y_{cm}$ is the rapidity of the c.m.,  
$y_{tgt}$ is the target rapidity and $dN_p/dy$ is the proton rapidity 
distribution (a similar formula exists relative to the projectile
rapidity $y_p$ with the integration limits interchanged).
This observable was also determined in some works of our Collaboration,
\cite{hong98,hong02}.
The {\it inelasticity K} was used by NA49 \cite{stroebele09p} and is defined by
$ K=E_{inel}/(\sqrt{s}/2-m_p)$ with $E_{inel}$ being equal to the average
energy loss of incident nucleons with rest mass $m_p$.
The common feature of these definitions is that one assesses stopping
relative to {\it initial} conditions: the conclusion is then
that a high degree of stopping is reached in collisions all the way up to
at least SPS energies, especially if it is assessed in terms of energies.
Our  stopping variable has a different goal: we want to assess the
difference relative to a completely stopped scenario, an assumption that
is frequently assumed in (one-fluid) hydrodynamic codes for 
heavy ion collisions because of its inherent simplicity of a needed starting
point of the calculation. 
Our observable, {\it varxz}, is important to assess the partial memory
of the original (accelerator induced)  counterflow of two fluids
 and hence deviations from {\it local} equilibrium.
It is also of value to help determine viscosity properties of nuclear
matter \cite{gaitanos05}.


\subsection{Isospin tracing}
We invariably find that longitudinal rapidity distributions are broader
than transverse rapidity distributions, see section \ref{rapidity}.
In principle, for a given system, one cannot exclude that this phenomenon
is caused by a rebound opposite to the incident direction after a complete
stop.
The system size dependence of the effect, see namely Fig.~\ref{dndy-400c1Z1A2},
can be  used to strongly argue against this interpretation which would
imply an unlikely stronger rebound in smaller systems.
An alternative method to demonstrate partial transparency was introduced
by our Collaboration \cite{rami00} and 
consists in so called {\it isospin tracing}.
We refer the reader to the original publication for further explanations.
Briefly, we combine rapidity distribution information of four systems
involving the isotopes $^{96}$Zr and $^{96}$Ru~: Ru+Ru, Zr+Zr, Ru+Zr
and Zr+Ru, where the mixed systems are merely technically different in the
sense that beam and target are inverted.

\begin{figure}
\begin{center}
\epsfig{file=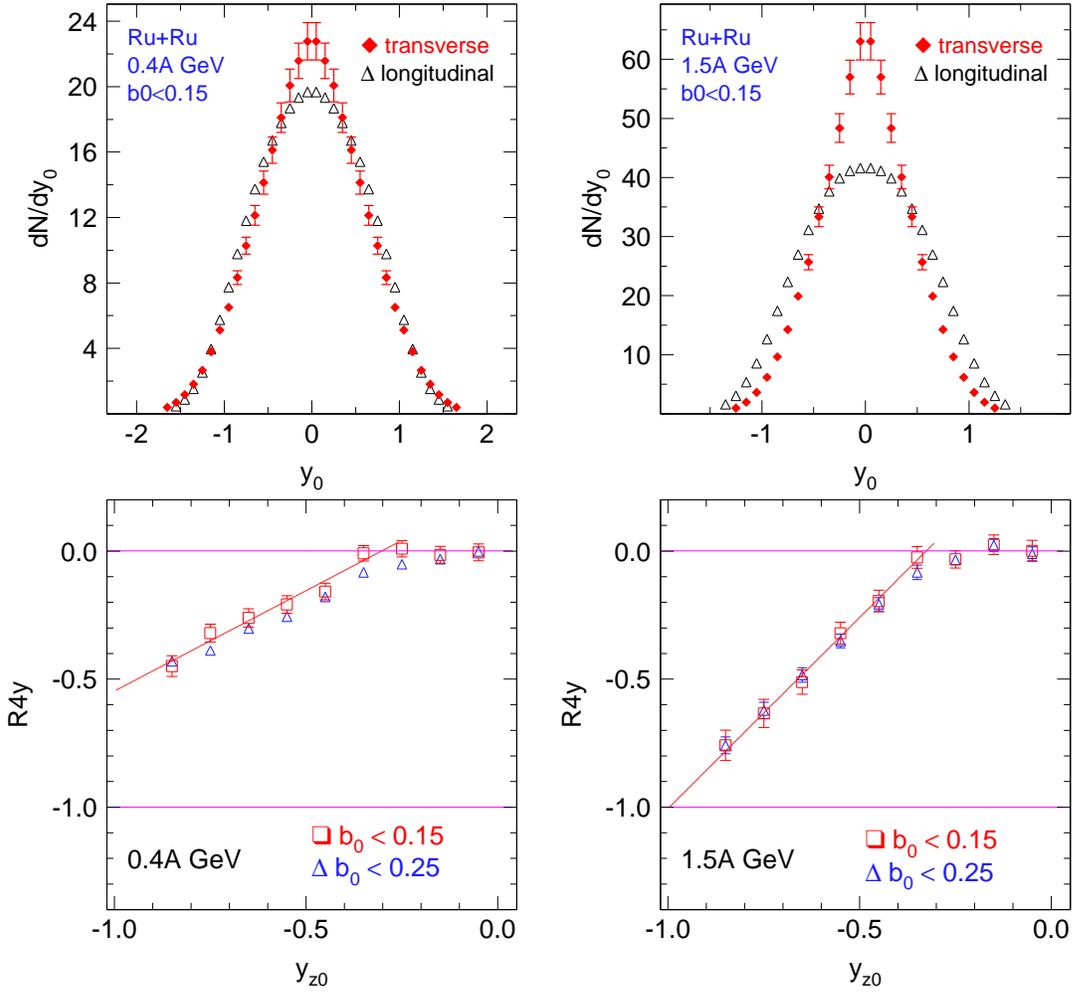,width=150mm}
\end{center}
\caption{
Top panels: Longitudinal and transverse rapidity distributions of protons in
collisions of Ru+Ru at $0.4A$ GeV (left) and $1.5A$ GeV.
Bottom panels: The corresponding $R_{4y}$ observables for two
slightly different indicated centralities as a function of the longitudinal
rapidity $y_{z0}$.
}
\label{r4y-400a-r4stop}
\end{figure}

%
In the following,  abbreviating $ N^{RuZr}_y(p)$ 
 the rapidity distribution of protons in the reaction Ru+Zr, etc, we define\\
$R_{4y}(p) = [N^{RuZr}_{y}(p) - N^{ZrRu}_{y}(p)] / 
             [N^{RuRu}_{y}(p) - N^{ZrZr}_{y}(p)]$ \\
This observable is a slight variant of the observable $R_Z$
 used in \cite{rami00}:
it is an average of the two opposite-sign branches obtained by switching
target and projectile in the mixed system. An arbitrary (here negative)
equal sign is defined (the two branches of $R_Z$ in the chosen symmetric-mass
systems must be equal except for sign).  
In Fig.~\ref{r4y-400a-r4stop} we show $R_{4y}$ for central collisions
at $0.4A$ and $1.5A$ GeV together with rapidity distribution plots.
In case of complete mixing with loss of memory 
(both alternatives: rebound as well as partial transparency)
of the incident geometry, the rapidity dependences of $R_{4y}$ should be flat
at zero value. 
There is indeed a small flat part around $y_{z0}=0$,
especially for the most central ($b_0<0.15$) selection,
 but for higher $|y_{z0}|$
there is an increasing deviation from zero, the sign of which can be
unambiguously associated with transparency, rather than rebound.
The effect is definitely more pronounced at the higher energy in
full accordance with the information from the variances of the rapidity
distributions (see the upper panels in the figure).
Only statistical errors are shown in the lower panels of
Fig.~\ref{r4y-400a-r4stop}.
A more detailed presentation and discussion of the new isospin tracing data
will be published elsewhere \cite{yjkim10}.

\subsection{Global stopping and the EOS}
\begin{figure}
\begin{center}
\epsfig{file=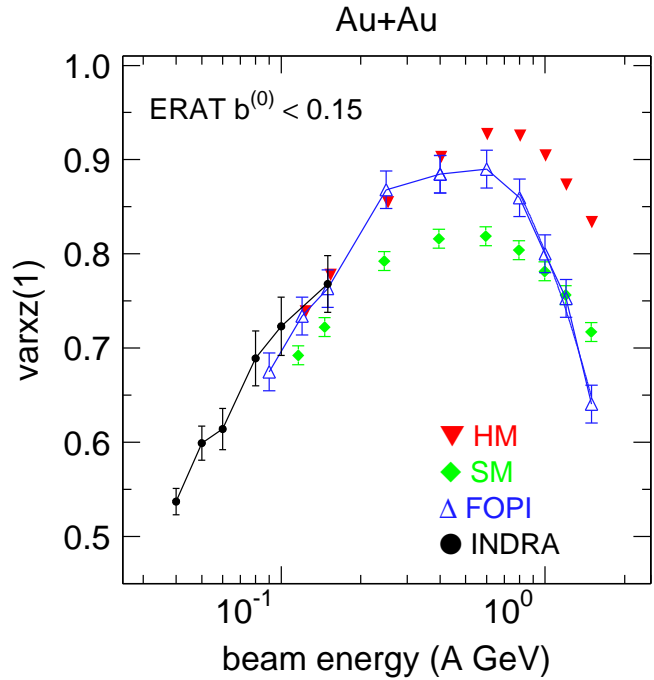,width=100mm}
\end{center}
\caption{
Excitation function of the stopping observable {\it varxz(1)} 
in central Au+Au collisions ($b_0<0.15$) \cite{reisdorf04a,andronic06} .
 FOPI data: open blue triangles, INDRA data: black dots.
These experimental data points are joined by straight line segments to
guide the eye. IQMD simulations extending from $0.12A$ to $1.5A$ GeV:
red full triangles (HM) and green full diamonds (SM).
}
\label{vartl-indra}
\end{figure}

By adding up  the measured rapidity distributions of the different ejectiles
one can try to obtain a global system stopping. 
The excitation function for Au on Au, first published in \cite{reisdorf04a},
was later slightly revised and combined with data from the
INDRA collaboration \cite{andronic06}.
The combined data are shown again in Fig.~\ref{vartl-indra},
 but this time together with IQMD
simulations using alternatively a hard ($K=380$ MeV, HM) or a soft
 ($K=200$ MeV, SM) EOS with momentum dependence \cite{aichelin87}.
As can be seen from the Figure, the difference between the HM and the SM
simulation is significantly larger than the error bars in the data.
This is an important observation as it opens up the possibility to
observe EOS sensitivity also in {\it central} collisions in contrast
to directed and elliptic flow which are linked to off-centrality
(and in general collisions with less achieved compression).
However, it is also clear that the 'residual' interaction, i.e. the
explicit collision term, influences the outcome.
The present parameterization of IQMD as used here is obviously not
able to reproduce the data, in particular the rapid drop of {\it varxz(1)}
beyond $0.8A$ GeV is not reproduced.
A fair reproduction of a portion ($0.25A$ to $1.0A$ GeV) of the excitation
function was achieved in \cite{gaitanos05}. 

Clearly, {\it varxz} data should prove useful also to constrain
the viscosity of nucleonic (and more generally of hadronic) matter if
compared to transport model simulations.
It is tempting to conjecture that the rising part of the excitation function
is dominated by the decreasing importance of Pauli blocking as the
rapidity gap exceeds the Fermi energy.
Particle creation starts becoming important beyond $1A$ GeV \cite{reisdorf07}
(see also our Fig.41 in the present work).
But the decreasing trend of the {\it varxz} data seen in Fig.~\ref{vartl-indra}
suggests that this new channel does not compensate for the decrease and
forward focusing of elastic nucleon-nucleon processes.
The rapid decrease of {\it varxz} at the high SIS energy end is continued
at AGS and SPS as we shall show later for net protons.
 
\subsection{p, d, t stopping: a hierarchy}
\begin{figure}
\begin{center}
\epsfig{file=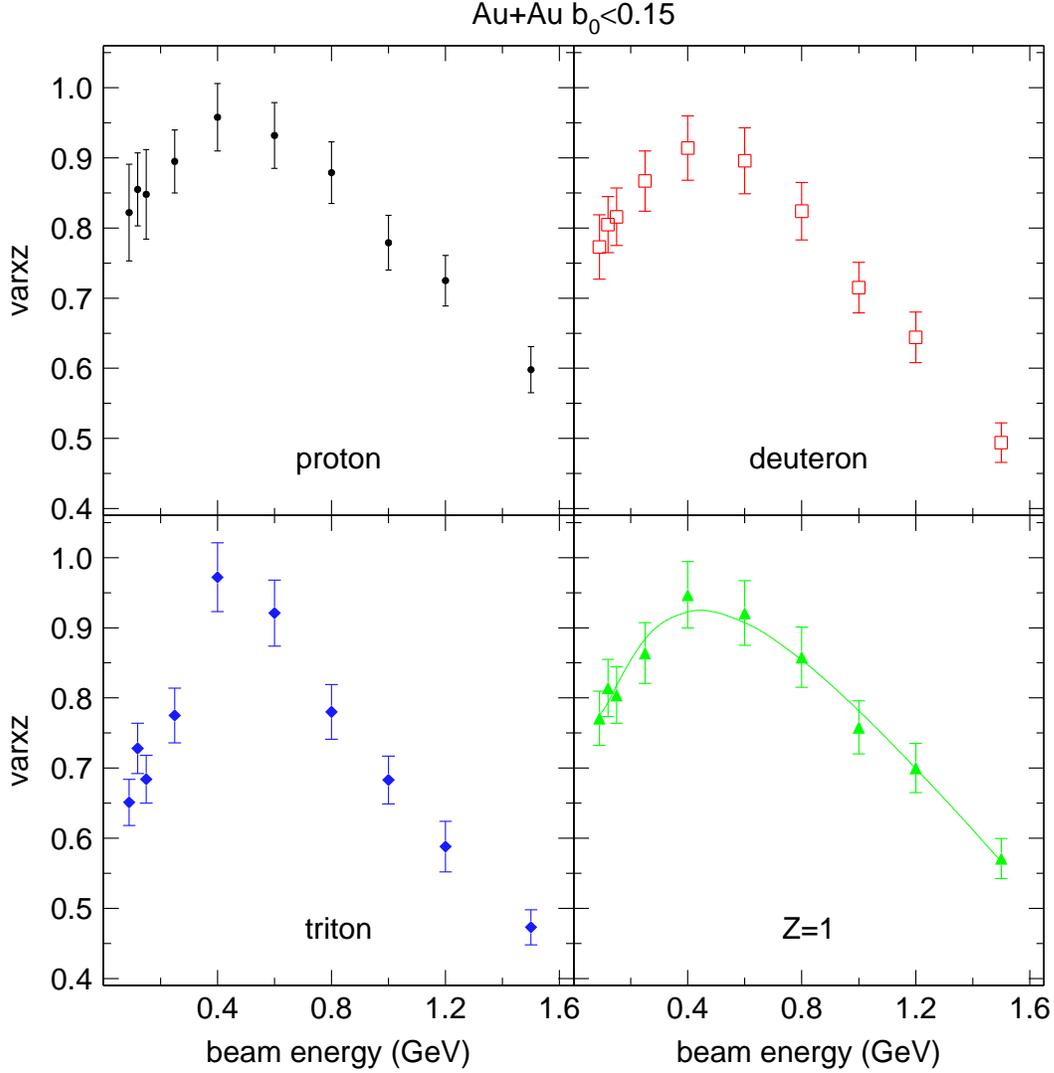,width=150mm}
\end{center}
\caption{
Excitation function of the stopping observable {\it varxz}
in central Au+Au collisions ($b_0<0.15$).
As indicated in the various panels the excitation functions are shown
separately for the three hydrogen isotopes.
The right-lower panel allows to compare relative to the
average weighted Z=1 fragments. 
The smooth curve is just a polynomial fit. 
}
\label{var-aupdt-h}
\end{figure}

\begin{figure}
\begin{center}
\epsfig{file=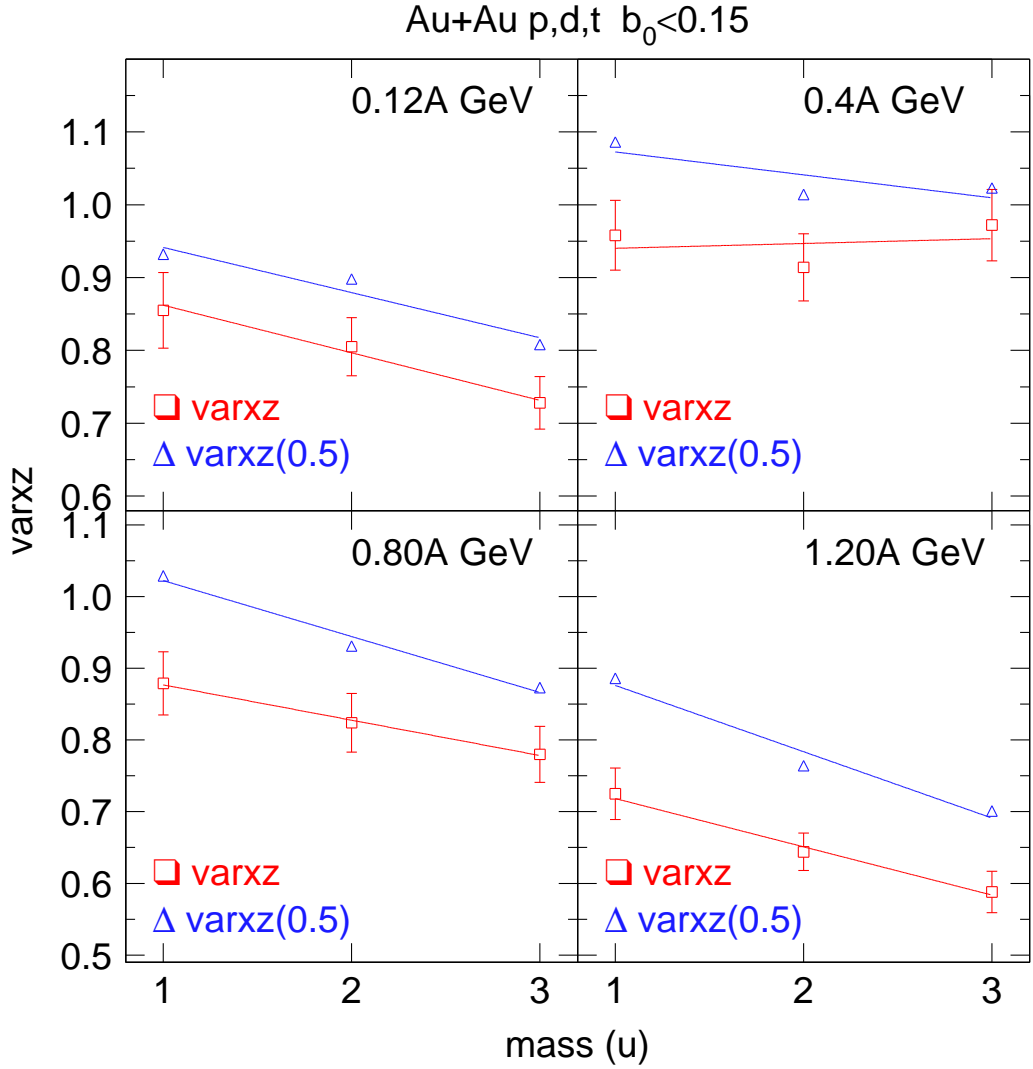,width=150mm}
\end{center}
\caption{
Mass number dependence of the stopping observable {\it varxz} of hydrogen
fragments in central Au+Au collisions ($b_0<0.15$).
Data are shown for 4 representative energies indicated in the panels 
(notice the one order of magnitude span). 
For comparison data are also shown for the 'constrained' stopping
observable {\it varxz(0.5)} (blue open triangles, error bars omitted). 
The lines are linear fits.
}
\label{varauc1pdt-4}
\end{figure}

By plotting {\it varxz} separately for identified fragments further
insights (and constraints to simulations) can be gained.
This is shown in Fig.~\ref{var-aupdt-h}.
A remarkable feature is the similarity in the behaviour for the three
isotopes, that suggests the collectivity of the phenomenon.
There is a well defined maximum around $0.4A$ GeV for each of the isotopes
qualitatively similar to the observed \cite{reisdorf04a} global stopping.
On the other hand a closer look reveals that there are some subtle differences
in the shapes of the excitation functions.
Except around the maximum, there is a hierarchy in the degree of stopping.
This is illustrated in Fig.~\ref{varauc1pdt-4}.

On the average clusters have 'seen' less violent (less stopped) collisions
when the stopping is incomplete for single emitted nucleons (away from the
maximum around $0.4A$ GeV).
This effect may be a surface effect but is not a 'trivial'
 spectator-matter effect, as the rapidity
distributions in these most central collisions do not exhibit well
defined 'spectator ears' around $|y_{z0}|=1$:
 there are no spectators in the strict sense here.
The constrained stopping {\it varxz(0.5)} (blue open triangles in the figure) 
is always somewhat larger because the 
transverse rapidity distributions are broadened when a cut around longitudinal
rapidity is applied. 
The trends with mass number and incident energy are similar however.


\subsection{System size dependence of stopping}
As remarked earlier, the system size dependence of the observable
{\it varxz} is an important constraint to the question of
transparency versus rebound dominated collision scenario,
 especially when only the
most central collisions are compared in scaled units of centrality.
All our data support transparency dominance.
This does not mean that the degree of stopping is small, but it is
significantly less than expected from ideal one-fluid hydrodynamics.

\begin{figure}
\epsfig{file=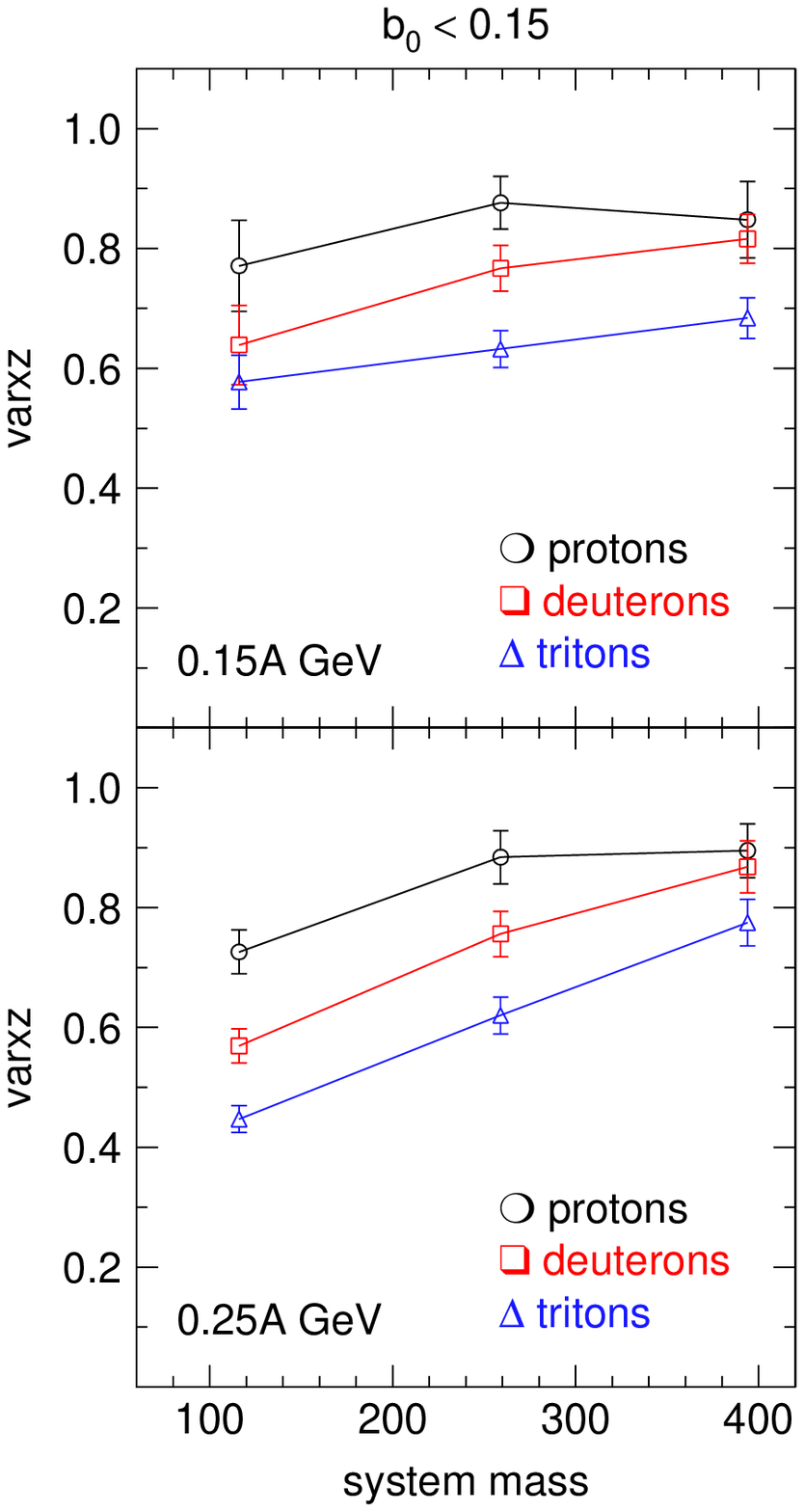,width=80mm}
\epsfig{file=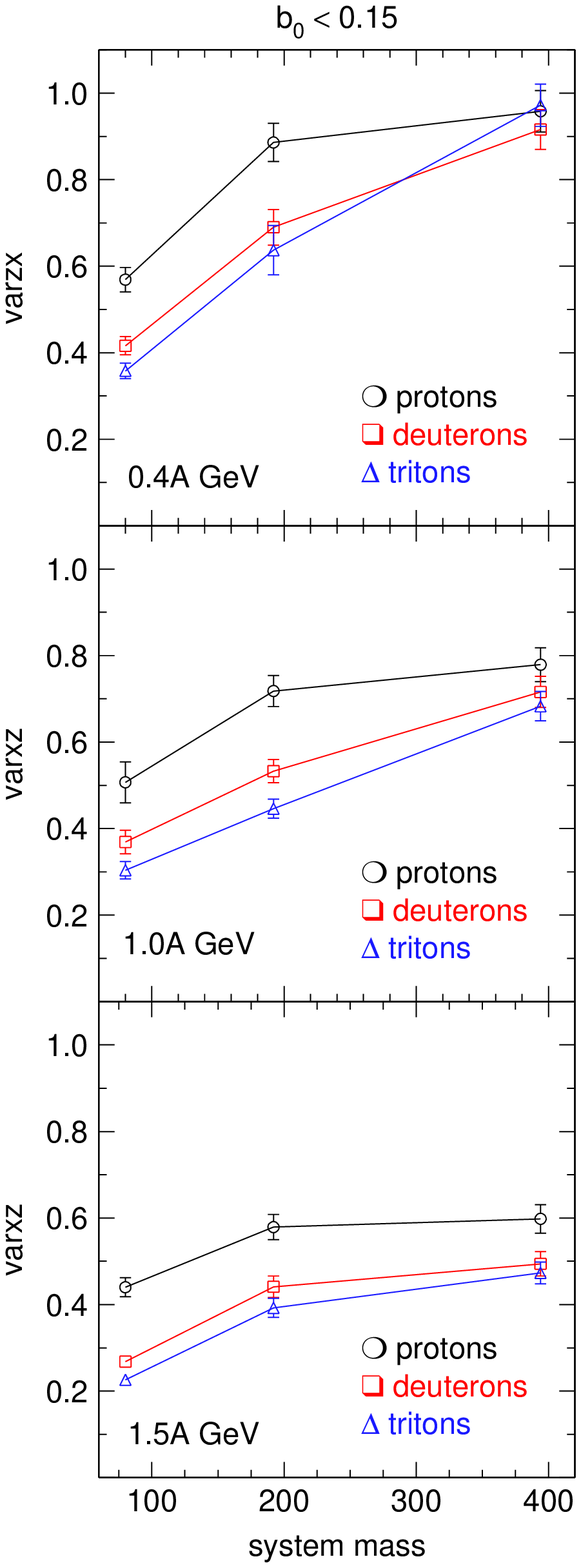,width=80mm}

\caption{
System size dependence of {\it varxz} for protons (black circles),
deuterons (red squares) and tritons (blue triangles) at various
indicated beam energies.
}
\label{var400c1pdt}
\end{figure}

 Figure \ref{var400c1pdt} shows a rather complex behaviour
 and will be a challenge to
ambitious microscopic simulations.
One remarkable feature is the apparent saturation of {\it varxz} at
a relatively low value for the highest incident energy.
This was already seen earlier, \cite{reisdorf04a}, for global stopping.
This suggests the beginning of a phenomenon seen in a more spectacular
way recently at the SPS \cite{stroebele09p} ($E/A=158A$ GeV):
 at midrapidity there is
 no non-trivial increase of population  in Pb+Pb relative to p+p collisions. 

\subsection{Isospin and stopping}
\begin{figure}
\begin{center}
\epsfig{file=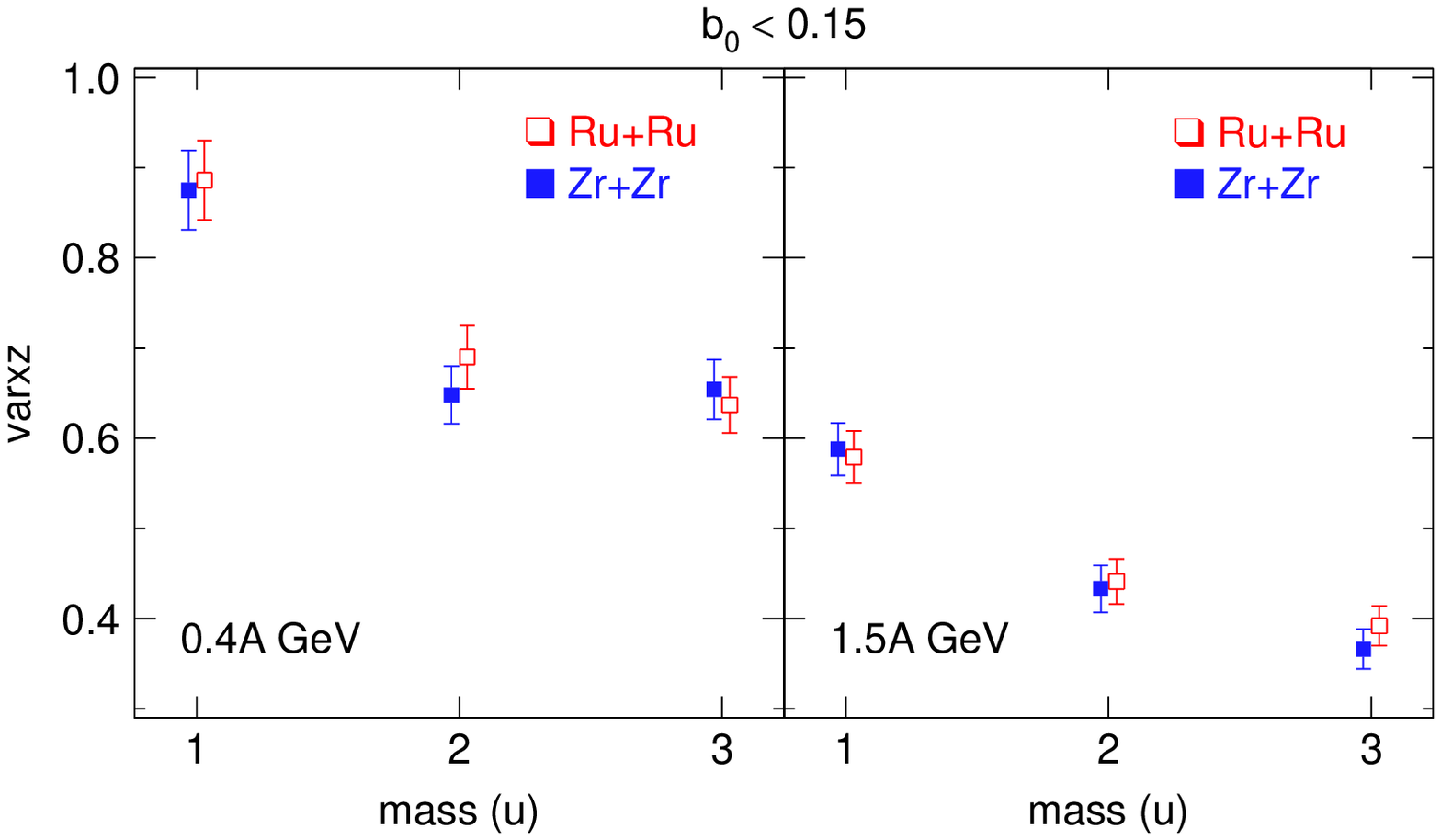,width=140mm}
\end{center}
\caption{
Mass dependence of {\it varxz} for hydrogen isotopes
in central ($b_0<0.15$) Ru+Ru  and Zr+Zr
collisions  at incident
beam energies of $0.4A$ GeV (left panel) and $1.5A$ GeV (right panel).
}
\label{varruzrc1pdt}
\end{figure}

One expects isospin dependences because the free nucleon-nucleon cross
sections $\sigma_{nn}$ and $\sigma_{np}$ are different and also
because the mean fields, shown earlier to influence global stopping,
are expected to be isospin dependent.
However the data show no convincing effect considering the error limits.
This is shown in Fig.~\ref{varruzrc1pdt} where a comparison of two systems with
the same mass, but different composition (Ru+Ru, $N/Z=1.182$ and 
Zr+Zr, $N/Z=1.400)$ is shown.
In this context it is interesting to note that very recently a similar
observation was made \cite{lehaut10} at lower energies ($32-100A$ MeV)
for Xe+Sn using various isotopes. 

\begin{figure}
\hspace{\fill}
\epsfig{file=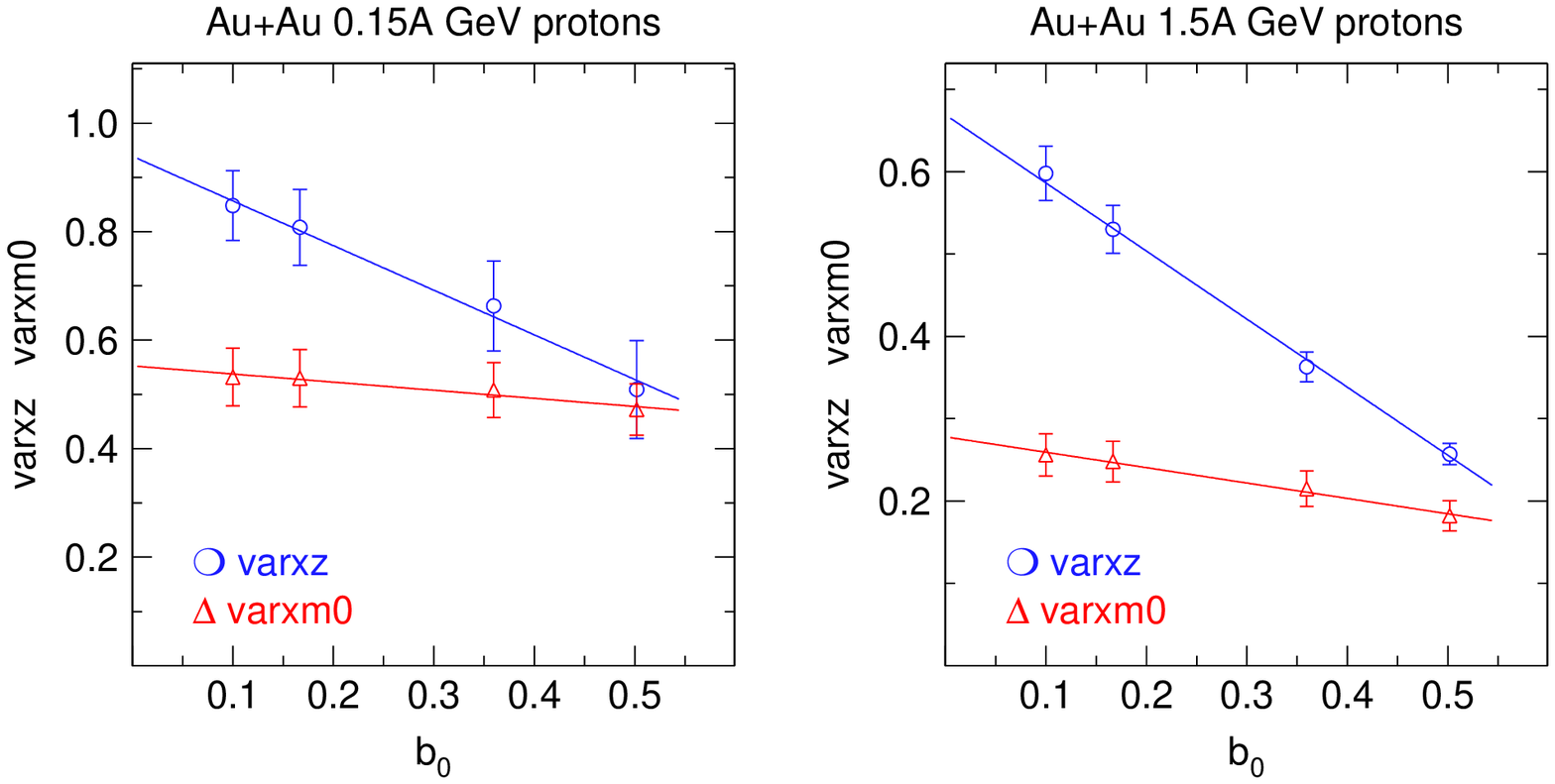,width=140mm}
\hspace{\fill}
\caption{
Centrality dependence of proton stopping ($varxz$) and mid-rapidity constrained
scaled transverse variance ($varxm0$) at 0.15 and $1.5A$ GeV.
}
\label{varzx-au1500}
\end{figure}

\subsection{Centrality dependence of stopping}
For comparisons with simulations it is useful to check how sensitively
the observable {\it varxz} reacts to matching the centrality correctly
with that of the experiment.
The relatively strong dependence of {\it varxz} on centrality $b_0$ is
illustrated in Fig.~\ref{varzx-au1500} for Au+Au at two rather different
incident energies.
 The linear fits allow to estimate a limit for $b_0=0$.
Since {\it varxz} is the ratio of {\it varx} and {\it varz}
(the zero suffix is immaterial here since the scaling cancels out ), it is
interesting to explore what causes the strong centrality dependence. 
The figure also illustrates the comparatively modest change of the scaled
variance, {\it varxm0} of the {\it constrained} transverse
rapidity distribution  (which we remind is obtained
 with a  with  a cut $|y_{z0}|<0.5$ 
 on the scaled longitudinal rapidity).

\begin{figure}
\begin{center}
\epsfig{file=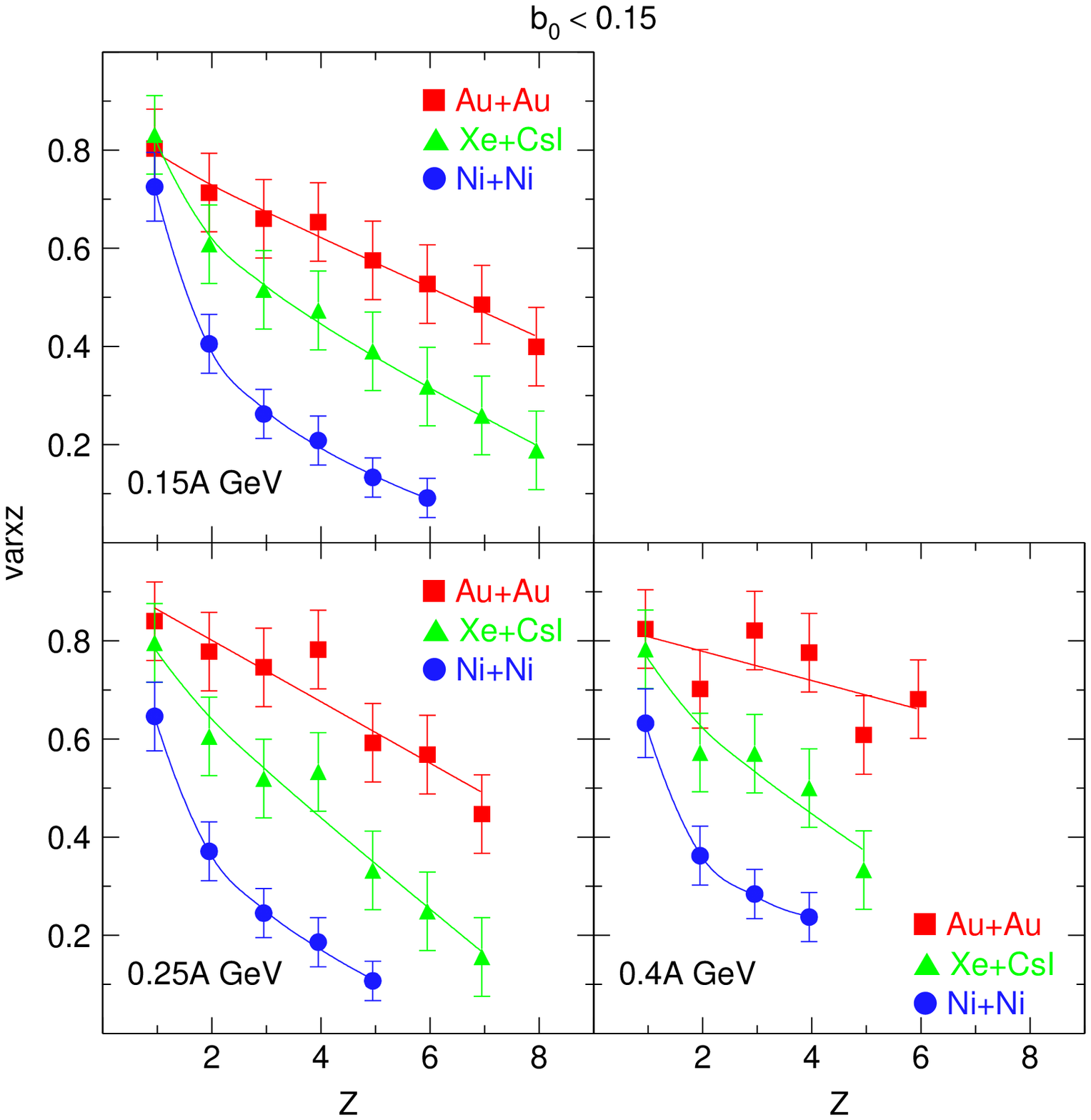,width=140mm}
\end{center}
\caption{
Charge number and system size dependence of the stopping observable
 {\it varxz} for 
fragments emitted in central Au+Au collisions ($b_0<0.15$).
Data are shown for 3 representative energies and three systems 
indicated in the panels.
The smooth curves, fitted to $a_0+a_{-1}Z^{-1}+a_1Z$, serve to guide the eye. 
}
\label{var-150-a}
\end{figure}

\subsection{Heavy clusters}
Since at lower energies we have been able to get information also for
fragments with $Z>2$, we can extend the hierarchy of stopping to heavy
clusters.
The results are put together for three incident energies and three systems
in Fig.~\ref{var-150-a}.
This confirms the existence of a stopping hierarchy in a very systematic
way. Also a significant system-size dependence is evidenced.
Note that the Z-dependence is flattest for Au+Au at $0.4A$ GeV,
 close to the maximum
of global stopping shown before in Fig.~\ref{vartl-indra}.
An extension, for $0.15A$ GeV, all the way to Z=20 using INDRA data has
been shown earlier in \cite{andronic06}, demonstrating also a reasonable
 consistency between data of FOPI and INDRA.
Quantitatively,
such data are presently outside the capabilities of {\it microscopic}
simulations, but present a challenge for our theoretical understanding.

\subsection{Joining up to higher energies}
As mentioned at the beginning of this section, stopping at higher
incident energies has not been characterized by {\it varxz}.
The reason is connected with the increasing difficulty to measure  and 
reconstruct the distributions
 $dN/dy_z$ and $dN/dy_x$ over the full range of rapidities that one needs to
know.
Often transverse momentum spectra in narrow $y_z$ bins, preferably around 
$|y_{z0}|=0$, are fitted
individually to thermal ansatzes.
 We show therefore the ratio of variances using in the numerator the variance
 inferred from
inverse slopes $T_{inv}$ determined from transverse momentum spectra in a bin 
$\Delta y_{z0}$ around (longitudinal) mid-rapidity.
In terms of the notation introduced at the beginning of this section,
this is likely to give an upper limit, {\it varxz$(\sim 0.1)$},
 to the unconstrained {\it varxz}, as suggested by Fig.~\ref{varauc1pdt-4}. 

\begin{figure}
\begin{minipage}{80mm}
\epsfig{file=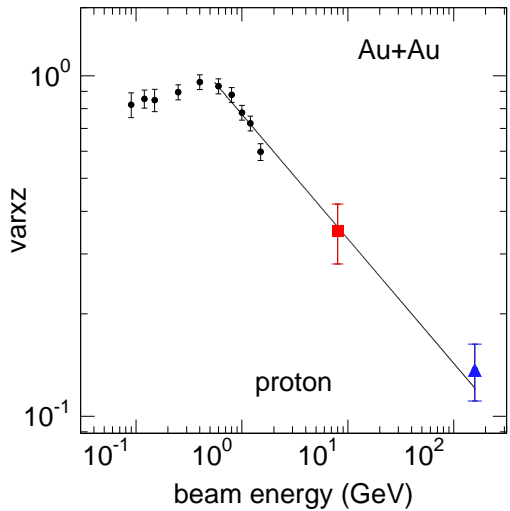,width=78mm}
\end{minipage}
\begin{minipage}{80mm}
\caption{
Excitation function of the stopping observable {\it varxz} for protons
in central Au+Au collisions.
In addition to FOPI data, obtained at SIS, we show here estimates
 from AGS (red full square) and SPS (blue full triangle) data.
See text for further details.
}
\label{var-aupdt-p}
\end{minipage}
\end{figure}

With these restrictions in mind,
we show in Fig.~\ref{var-aupdt-p} an extension of our proton data on stopping
to higher energies.
The AGS (Alternating Gradient Synchrotron at Brookhaven) data 
point was constructed
from ref. \cite{back01}, the SPS (CERN Super Proton Synchrotron) 
data point from ref. \cite{appelshauser99} (NA49).
A rather dramatic decrease of stopping, as defined in the present work,
is evidenced, suggesting a fair amount of residual counterflow if
one takes  transverse fluctuations as scale.
The straight line is a fit joining  to SIS data starting at $0.6A$ GeV.
However for the interpretation of stopping excitation functions of one
specific particle, here protons, extending over more than three orders in
the magnitude of the incident energy, it is important to realize that
they can be misleading on the {\it global} system trend.
The reason is that stopping hierarchies are omnipresent.
At the low energy end protons are the most stopped particles, while
at the high energy end {\it net} protons (the difference between protons
 and antiprotons) tend to be rather the least stopped
particles.
We are lead to expect stopping hierarchies in the higher energy regimes
similar to the ones described here, but shall not pursue this as it would
far exceed the scope of the present work.
As consequence of incomplete stopping, the time the system spends at maximum
 compression (and heat) is shortened.
This could prevent the full development of critical fluctuations expected
theoretically in certain areas of the phase diagram.


\section{Radial Flow}\label{radial}
One of the first attempts to understand radial flow in central heavy ion
collisions in the framework of microscopic transport theory
was published in \cite{danielewicz92}.
There, the  transverse flow energy $E^f_t$ per nucleon was defined by
$$ E^{f}_t = \int\!\!\int d\vec{r} d\vec{p} f [\frac{1}{2} m (v^{f}_t)^2] /
   \int\!\!\int d\vec{r} d\vec{p} f $$
and the transverse flow velocity $v^f_t$ was calculated locally, using only
those particles from the environment that have participated in the collisions
and stopping the integration when the density $\rho_0/8$ was reached in the
expansion stage ($f$ is the Wigner distribution function).
Evidently this kind of radial flow is not a direct observable.  Rather,
it is usually inferred from the mass dependence of average kinetic energies
of particles emitted around mid-rapidity (in cylindrical coordinates) or
around $90^\circ$ c.m. (in spherical coordinates).
This implies an interpretation of the data suggested
by hydrodynamical thinking: the system's quasi-adiabatic expansion leads
to cooling, 
part of the energy being converted to (radial) flow.
As we shall see this interpretation can be put to a critical test by studying
the dependence of the apparent radial flow on the centrality and the
 system size and also by inspecting 
the correlation of the radial flow with the degree of clusterization
taking the latter as a measure of the
degree of cooling.

Relevant data for the SIS (or Bevalac) energy range have been published
by Lisa et al. \cite{lisa95} and in a number of papers from our collaboration
\cite{reisdorf97,poggi95,hong98,jeong94,petrovici95}.
Azimuthal dependences of radial flow in the same energy range were studied
in refs. \cite{stoicea04,swang96} and shown to be sensitive to the EOS, while
little sensitivity was suggested for the azimuthally averaged radial flow.

\begin{figure}
\begin{center}
\epsfig{file=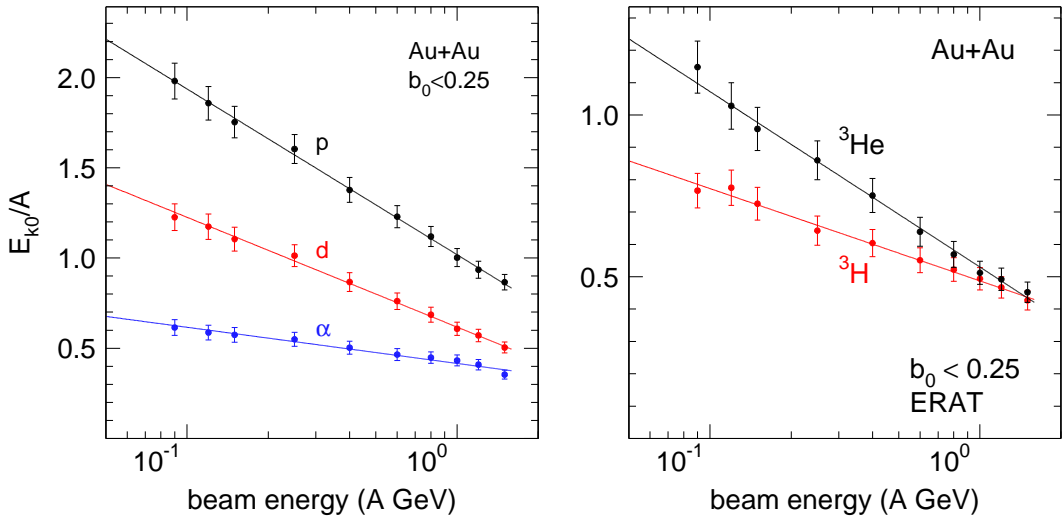,width=150mm}
\end{center}
\caption{
Average kinetic energy per nucleon (scaled units) of protons,
 deuterons and $^4$He
(left panel) and of $^3$H and $^3$He (right panel)
emitted in $90^\circ \pm 10^\circ$ in the c.m. 
in central collisions of Au+Au as function of beam energy.
The excitation functions follow a linear dependence (straight lines) if plotted
on a logarithmic energy scale.
}
\label{ekinavc4-r}
\end{figure}

We shall start by taking a look at the systematic evolution with incident
energy and ejectile mass (hydrogen and helium isotopes) of average scaled
kinetic energies per nucleon, $E_{k0}/A$.
This is shown in Fig.~\ref{ekinavc4-r} for collisions with $b_0<0.25$
 and emissions into c.m. angles ($90\pm 10^{\circ}$).
A striking feature is the extremely smooth evolution with beam energy:
the point to point errors seem to be smaller than the indicated systematic
errors.
In these scaled units the energies gradually decrease as the
available energy must be shared by an increasing number of emitted particles.
The regularity suggests a very continuous change without any dramatic fast
transition to some higher entropy phase.
Note that at the low end we observe protons emitted {\it perpendicular}
to the beam axis with {\it on the average} twice the incident beam energy
 per nucleon.

Another feature, demonstrated in the right panel of the figure is the difference
between the two equal mass isotopes $^3$He and $^3$H.
Starting at the low beam energy end with a difference of kinetic energies
well in excess of the expectations from Coulomb effects \cite{poggi95},
 this difference gradually converges to zero {\it in these scaled units}.
We have already discussed this 'anomaly' in section \ref{rapidity} for
the special case of an incident energy of $0.4A$ GeV and have given there
a tentative interpretation. Here we see the 'anomaly' apparently
disappearing as production of clusters heavier than mass three drastically
decreases (see also Fig.~\ref{y-auc1-he}).

\begin{figure}
\begin{center}
\epsfig{file=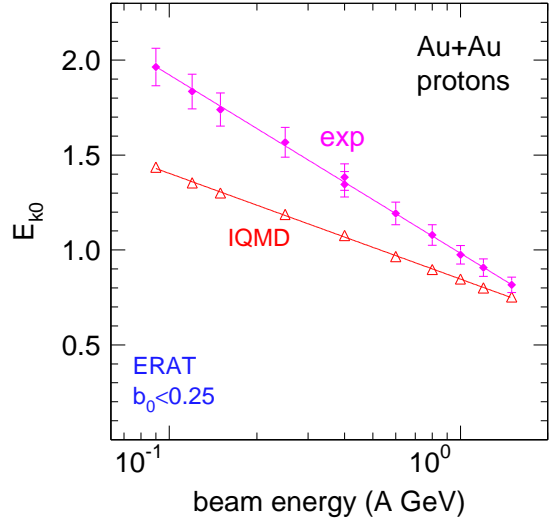,width=80mm}
\end{center}
\caption{
Average kinetic energy (scaled units) of protons emitted around
$90^\circ$ c.m. in central collisions of Au+Au as function of the beam
energy. 
Comparison of experimental data with simulations using IQMD-SM.
}
\label{ekinavc4p}
\end{figure}

Another demonstration for the need to reproduce correctly the
degree of clusterization in order to understand the kinetic energies
of emitted particles is given in Fig.~\ref{ekinavc4p} for protons.
Much like the experimental data the calculated data are reproduced with 
high accuracy ($0.7\%$)  by 
a linear dependence on $\log(E/A)$.
The slope characterizing the calculated trend is different however.
The reason is connected with a lack of sufficient clusterization at
the low energy end in the simulation.
Too many single protons, together with the necessity to conserve the
total energy (and mass), lead to smaller energy per nucleon.
This stresses the need to reproduce the degree of
clusterization for quantitatively correct comparisons with experimental
observables pertaining to specific identified particles.
In particular radial flow is connected with cooling leading to clusters.
The two observables, cluster yields and radial flow, are therefore 
interrelated.

\begin{figure}
\begin{center}
\epsfig{file=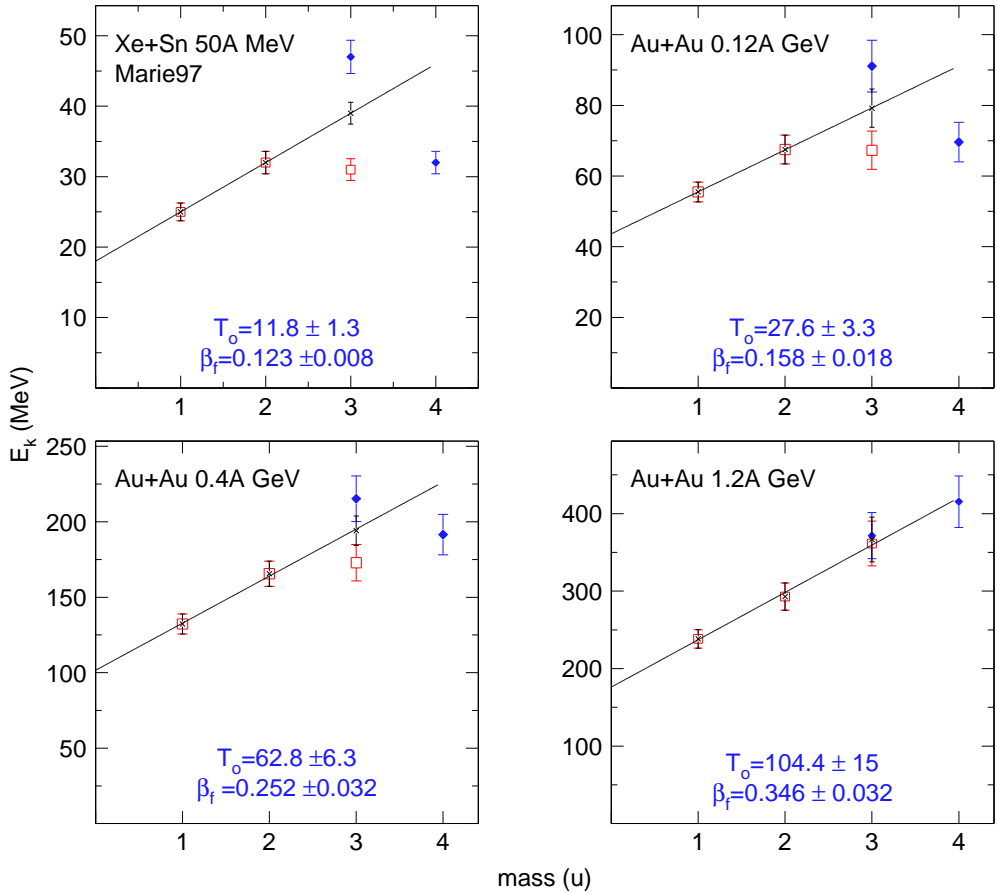,width=140mm}
\end{center}
\caption{
Average kinetic energies $E_k$ of H and He isotopes emitted at $90^{\circ}$ c.m.
in Au on Au collisions with centrality $b_0 < 0.25$.
Full (blue) diamonds: $^{3,4}$He. 
The data are obtained at various indicated beam energies.
The straight line is a linear fit up to mass 3.
The data point at mass 3 ($\times$ symbol) was taken to be the average
 of $^3$H and $^3$He.
The derived apparent temperatures, $T_o$, and (radial) flow velocities,
$\beta _f$, are indicated.
The upper left panel illustrates the use of the same method to INDRA
data \cite{marie97} for Xe+Sn taken at $50A$ MeV.
}
\label{ekinav-au120c4}
\end{figure}

In Fig.~\ref{ekinav-au120c4} we show average kinetic energies 
of H and He isotopes
emitted at $90^{\circ}$ c.m. in Au on Au collisions with centrality
$b_0<0.25$.
The data, plotted versus fragment mass for various incident beam energies,
show pronounced structures that tend to decrease with incident energy:
by adding INDRA data \cite{marie97} for Xe+Sn at 0.05 GeV a very large range
of incident energies is spanned.
The structures prevent a straight-forward determination of a linear slope
with mass that could be tentatively associated to a collective radial flow.
The straight lines plotted in the various panels join the proton and
deuteron data to the average kinetic energy of $^3He$ and $^3$H and ignore
the $^4He$ data. 
As can be seen, the average mass three value continues the proton-deuteron
trend with a remarkable accuracy, an observation that we found to be true
for all studied system-energies and centralities (see also
Fig.~\ref{dndym-150c1lcp-teq}), allowing therefore to
postulate a well defined observable, the slope of the corresponding straight
line, that we shall call 'radial flow with the A3 method', although
this interpretation  is clearly subject to caution.

Defining
$ E_k=E_f\times A+E_o$  with  $E_f=(\gamma _f - 1)\,m_N$
($m_N$ nucleon rest mass), we
convert the mass slope $E_f$ from the just introduced A3 method to a
(radial) flow velocity $\beta_f$\,, derived from the average flow factor
$\gamma_f$\,, and the offsets $E_o$
 in Fig.~\ref{ekinav-au120c4} to an
'offset temperature' $T_o$ using relativistic formulae and ignoring mean field
contributions to $E_o$\,, in particular Coulomb fields.
This allows us to compare to the analysis of 
\cite{lisa95},
although the authors seem to have handled the data structures differently.

\begin{figure}
\begin{center}
\epsfig{file=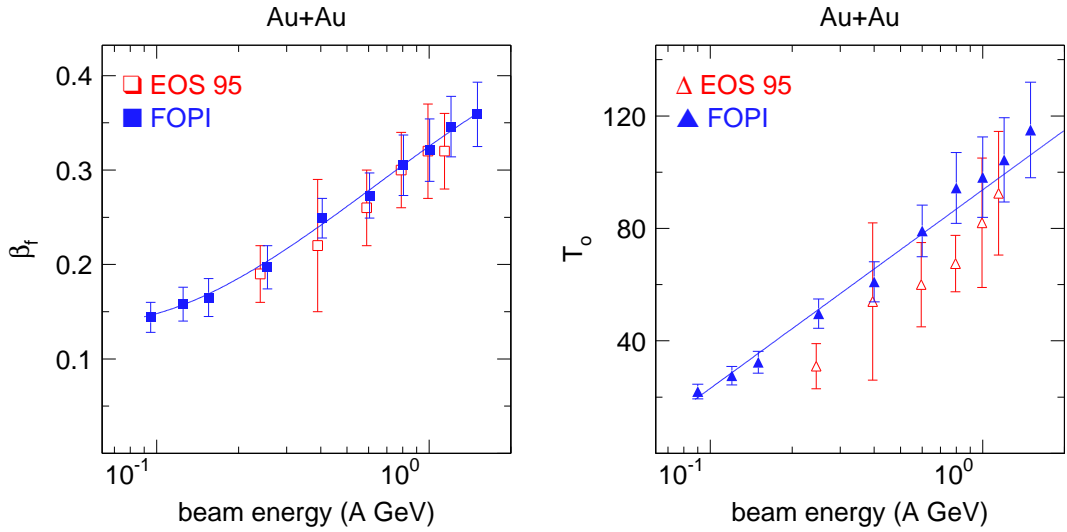,width=150mm}
\end{center}
\caption{
Radial flow velocities $\beta_f$ (left panel) and offset temperatures $T_o$
(right panel) for Au+Au collisions deduced from $90^\circ$ kinetic energies of
mass 1-3 ejectiles ($A3$ method).
The present data (blue full symbols) are compared to data from
 ref. \cite{lisa95}. The smooth curves are fitted to the present data only and
serve to guide the eye.
}
\label{radialflow-au-lisa}
\end{figure}

This is shown in Fig.~\ref{radialflow-au-lisa}.
The two datasets with nearly matched centralities are compatible,
 although our offset temperatures are
 systematically somewhat higher. 
Our data extend over a larger energy range and follow a more regular trend
fixed by smaller error bars.
The interpretation of the data in terms of a collective flow and a common
'local' temperature is probably too naive: the stopping hierarchies evidenced
in section \ref{rapidity} make a perfectly common flow improbable.
We expect further the apparent offset $T_o$ parameters to be influenced
both by Coulomb fields (especially at the low energy end) and by the
repulsive nuclear fields built up by the compression
 (see Fig.~\ref{vartl-indra}).
 

\subsection{System size dependences}
\begin{figure}
\begin{minipage}{80mm}
\epsfig{file=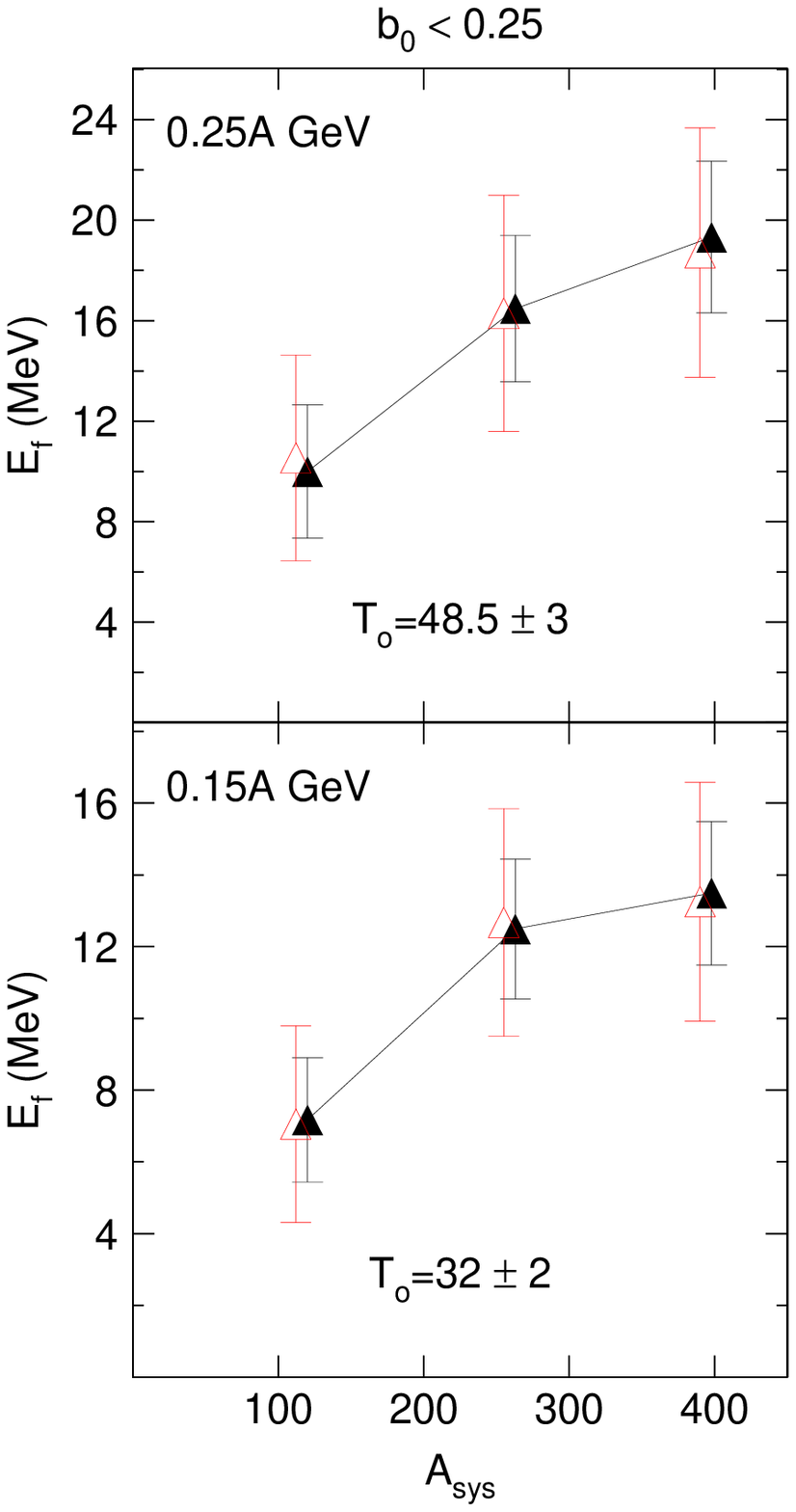,width=78mm}
\end{minipage}
\begin{minipage}{80mm}
\epsfig{file=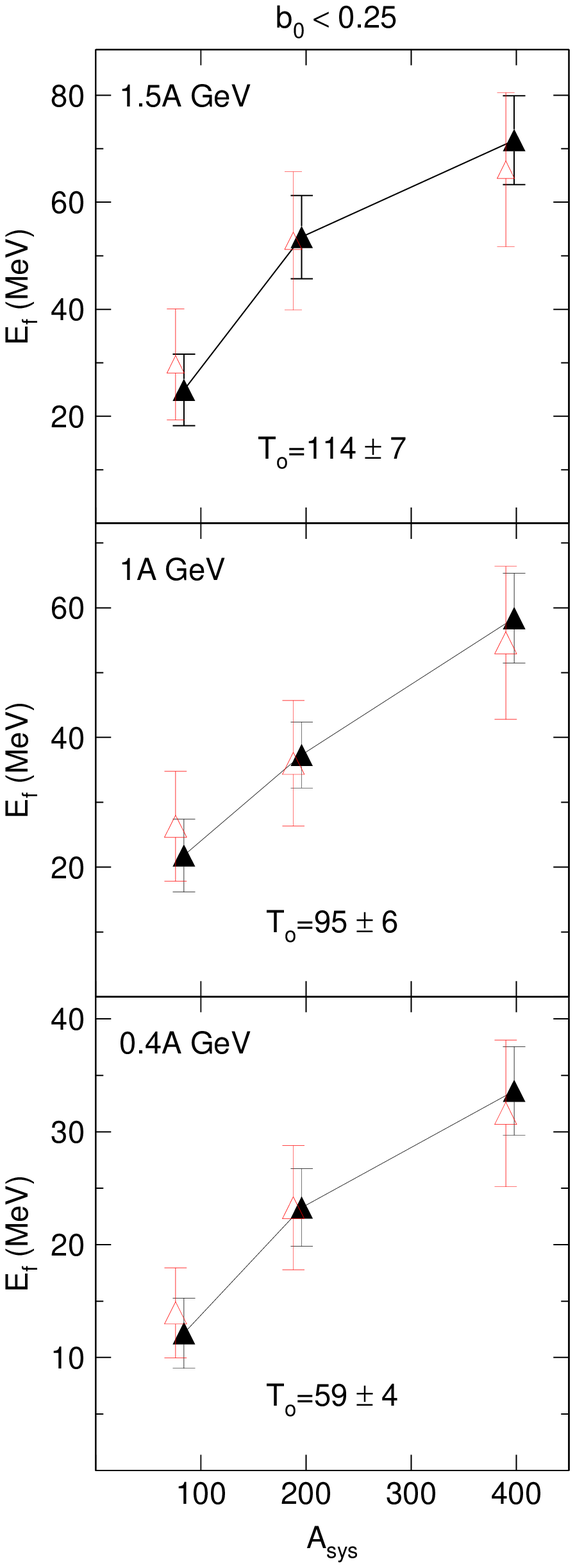,width=78mm}
\end{minipage}
\caption{
Radial flow energy in central collisions ($b_0 < 0.25$) as a function
of system size. 
This observable is deduced from the mass dependence (A=1,2,3) of average
kinetic energies at $(90\pm 10)^{\circ}$ c.m..
The open (red) data points are from separate fits (see previous figures)
varying both the offset and the slope per mass unit (i.e. the
flow parameter). 
The full (black) data points are obtained requiring a common offset parameter
(indicated in the panels as offset temperature $T_o$).
The  incident beam energies, varied over one order of magnitude,
are indicated in the upper left corner of each panel.
}
\label{ekinav-150c4}
\end{figure}

In Fig.~\ref{ekinav-150c4}
 we show the system size dependence
of radial flow varying the incident energy by one order of magnitude.
The increasing trend with system size is striking and characteristic for
all energies.
In each panel two sets of data are shown: 
the data points plotted with the open symbols and the
larger error bars were obtained fitting two parameters separately to the data
for one incident energy, the flow (slope) and the offset $T_o$.
The second set of data points were obtained using a common offset
 $T_o$ indicated in the
panel, a constraint allowing to fix the flow value better.

The difference between the two curves is small.
However we find, when allowing  two adjustable parameters for each system,
  that the $T_o$ values is not completely independent of the 
system size: from the lightest system to the heaviest (Au+Au) there is an
increase by about $20\%$. 
This could be a 'relic' of the repulsive mean fields 
due to increased compression in heavier systems.
Unfortunately the uncertainty of $T_o$ is relatively large. 

\subsection{Comparison to IQMD}
\begin{figure}
\begin{center}
\epsfig{file=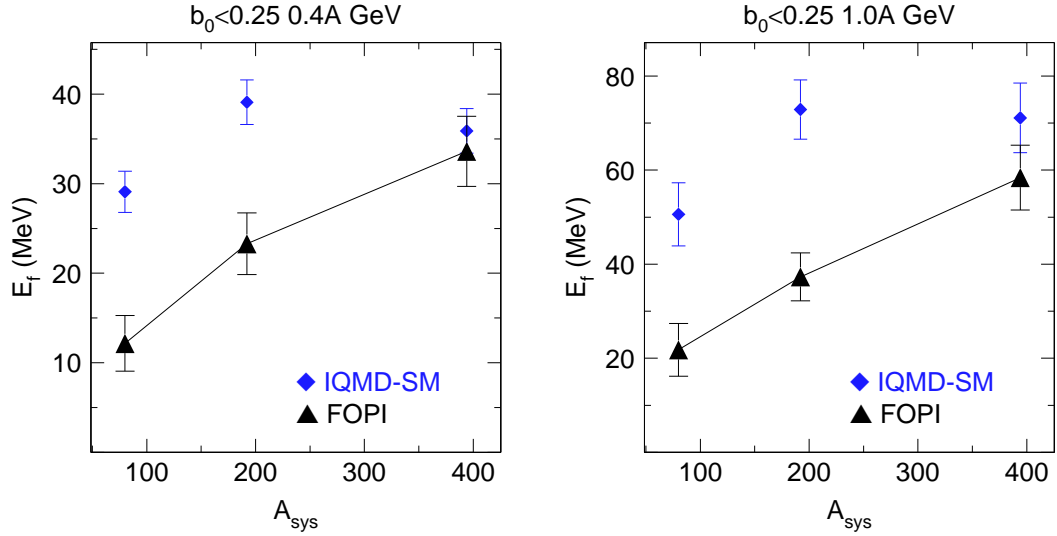,width=150mm}
\end{center}
\caption{
System size dependence of radial flow in central ($b_0<0.25$) collisions.
The experimental data (black full triangles joined by straight lines) are
compared with IQMD-SM predictions for $0.4A$ Gev (left) and $1.0A$ GeV (right)
incident energies. 
}
\label{ekinav-1000c4q}
\end{figure}

The observed system size dependence of the radial flow is not trivial:
a comparison with the simulation using IQMD-SM, see
Fig.~\ref{ekinav-1000c4q}, reveals that the strong dependence of the data on
system size is not correctly reproduced.
In \cite{hartnack01}, using IQMD, the authors conclude that radial flow is not
sensitive to the EOS: they find that the sum of two contributions to this kind
of flow, namely the mean field EOS and the nucleon-nucleon scatterings, tends
to be the same for a stiff, resp. soft EOS. In the soft case the scatterings
dominate because of the larger compression reached but this is compensated by
the less repulsive mean fields.
In view of the failure to reproduce the observed size dependence it is
possible to question this conclusion.
The increased production of clusters in larger systems, correlated with
increased radial flow and stopping could well be a memory of the original
compression (cooling by compression followed by expansion). 
In \cite{reisdorf04b} it was found that softer EOS favours cluster formation
at freeze out.  
This interpretation is supported by the theoretical work of 
ref. \cite{santini05}.


\subsection{Heavier clusters}
\begin{figure}
\begin{center}
\epsfig{file=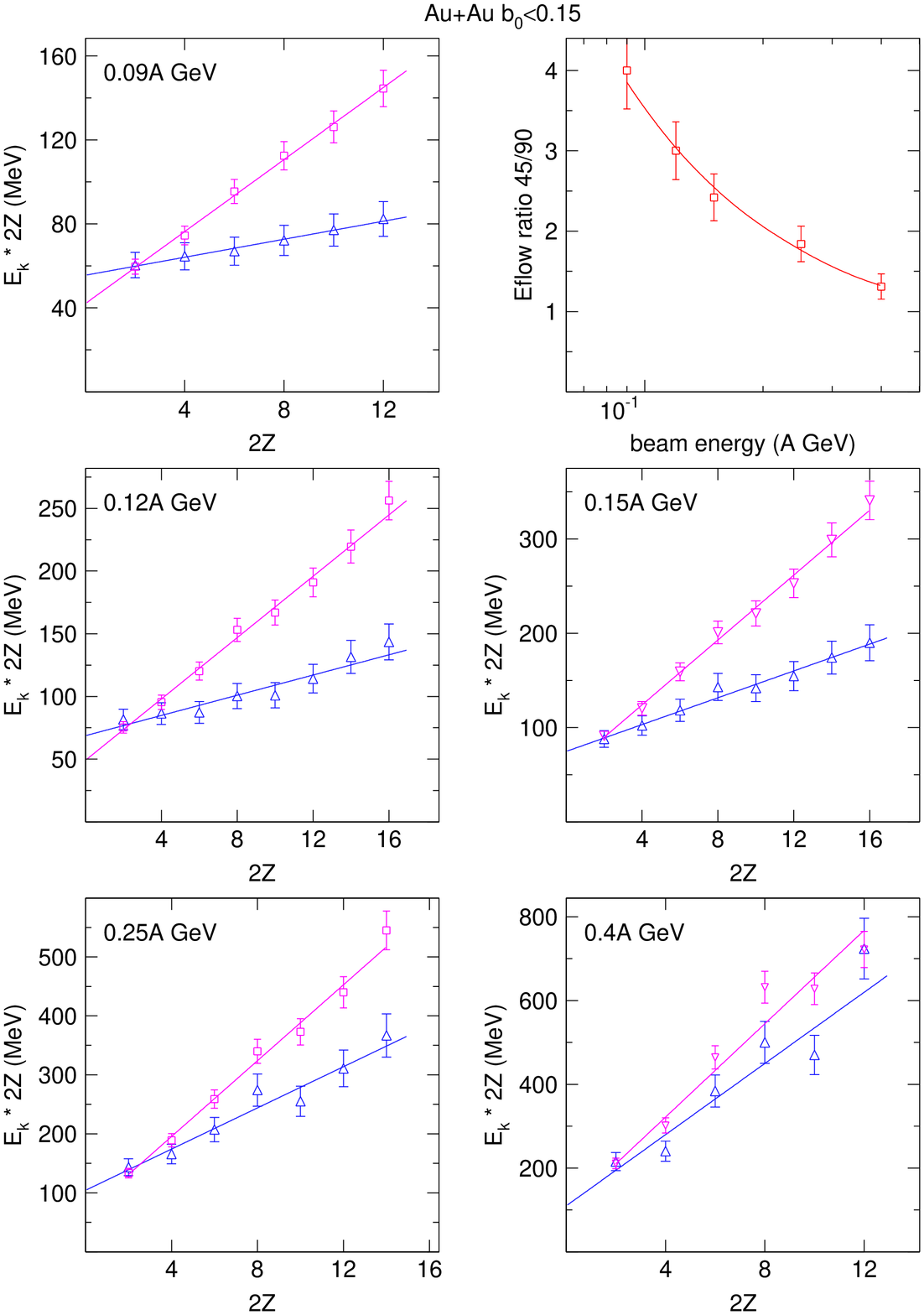,width=140mm}
\end{center}
\caption{
Nuclear charge dependence of kinetic energies for various indicated
beam energies in central collisions of Au+Au.
The measured kinetic energies per nucleon are multiplied by twice the
nuclear charge (i.e. by $2Z$) and plotted versus $2Z$.
Pink open squares: polar angles 5 to 45 degrees, blue open triangles:
polar angles 80 to 100 degrees.
The straight lines are linear fits.
The upper right panel shows the ratio of the deduced slopes (i.e. 'flow'
energies) for the two types of polar angle intervals versus incident
energy.
}
\label{ekinav-au90c1-6}
\end{figure}

For incident energies below $0.4A GeV$ we were able to obtain
 data for fragments up to nuclear charge 6-8, depending somewhat on
the incident energy. 
Although such data, not separated by isotope, and more limited in acceptance
(see Fig.~\ref{ecmtcm-au90c1})  have been published 
('PHASE I') and extensively
discussed earlier by our Collaboration, 
\cite{reisdorf97,jeong94,petrovici95,kuhn93,petrovici00}, 
it is  worthwhile to look again at these data in the light
of the present more complete 'PHASE II' information and new insights.
The relevant information is summarized in Fig.~\ref{ekinav-au90c1-6}
 where we show the
nuclear charge dependence of kinetic energies in central collisions of Au+Au
for five incident energies varying from 0.09A GeV to 0.4A GeV.
An earlier evaluation of these data, showing similar features, has been
presented in \cite{petrovici00}. The results cannot be compared directly:
in ref.~\cite{petrovici00}
 a coordinate system rotated into the flow axis was used and in-plane
fragments were suppressed. 
The datasets presented in each panel of Fig.~\ref{ekinav-au90c1-6},
 least squares fitted by a linear
function, hold for c.m. polar angles of $(5-45)^{\circ}$ and
$(80-100)^{\circ}$, respectively. 
To allow a more direct approximate comparison with the mass-dependent data
discussed before, we have multiplied the measured kinetic energies per nucleon
by $2Z$ and plotted them versus $2Z$.
Obviously the slopes of the fitted lines ('$2Z$ method') 
are very different for the two
polar angle ranges (in agreement with \cite{lisa95}),
 but they tend to gradually align as the energy is raised
(see upper right panel in the figure),
a confirmation of the global trend of increasing stopping in this range of
incident energies (Fig.~\ref{vartl-indra}).
Increased stopping leads to a more isotropic distribution
of kinetic energy, that was also visible for the
 light charged particles
(Fig.~\ref{varauc1pdt-4}, upper right panel showing the $0.4A$ GeV data).
It is fair to remind at this stage, that while the data below $45^{\circ}$
are well covered by our acceptance, the data around $90^{\circ}$ are
somewhat less reliable due to the necessity for a two-dimensional
 extrapolation (section \ref{reconstruction}).
We are confident that the major conclusions are not seriously affected
as, below $0.2A$ GeV, our analysis was found to be consistent with the
more complete INDRA data \cite{andronic06}.

\begin{figure}
\begin{minipage}{80mm}
\epsfig{file=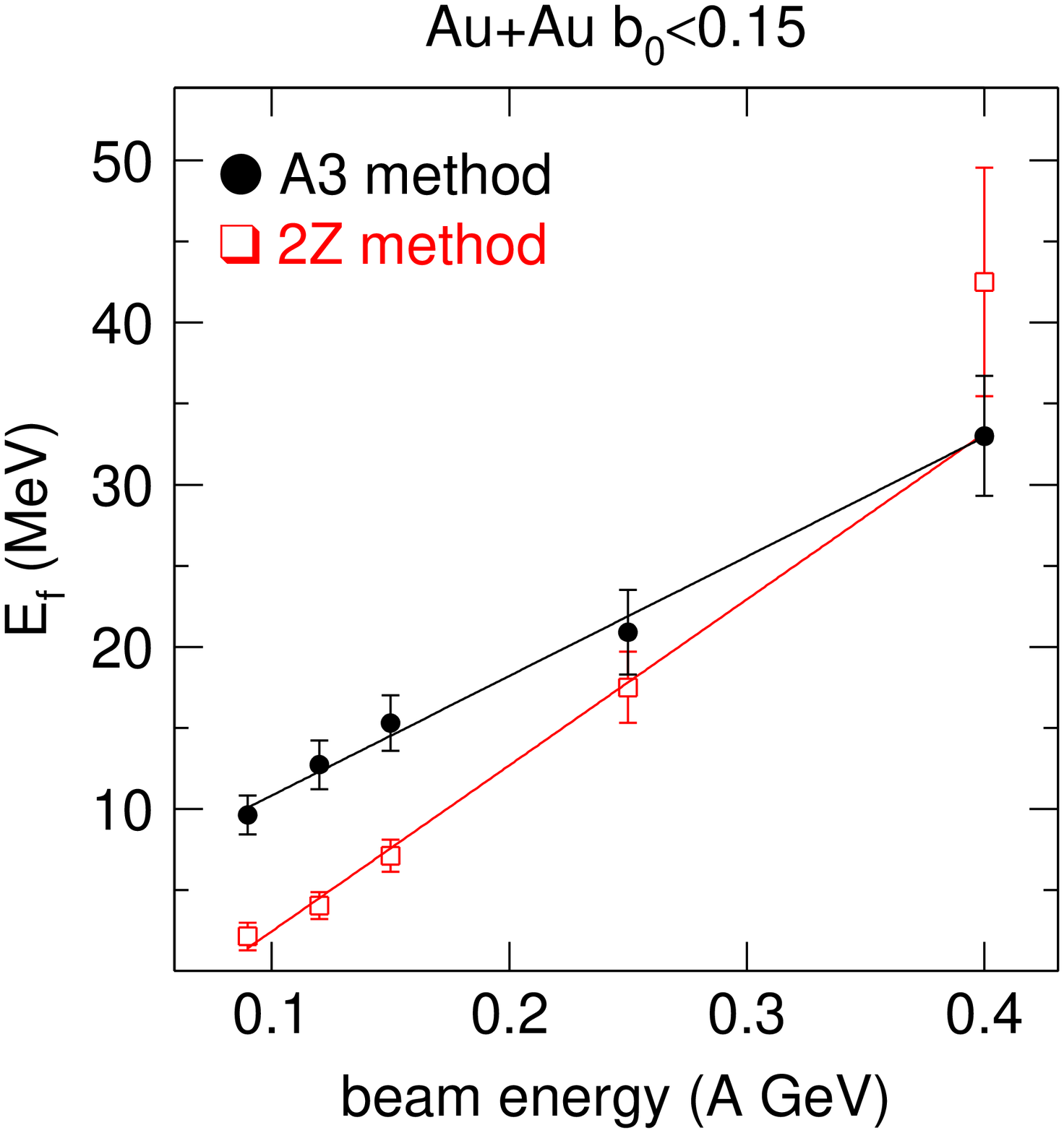,width=78mm}
\end{minipage}
\begin{minipage}{80mm}
\caption{
Radial flow versus incident kinetic energy.
Comparison of analyses results obtained with $A3$ and with $2Z$ methods.
The straight lines are linear fits.
}
\label{ekinav-au90c1-A3-2Z}
\end{minipage}
\end{figure}

It is interesting to compare the apparent radial flow deduced by the '2Z
method', Fig.~\ref{ekinav-au90c1-6}, with that deduced by the 'A3 method',
 Fig.~\ref{ekinav-au120c4}.
This is shown in Fig.~\ref{ekinav-au90c1-A3-2Z} for centrality $b_0<0.15$
(note that in order to match the centrality used in the analysis illustrated
by Fig.~\ref{ekinav-au90c1-6},
 the centrality is higher than in Fig.~\ref{ekinav-au120c4}
which is constrained to $b_0<0.25$ and therefore yields $E_f$ values lower
by 4 to 7 $\%$).
One sees the convergence of the two methods at $0.4A$ GeV where
stopping is maximal, but at lower incident energies the A3 method yields
higher values.
We suggest to correlate this with the pronounced stopping hierarchy found
at the lower energies (section \ref{stopping}).
Clearly then, one needs to correct the apparent mass dependences for this
effect when converting them to radial flow. 
This requires transport codes that reproduce the observed clusterization
and stopping hierarchy.

At this stage we wish to make a comment to our flow analysis published
in 1997, \cite{reisdorf97}.
In the light of the present extensive information on incomplete stopping
and hence strong deviations from isotropy, the earlier analyses assuming
spheric expansion and neglecting partial transparency
 must be revised in the sense that the flow values
found using the limited  acceptance of the PHASE I of FOPI included
a significant longitudinal (memory) component.
Thus the analysis included some counter-streaming effects of the nucleons
due to incomplete stopping (see Fig.~\ref{var-150-a}).
As the method used in \cite{reisdorf97} 
was strictly constrained by energy conservation, the
sum of radial and two-fluid counter flow  turned out to be correct, but the
interpretation of the mass or charge dependence of the kinetic
energies at these forward angles as just radial (spheric) flow cannot be
upheld.
The data themselves remain valid, except for a subtle effect on the
topology of the two-dimensional ($u_t$ vs rapidity $y_z$) spectra:
applying the ERAT criterion on the apparatus limited in PHASE I to
laboratory angles forward of $30^{\circ}$ lead for the most central collisions
(i.e. in the tails of the ERAT distributions)
to an underestimation of the anisotropy as the selection criterion favoured
events with a higher hit rate near the large angle apparatus limits,
increasing  the apparent isotropy.
In the present work, with the significantly larger PHASE II  acceptance,
 this effect, caused by event-by-event fluctuations, is minimized.


\subsection{Sytematics of equivalent temperatures}
\begin{figure}
\begin{center}
\epsfig{file=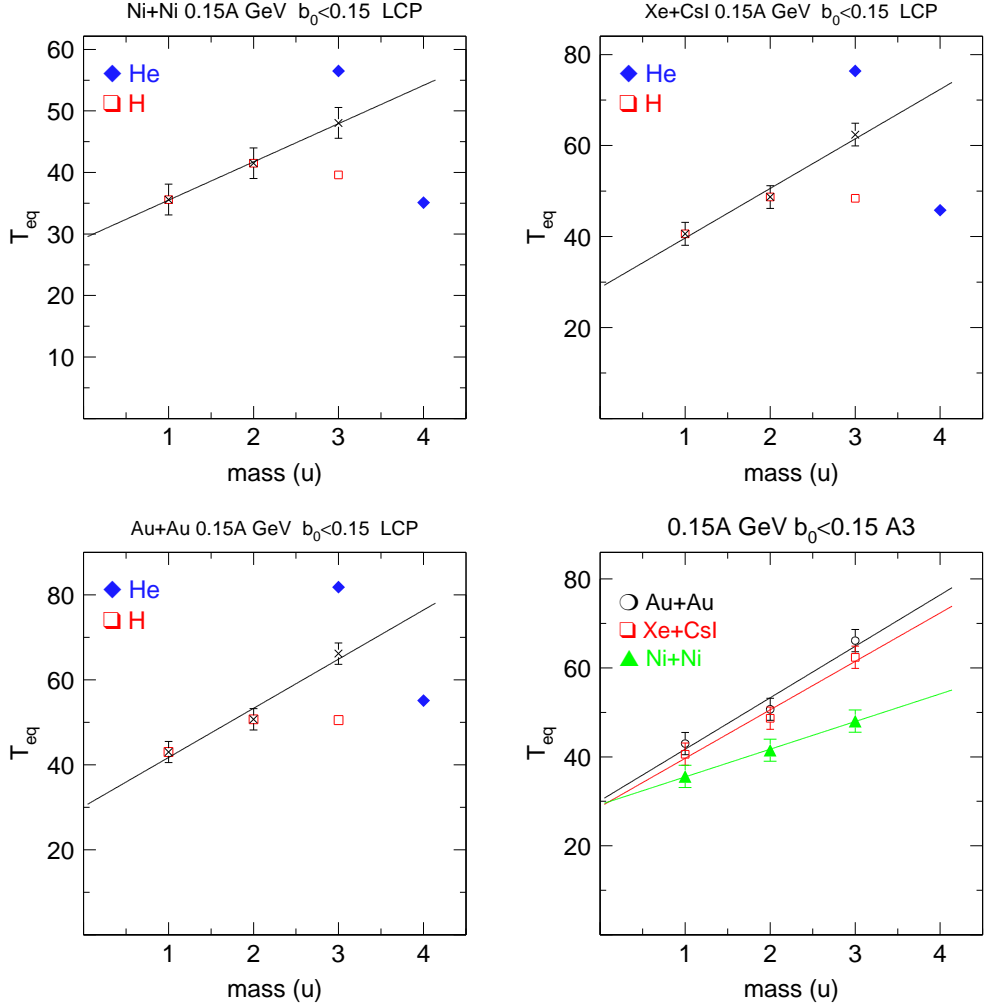,width=140mm}
\end{center}
\caption{
Equivalent transverse temperatures of hydrogen (open red squares) and
helium isotopes (full blue diamonds) as function of their mass number
 in central Ni+Ni
(upper left panel), Xe+CsI (upper right panel) and Au+Au collisions
(lower left panel). 
A linear fit (A3 method) is also shown.
The lower right panel allows to compare the three systems directly using
the data points of the A3 method.
}
\label{dndym-150c1lcp-teq}
\end{figure}

In section \ref{rapidity} it was shown that the constrained rapidity
distributions could be well reproduced with a thermal ansatz, giving
an equivalent temperature $T_{eq}$.
This represents an alternative way of systematizing mass dependences
of phase space distributions.
A systematics of $T_{eq}$ is shown in the next four figures that close
this section.
We find a  similarity with the systematics of average kinetic energies
of fragments emitted around $90^{\circ}$ : 
compare Fig.~\ref{ekinav-au120c4} with Fig.~\ref{dndym-150c1lcp-teq}.
However, since the phase-space cuts are different and the stopping is
incomplete (non-spherical topology),
 the deduced parameters are not substitutes for each other.
While we have started with $90^{\circ}$ kinetic energies
 primarily to join up to some
of the earlier literature \cite{stoicea04,poggi95,lisa95,petrovici00},
 the $T_{eq}$ characterizing the constrained rapidity distributions
allow us to join up to higher energy heavy ion data, including some of our own
work \cite{hong98}, which are
traditionally shown in (more appropriate) axially symmetric coordinate
 systems rather than spherical coordinate systems.
The variation of $T_{eq}$ with particle type evidently excludes a naive
interpretation in terms of a common (equilibrium) temperature.
However, the observations illustrated in
 Figs. \ref{dndy-au150c1Z1A1th} to \ref{dndym-au400c1he} go
beyond just defining average kinetic energies since they allow a statement on
the {\it shapes} of the distributions.

In Fig.~\ref{dndym-150c1lcp-teq}
 we show data for $0.15A$ GeV beam energy varying the system size.
The same strong structures in the data for hydrogen and helium isotopes are
visible and the A3 method introduced earlier, joining the proton and the
deuteron $T_{eq}$ with the average of the two mass three ($^3$He and $^3$H)
data points is seen to define an accurate slope that can be again taken to
quantify radial flow.
The lower right panel allows to assess the system size dependence.

\begin{figure}
\begin{center}
\epsfig{file=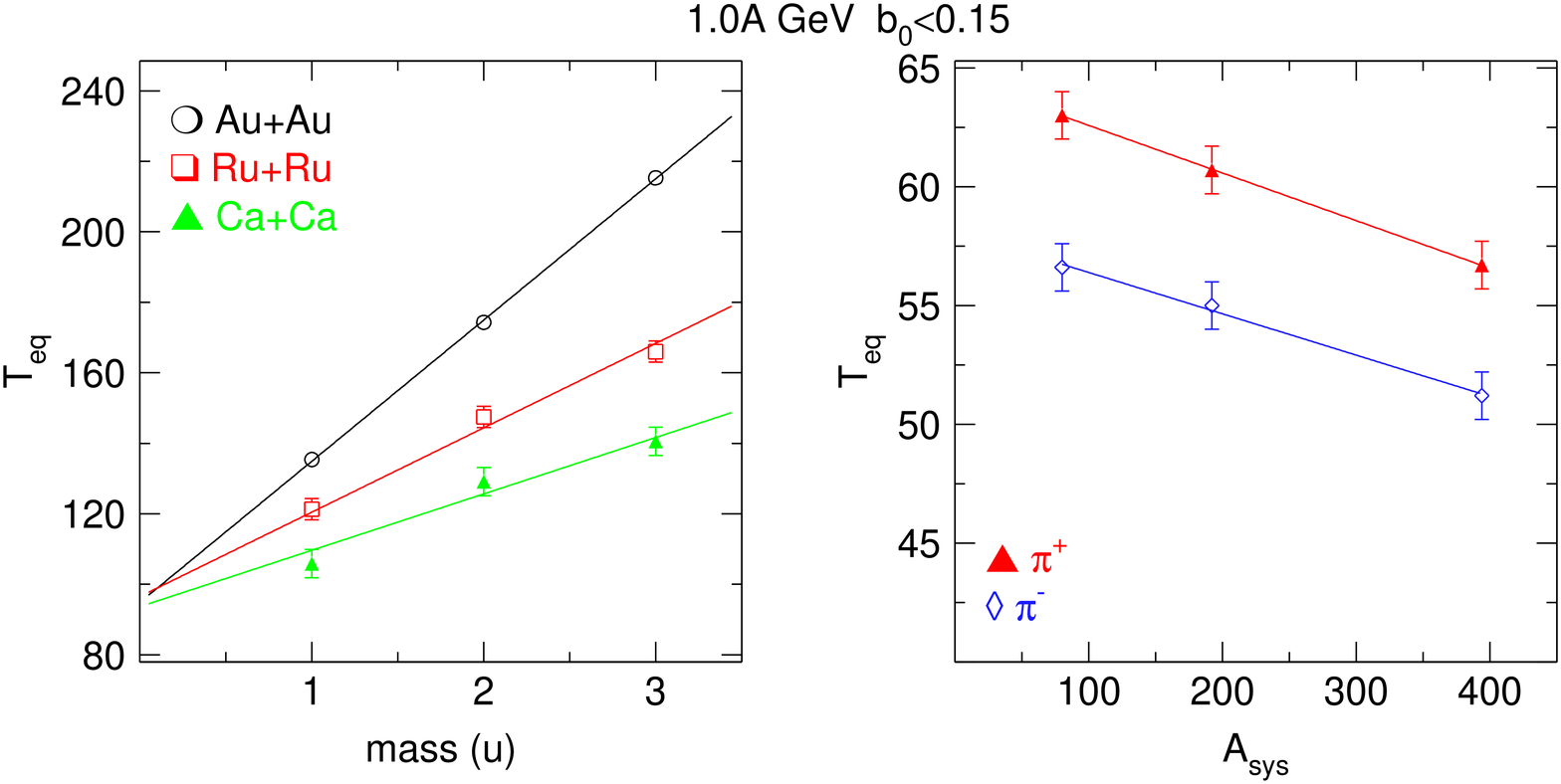,width=140mm}
\end{center}
\caption{
 Equivalent transverse temperatures for symmetric
central collisions at $1A$ GeV beam energy.
Left panel: data for hydrogen isotopes versus mass for three indicated
systems.
Right panel: data for pions versus system mass.
Note the tendency 
(for the hydrogen isotopes) to converge to a common apparent offset
'temperature' $T_o$ (left) and the systematic difference between 
the two charged pions, as well as their weak {\it decreasing} trend 
as a function of $A_{sys}$.
}
\label{dndym-1000c1lcp-teq-2}
\end{figure}

In Fig.~\ref{dndym-1000c1lcp-teq-2} for data obtained at $1.0A$ GeV, we confront
the systematics for hydrogen isotopes (left panel) with that for pions
\cite{reisdorf07} of both charges.
Note the strikingly different system size dependence of nucleonic clusters
and of pions: while the $T_{eq}$ of pions slowly (the data are plotted with
an ordinate offset) decrease with system size, the contrary is true for
the hydrogen isotopes.
Again, as suggested earlier, this is correlated to the different
consequences of compression (creation of pions) and decompression-cooling
(reabsorption of pions, creation of clusters from nucleonic matter).

Apparent temperatures for protons and deuterons in Ni+Ni collisions
have already been published earlier by our Collaboration \cite{hong98}.
If we interpolate the data in Fig.~\ref{dndym-1000c1lcp-teq-2} to estimate
the Ni+Ni values we obtain $T_{eq}$ values exceeding those of \cite{hong98}
by about $15\%$. It is unlikely that the difference is due to a different
analysis method. One possible explanation is the different methods used
to select the most central collisions.
 In \cite{hong98} the multiplicity was maximized between
the polar laboratory angles $(7-30)^{\circ}$, while in the present work
 ERAT, determined using the full apparatus acceptance, was maximized.

\begin{figure}
\hspace{\fill}
\epsfig{file=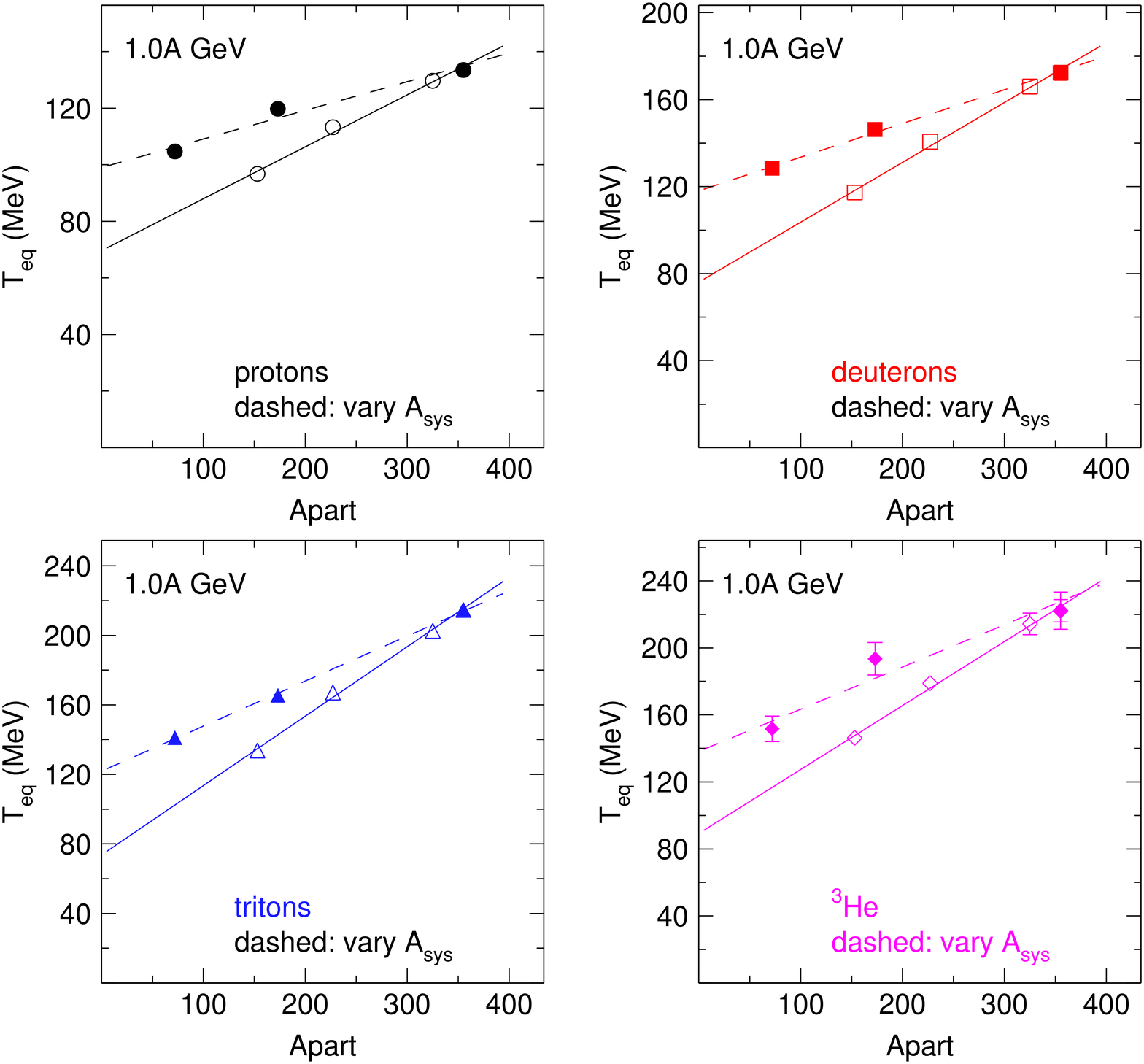,width=150mm}
\hspace{\fill}
\caption{
Equivalent transverse temperatures of protons (upper left panel), deuterons
(upper right panel),  tritons (lower right panel) and $^3$He as function of
participant number $Apart$ for collisions at $E/A=1.0A$ GeV.
The straight lines are linear fits.
Open symbols and solid lines are for a variation of the impact parameter
using Au+Au, full symbols and dashed lines are for a variation of
the system size (Ca+Ca, Ru+Ru and Au+Au) at fixed reduced impact parameters
($b_0<0.15$).
}
\label{dndym-au1000c1pdt-Teq}
\end{figure}

In Fig.~\ref{dndym-au1000c1pdt-Teq} we show an observation typical for
the SIS energy range: the presence of cold spectator matter in non-central
collisions tends to cool locally the participant matter.
We compare collisions of different systems at the same estimated
value $Apart$ of the participant matter.
This effect can be studied in more detail by varying the angle relative
to the reaction plane \cite{stoicea04}, but this will not be done here.

\begin{figure}
\begin{center}
\epsfig{file=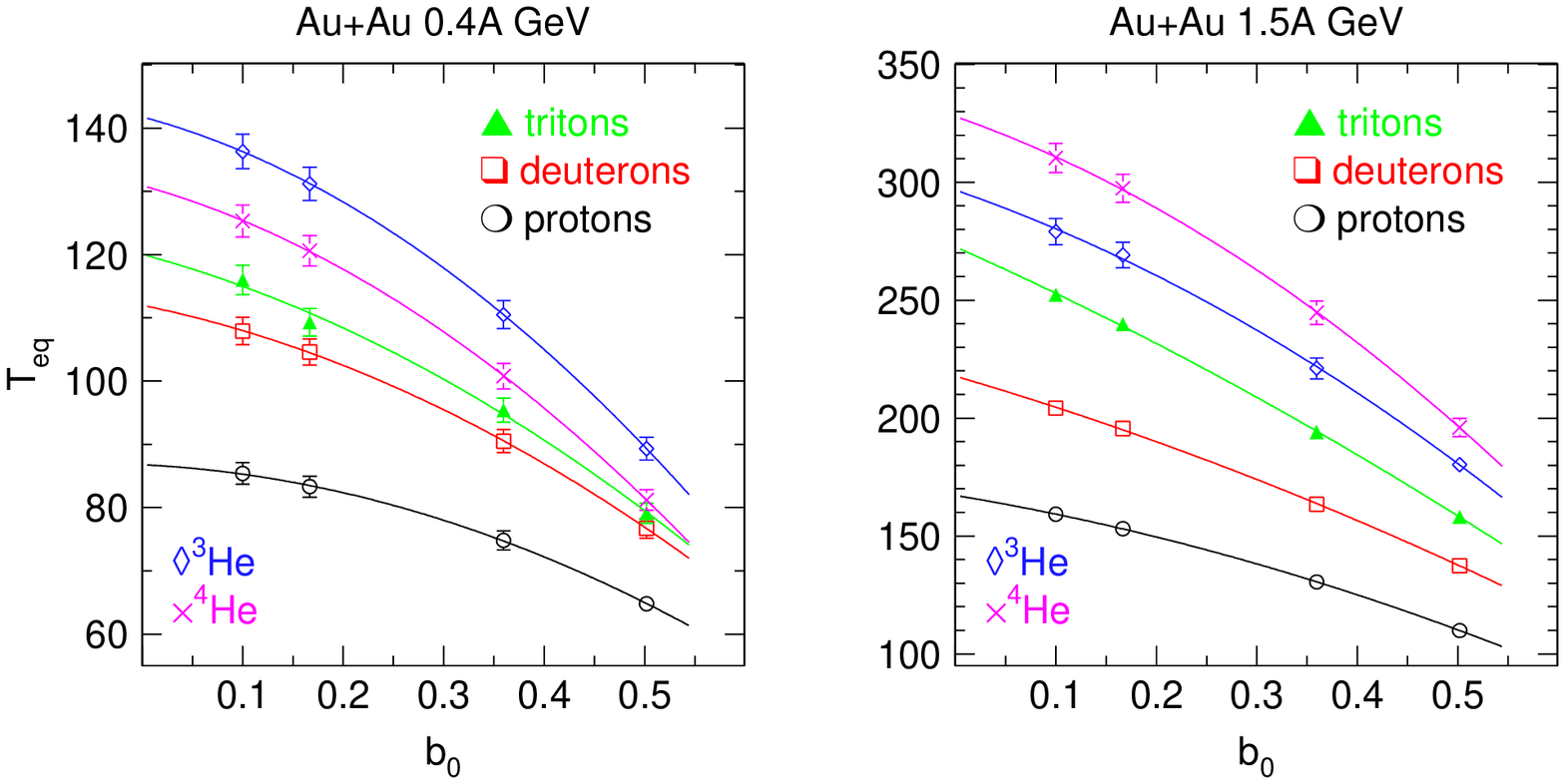,width=150mm}
\end{center}
\caption{
Centrality dependence of equivalent transverse temperatures for
protons (circles), deuterons (squares), tritons (full triangles),
$^3$He (diamonds) and $^4$He clusters in Au+Au collisions.
Left panel: 0.4A GeV incident beam energy, right panel: 1.5A GeV.
The smooth lines guide the eye and allow an estimate for $b_0=0$.
}
\label{varau400}
\end{figure}

Finally, we present in Fig.~\ref{varau400} for Au+Au at two different
incident energies ($0.4A$ and $1.5A$ GeV) a complete systematics
of $T_{eq}$ for all five LCP varying the centrality.
From the smooth trends the limits for $b_0=0$ can be deduced.
Note the inversion of the hierarchy for the two He isotopes
when passing from the higher to the lower energy.

\section{Chemistry}\label{chemistry}
The $4\pi$ reconstructions, explained in detail in section \ref{reconstruction},
allow us to assess the integrated yields of all the ejectiles that
constitute the bulk of the colliding system.
For the most central collisions we present these yields in tabulated form
in an appendix for the 25 system-energies studied in
this work.
Except for the two lowest incident energies ($0.09A$ and $0.12A$ GeV),
we account for the total system charge with an accuracy of about $5\%$ or
better.
Note that we include in this balance also the pions, \cite{reisdorf07}.
Other created particles, such as kaons etc, are negligible on
the $5\%$ level (but are of course interesting for many other reasons).
For the two lowest energies we miss significant contributions from
heavy clusters beyond Z=8.
The interested reader might consult INDRA-ALADIN works 
\cite{andronic06,lefevre04,lavaud01} 
for more complete distributions  in the energy range at and below $0.15A$ GeV.

 Such data are often analysed using
models assuming (local) chemical equilibrium with a unique 'chemical'
 temperature .
An early, instructive presentation of the arguments in favour of a conjectured
chemical equilibrium at freeze-out can be found in \cite{mekjian78}.
Even if one does not adopt the equilibrium assumption, (many of the observations
presented in this work do not favour the equilibrium assumption)
the degree of clusterization quantized by these yields can be used
to constrain the outgoing (non-equilibrium) entropy 
\cite{kuhn93,doss88}.
This information, together with our extensive data on stopping,
section \ref{stopping}, can be used to assess the viscosity to
entropy ratio that is currently intensely investigated in connection
with the quark-gluon phase thought to be copiously produced at the
highest currently available energies \cite{QM05,QM06,QM08}.
The task of analysing our chemistry data using
a modern thermal code that bridges the low-energy multifragmentation regime
with the high energy particle-producing regime using a realistic
approach for late decays  is delayed to  later publications.
In \cite{reisdorf97} it was shown that a purely statistical interpretation of
the non-collective part of the available energy using various statistical
codes that included late evaporations was not successful as it strongly 
underestimated the yields of heavy clusters, such as oxygen.
This conclusion is not touched by the now favoured interpretation of the 
collective energy as being to a sizeable part due to still two-fluid flow
caused by incomplete stopping.
In contrast to this failure one has to acknowledge a very reasonable 
rendering of the full nuclear charge distributions for Au on Au at 0.15 and
$0.25A$ GeV when using the AMD code (Antisymmetrized Molecular Dynamics)
\cite{ono99,ono00}. This transport code does not rely on the equilibrium
assumption.

Despite the different data analysis method, the yields listed in the
appendix for Au+Au at 0.15, 0.25 and $0.4A$ GeV are fully compatible with
those published earlier by our Collaboration \cite{reisdorf97,petrovici95}.
The possibility to compare these yields with those published by  other
authors are limited: besides the need to match centralities, one is generally
confronted with the lack of $4\pi$ estimates.
As shown partially in \cite{andronic06}, our data are nicely compatible with
INDRA data in the low energy range where the measurements overlap
and centralities have been carefully aligned.

The situation on cluster production as it existed in the mid-eighties
was described in \cite{csernai86}.
Well suited for a direct comparison are
yield ratios of light clusters (d, t, $^3$He, $\alpha$) to protons 
measured  \cite{doss88} with the Plastic Ball spectrometer
for the reactions Nb+Nb (which is close to one of our systems: Ru+Ru) and
Au+Au for incident beam energies of 0.15, 0.25, 0.40 and 0.65 GeV/nucleon and
were
plotted by the authors versus the so-called participant proton multiplicity,
$N_p$,
which was the total charge included in all the LCP cutting out some areas close
to projectile and target rapidity to exclude 'spectators'.
(Without these specifications the total charge would always be
trivially constant in each collision due to charge conservation).
Assuming Boltzmann like spectra, ratios were taken in limited regions
of phase space where isotopes were
identified and the momenta scaled with $(1/m)^{1/2}$.
If we compare these ratios taken at the highest $N_p$ with corresponding
ratios obtained from our Table for the most central collisions we 
get  excellent agreement, for most cases.
See the sample Table~\ref{tab-PB} below for Au+Au at $0.15A$, $0.40A$, and
Nb+Nb/Ru+Ru at $0.4A$ GeV. 
  
\begin{table}
\begin{center}
\caption{
Ratios  of light cluster yields to proton yields in central collisions.
Comparison of Plastic Ball data \cite{doss88} with FOPI data.
}
\label{tab-PB}
\bigskip

\begin{tabular}{c|cc|cc|cc}
\hline
  & \multicolumn{2}{|c}{Au+Au $0.15A$ GeV} &
\multicolumn{2}{|c}{Au+Au $0.4A$  GeV} &
\multicolumn{2}{|c}{Nb(Ru)+Nb(Ru) $0.4A$ GeV} \\
ratio & Plastic Ball & FOPI & Plastic Ball & FOPI & Plastic Ball & FOPI\\
\hline
d/p       & $0.80 \pm 0.08$  & $0.85 \pm 0.06$
          & $0.65 \pm 0.07$  & $0.65 \pm 0.04$ 
          & $0.50 \pm 0.05$  & $0.55 \pm 0.03$  \\
t/p       & $0.40 \pm 0.04$  & $0.63 \pm 0.06$  
          & $0.40 \pm 0.04$  & $0.37 \pm 0.03$ 
          & $0.28 \pm 0.04$  & $0.25 \pm 0.025$ \\
$^3$He/p  & $0.50 \pm 0.045$ & $0.35 \pm 0.04$  
          & $0.20 \pm 0.02$  & $0.18 \pm 0.02$ 
          & $0.18 \pm 0.02$  & $0.17 \pm 0.017$  \\
$^4$He/p  & $0.65 \pm 0.06$  & $0.69 \pm 0.07$  
          & $0.20 \pm 0.02$  & $0.22 \pm 0.02$ 
          & $0.16 \pm 0.02$  & $0.16 \pm 0.016$ \\
\hline
\end{tabular}
\end{center}
\end{table}
 
The Plastic Ball data were read off the figures in \cite{doss88} 
and the uncertainties somewhat
arbitrarily set at about $10\%$ looking at data point straggling and assessing
reading errors.
The agreement is not trivial as there are, besides the very different
apparatus used, some differences in the
experimental procedure:
In contrast to \cite{doss88} we have obtained $4\pi$ reconstructed yields and
no  cuts on the data were done.
Our procedure therefore does not involve the necessity to assume Boltzmann-like
spectra, all with  the same apparent temperature,
 nor, for these {\it very central} collisions, an arbitrary definition
of 'spectator' contributions. Section \ref{rapidity} shows that such
contributions would be difficult to identify in the rapidity distributions.
 The authors of ref. \cite{doss88}, comparing with the
predictions of a two parameter quantum statistical code, QSM, \cite{hahn88},
conjectured that the only serious discrepancy with the code
 (and as can be seen from
 Table~\ref{tab-PB}, also with our data),  the $t/p$ and 
$^3$He/p ratios for Au+Au at $0.15A$ GeV, were due to deviations from
the Boltzmann shapes.
Our data fully support this conjecture: at the lower energies the spectra
of $^3$He and $^3$H are very different, see also section \ref{rapidity}.
As mentioned before,
we shall not repeat statistical model calculations in the present work
although our data considerably extend the available information.
Rather, it is desirable that such data be reproduced by the same
transport model codes that are used to extract EOS information from the data:
 here the precise sorting with fixed cross sections
of ERAT, an observable which is not directly connected with multiplicity,
 should be conceptually advantageous.
 
\begin{figure}
\begin{center}
\epsfig{file=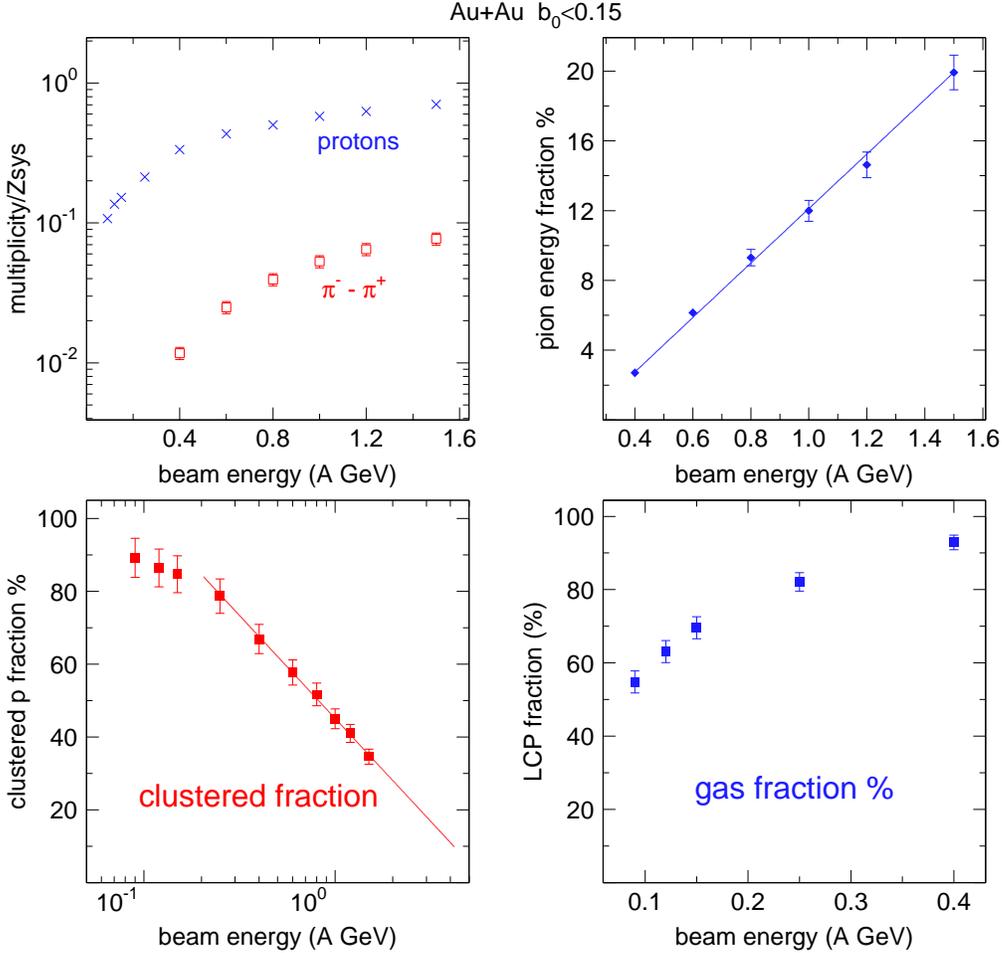,width=140mm}
\end{center}
\caption{
Some global characteristics of central Au+Au collisions as function of
incident energy.
Upper left panel: 
Reduced proton multiplicities and charge conversions by pion emission.
Upper right panel: fraction (in $\%$) of total energy emitted as pions.
Lower left  panel: $\%$ fraction of protons in clusters.
Lower right panel: measured nuclear charge fraction (in $\%$) emitted as LCP. 
}
\label{y-auc1-cluster-4}
\end{figure}

Even without a detailed comparison with statistical or dynamical codes
a number of interesting features can be deduced by just plotting some of
the information enclosed in our chemistry data.
We begin with some global features, shown in Fig.~\ref{y-auc1-cluster-4}.
As illustrated in the upper left panel,
in the SIS energy range the percentage of single protons emitted in central
Au+Au collisions starts at the lower end at roughly the $10\%$ level to
gradually increase to about $2/3$ of the available charge at $1.5A$ GeV.

The fraction complementary to single protons, the clustered fraction,
is plotted separately again in the lower left panel. Although it contains
no additional information, since it must complement the proton fraction to
$100\%$, the ordinate is now linear and the abscissa is now logarithmic,
allowing, in this representation, to illustrate a linear trend above
$0.2A$ GeV (see the straight line, that reproduces the data with an accuracy
of $1.5\%$) that suggests by extrapolation that the clustered fraction stays
above $10\%$ all the way up to $4A$ GeV.
This persistence of a significant probability to clusterize at freeze-out
up to an available energy per nucleon more than two orders of magnitude higher
 than typical nucleonic binding energies is a remarkable signal of
the local cooling process accompanying the fireball expansion and
serves as a strong constraint on the associated entropy.

In some of the literature, see for example \cite{rizzo07}, particles with
$Z\leq 2$ (LCP's) are supposed to be the ingredients of nuclear matter in the
gas state. 
These are the particles most easily evaporated from a hot liquid
and therefore forming the gaseous atmosphere above the liquid.
If one accepts this simple view then the 'fifty-fifty' crossing liquid-gas
takes place around $80A$ MeV beam energy.
Since 'gas' and 'liquid' should add up to $100\%$, one can also read off the
'liquid' fraction from the figure, which we call heavy clusters 
(or 'droplets' \cite{reisdorf04b}) and which is
called frequently IMF (intermediate mass fragments) in the literature.
Above $0.4A$ GeV the liquid fraction is  $5\%$ or less.
In the energy range considered here,  fully central collisions do not
lead any longer to fission-like (slightly less than half the projectile or
target mass) or compound-nuclear-like (half or more than the total mass)
 remnants.

The SIS energy regime is also the regime where particle creation,
very dominantly pions, varies from a 'perturbative' low level to a significant
fraction expected to heavily influence the global dynamics.
Thus, the energy fraction rises to the $20\%$ level as seen in the upper
right panel. The pertinent data were obtained from \cite{reisdorf07}.

The non-perturbative nature of pions at the high end of this energy regime
 is also
active for the isospin degree of freedom. 
According to the difference in $\pi^-$ and $\pi^+$ production, 
\cite{reisdorf07}, plotted also in the upper left panel, some eight protons
are converted to neutrons at freeze out, modifying substantially the
initial $N/Z$ (from 1.494 to 1.264), a feature that in turn will
modify the relative yields of the isotopes of hydrogen and helium.

\begin{figure}
\begin{minipage}{75mm}
\epsfig{file=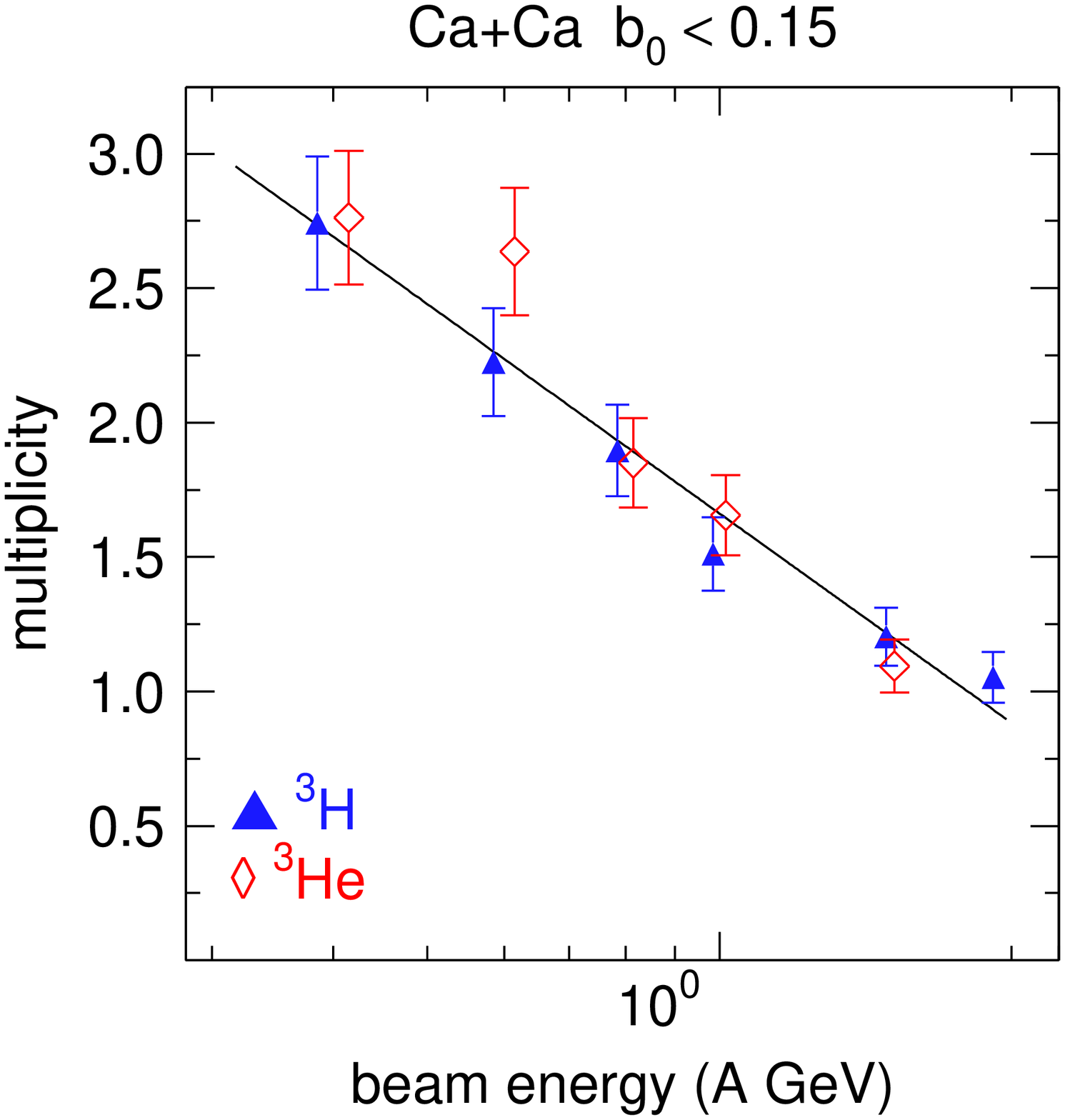,width=74mm}
\caption{
$^3$He and $^3H$ multiplicities in central $^{40}$Ca+$^{40}$Ca collisions
 as function of the incident beam energy.
 The straight line is a common fit linearly decreasing
with the logarithm of the energy.
}
\label{auca/y-cac1-a3}
\end{minipage}
%
\hspace{10mm}
\begin{minipage}{75mm}
\epsfig{file=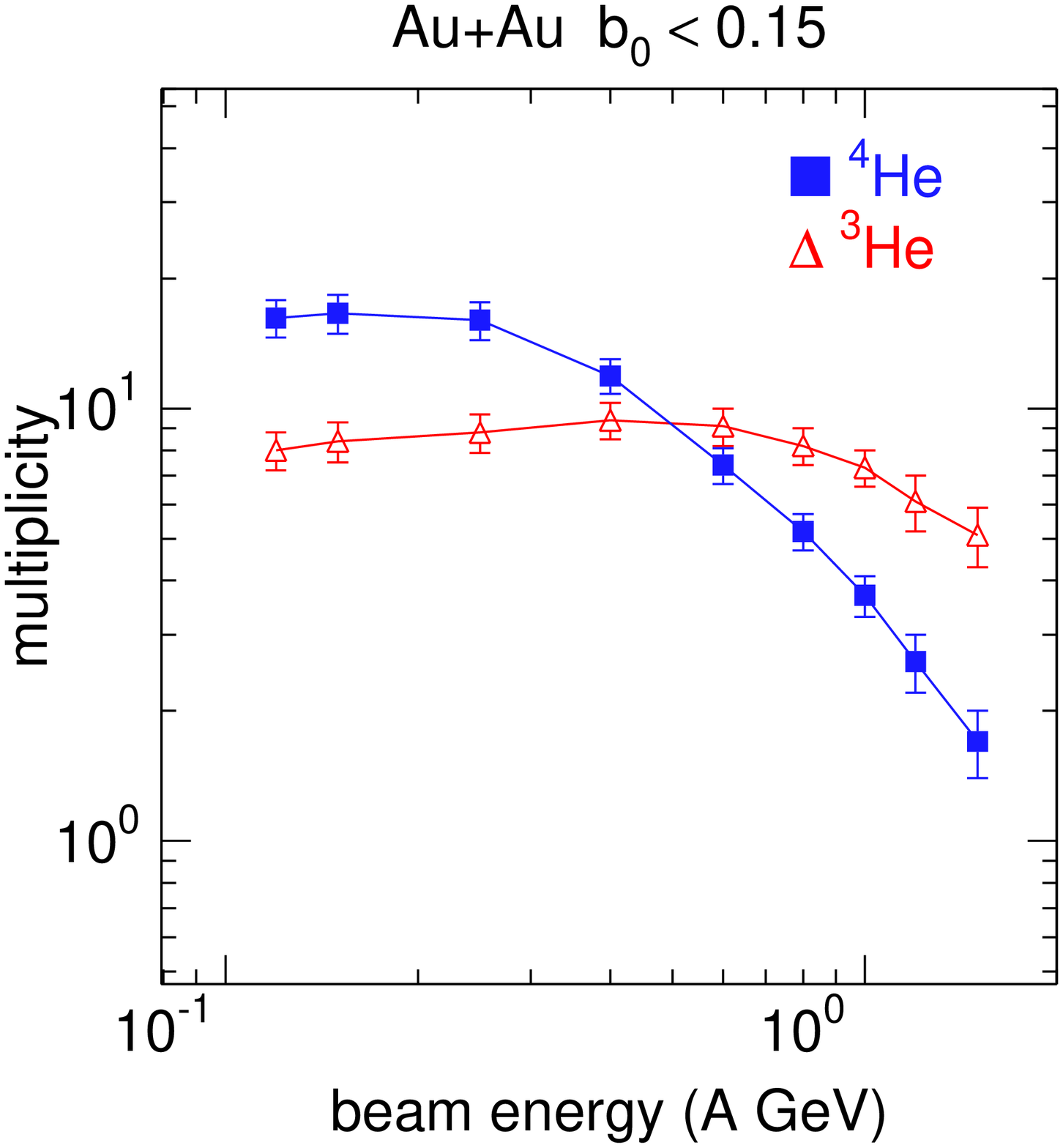,width=74mm}
\caption{
$^3$He and $^4He$ multiplicities in central Au+Au collisions as function of
the incident beam energy. The 'crossing' energy is $0.5A$ GeV.
}
\label{y-auc1-he}
\end{minipage}
\end{figure}

We will later compare yields of the isospin pair $^3$He and $^3$H.
It is therefore of some interest to check if the yields are equal in an
isospin symmetric system such as $^{40}$Ca+$^{40}$Ca.
In this system only Coulomb effects could be invoked to predict different
yields for the two isobaric clusters.
Within experimental uncertainties the yields are equal in the investigated
energy range (0.4 to $2A$ GeV), see Fig.~\ref{auca/y-cac1-a3}.

We have already discussed in section \ref{rapidity} the non-perturbative
character of clusterization. 
The high degree of clusterization seen in Fig.~\ref{y-auc1-cluster-4}
confirms this aspect.
In  Fig.~\ref{y-auc1-he} we show excitation functions for the two He isotopes
in central collisions of Au+Au. 
The two functions cross around 0.5A GeV.
Clearly a simple perturbative coalescence model \cite{swang95}
 could not explain this behaviour.

\subsection{Mid-rapidity chemistry}
In the present energy range a clean separation between 'participating'
(originally overlapping) matter and (peripheral) 'spectator' matter is
not possible as, especially in the most central collisions, there are
no readily identifiable 'spectators' moving undisturbed into angles close
to the beam direction.
Rather there is a superposition of various ejectiles having a more or less
violent collision history as suggested by the observed stopping hierarchies
(section \ref{stopping}).
To minimize the effects of less violent and less equilibrated
 surface collisions,
we concentrate our attention to (longitudinal) 
'midrapidity chemistry', defined here by a cut $|y_{z0}|<0.5$.
As explained already in section \ref{rapidity}, we remove some of
the arbitrariness of the $|y_{z0}|$ cut using the thermal model fits to
the constrained rapidity distributions applying the same cut and then 
correcting for the cut-out part using the fitted model data.
This gives us an estimate of mid-rapidity  yields extended to full rapidity.

\begin{figure}
\begin{center}
\epsfig{file=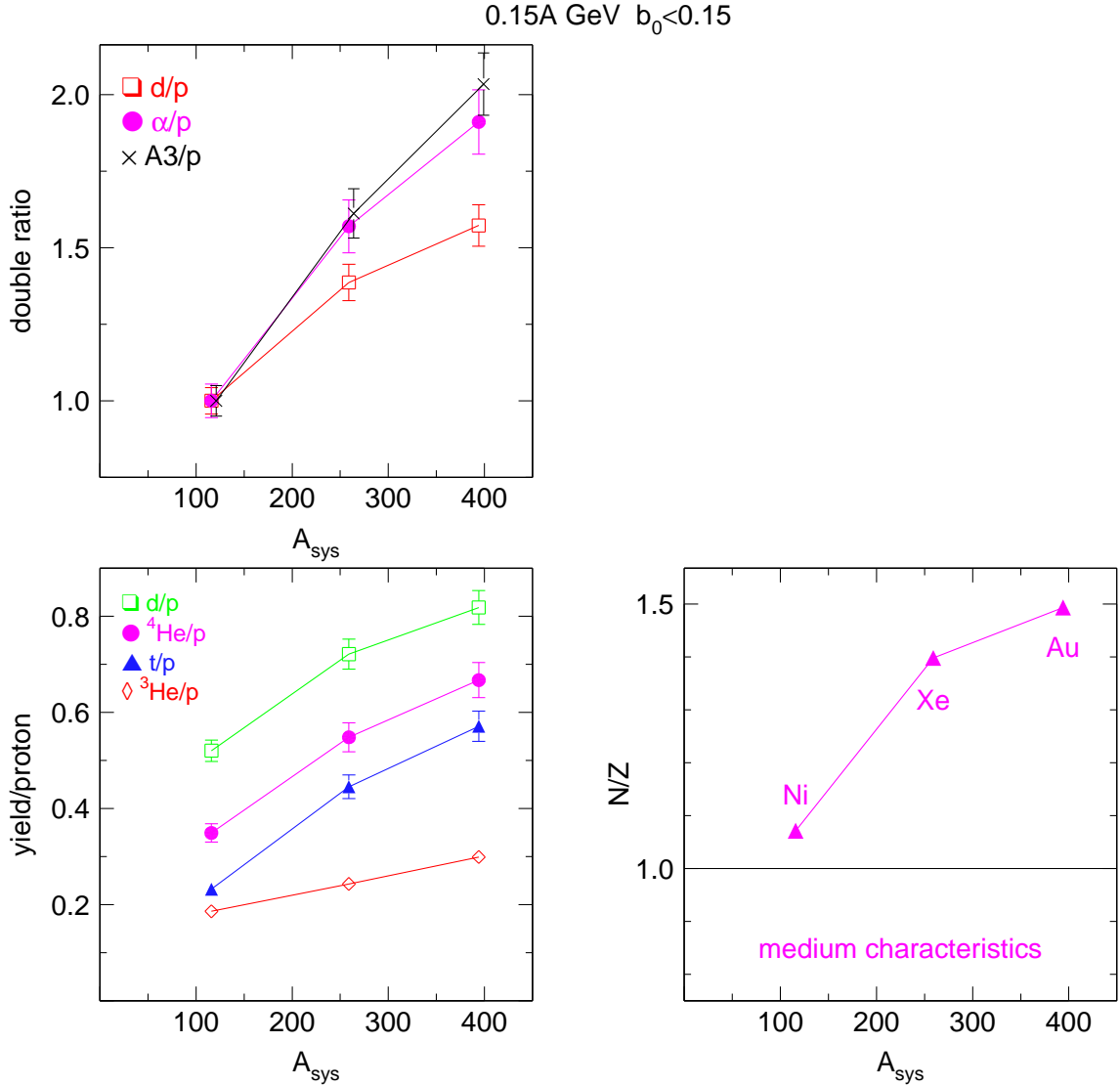,width=162mm}
\end{center}
\caption{
System size dependences for beam energies of $0.15A$ GeV.
Lower left panel: yield ratios relative to protons for
$^4$He (pink closed circles), $^3$He (red open diamonds), tritons
(blue closed triangles) and deuterons (green open squares). 
Upper left panel: system size dependence of mass two, three and four
yields relative to protons. The plotted
double ratios are relative to the corresponding ratio in the Ni+Ni system
$( A_{sys}=116)$.
The medium characteristics for the three studied systems in terms of $N/Z$
and total mass $A_{sys}$ are drawn in the lower right panel.
}
\label{dndym-150c1lcp-yield}
\end{figure}

\begin{figure}
\begin{center}
\epsfig{file=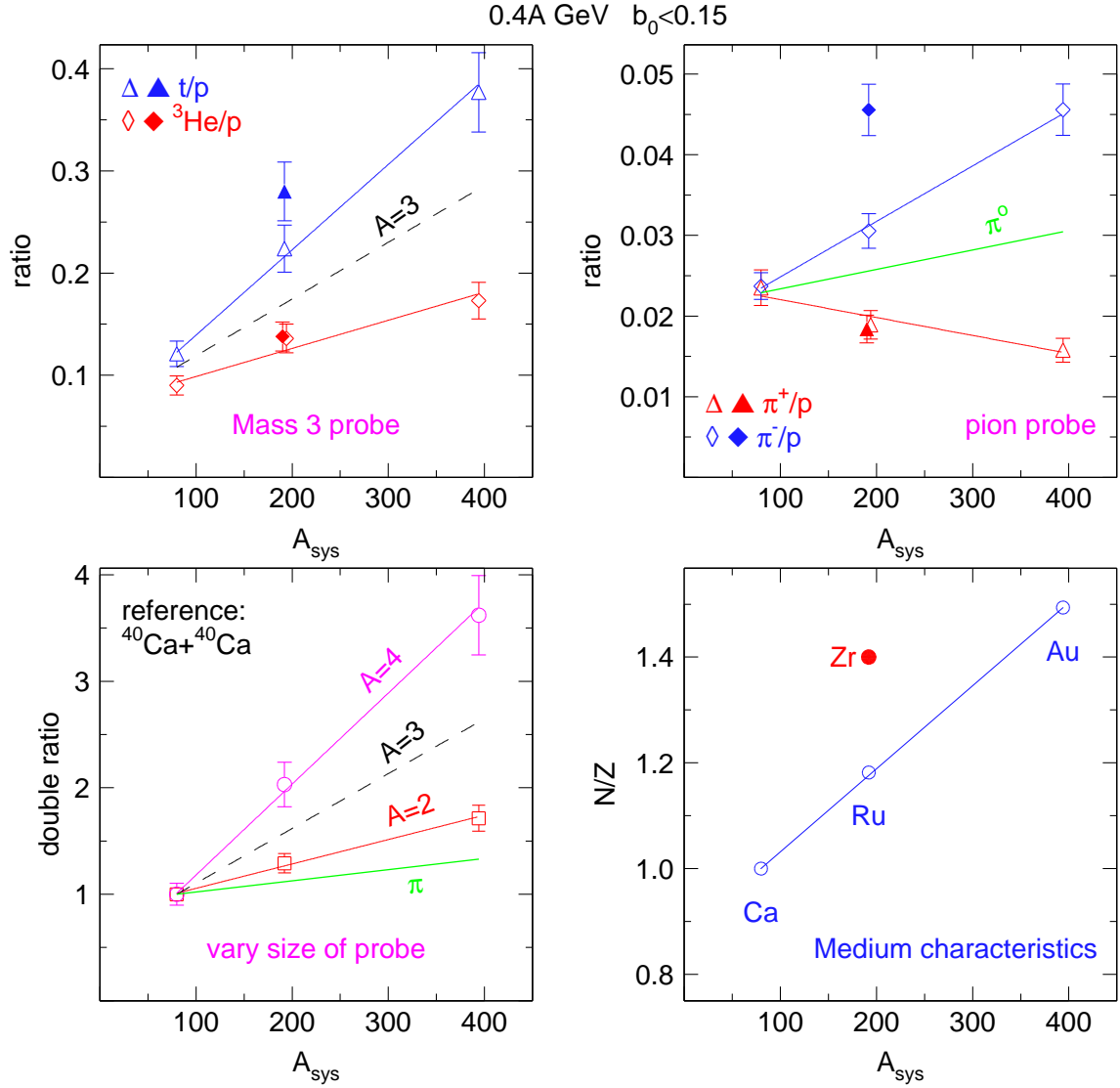,width=162mm}
\end{center}
\caption{
System size dependences for beam energies of $0.4A$ GeV.
Upper left panel: $t/p$ and $^3$He/p ratio, upper right panel: $\pi^+/p$
and $\pi^-/p$.
The straight lines are linear fits excluding the Zr+Zr data points which
are plotted with full symbols.
The resulting (linear) 'isospin averaged' trends are also drawn.
Lower left panel: double ratios (with respect to protons and 
relative to the Ca+Ca system, $A_{sys}=80$)
for mass 4, 3, 2 fragments and pions (upper to lower curves).
 The Zr+Zr data points are not included here.
The medium characteristics for the four studied systems in terms of $N/Z$
and total mass $A_{sys}$ are drawn in the lower right panel.
}
\label{dndym-400c1lcp-yield}
\end{figure}

\begin{figure}
\begin{center}
\epsfig{file=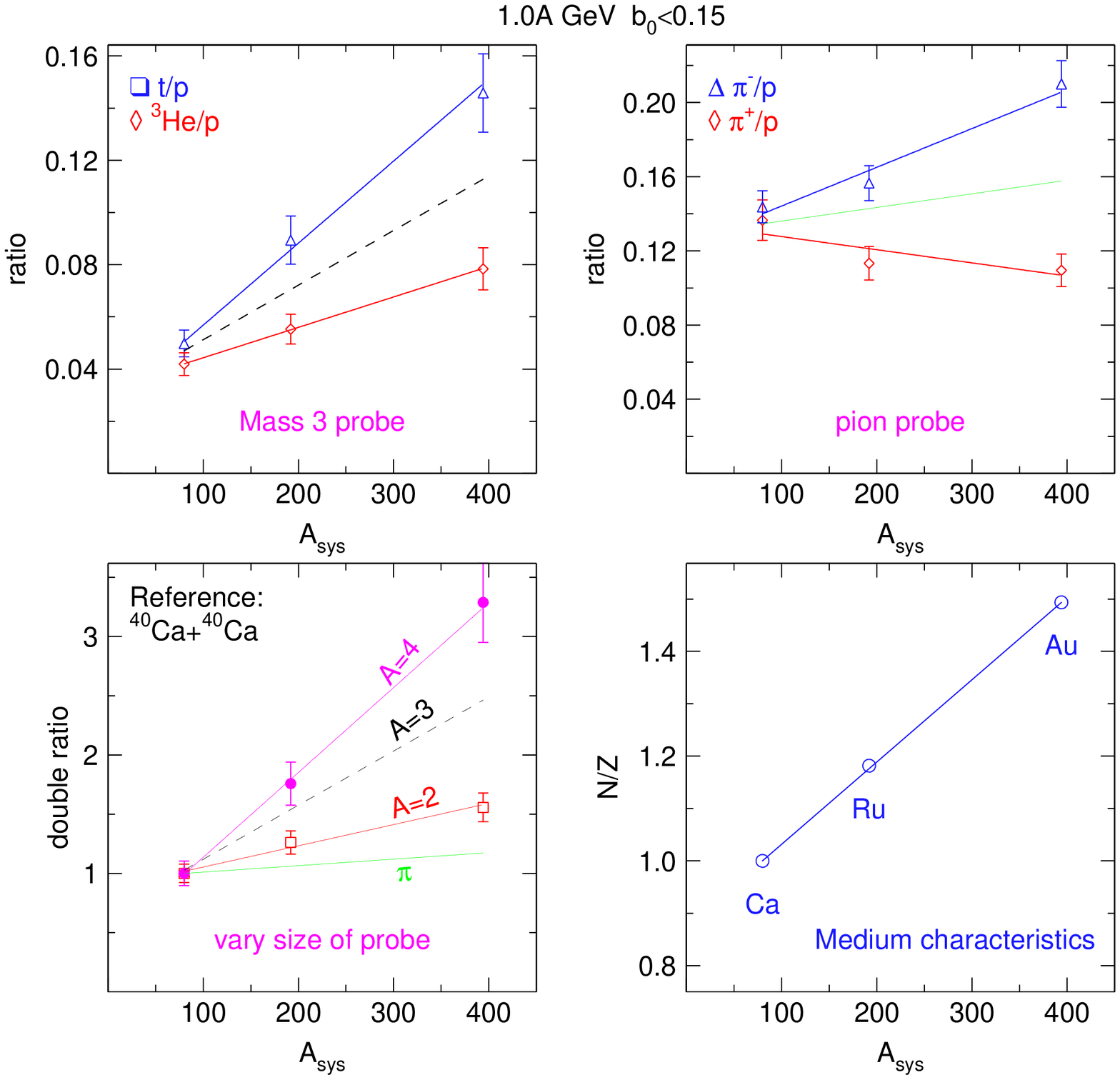,width=162mm}
\end{center}
\caption{
System size dependences for beam energies of $1.0A$ GeV.
Upper left panel: $t/p$ and $^3$He/p ratio, upper right panel: $\pi^+/p$
and $\pi^-/p$.
The straight lines are linear fits guiding the eye.                       
The resulting (linear) 'isospin averaged' trends are also drawn.
Lower left panel: double ratios (with respect to protons and 
relative to the Ca+Ca system, $A_{sys}=80$)
for mass 4, 3, 2 fragments and pions (upper to lower curves).
The medium characteristics for the three studied systems in terms of $N/Z$
and total mass $A_{sys}$ are drawn in the lower right panel.
}
\label{dndym-1000c1lcp-yield}
\end{figure}

\begin{figure}
\begin{center}
\epsfig{file=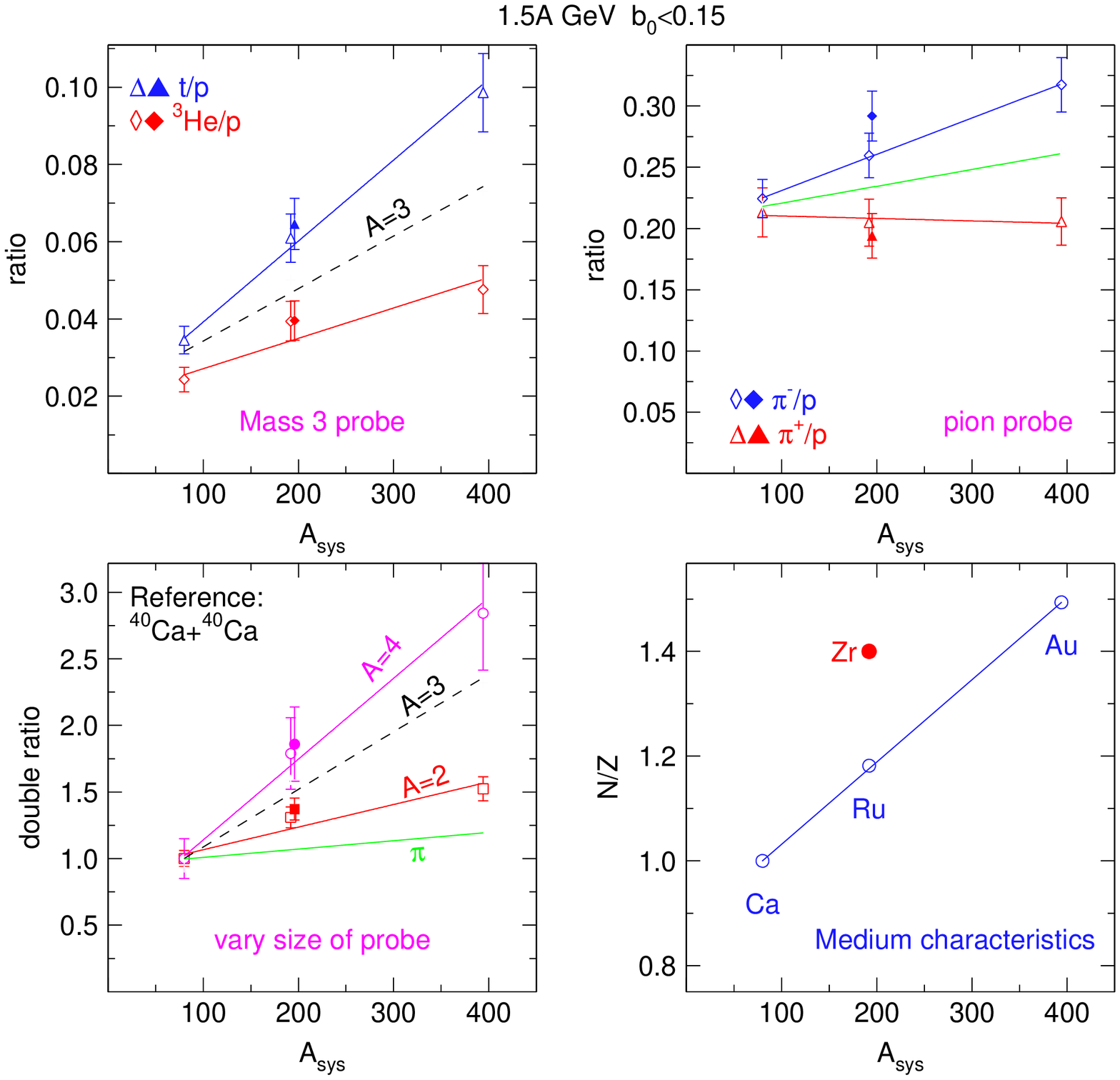,width=162mm}
\end{center}
\caption{
System size dependences for beam energies of $1.5A$ GeV.
Upper left panel: $t/p$ and $^3$He/p ratio, upper right  panel: $\pi^+/p$
and $\pi^-/p$.
The straight lines are linear fits guiding the eye.                       
The resulting (linear) 'isospin averaged' trends are also drawn.
Lower left panel: double ratios (with respect to protons and 
relative to the Ca+Ca system, $A_{sys}=80$)
for mass 4, 3, 2 fragments and pions (upper to lower curves).
Data for the Zr+Zr system are drawn as full symbols and are not included in the
fits.
The medium characteristics for the four studied systems in terms of $N/Z$
and total mass $A_{sys}$ are drawn in the lower right panel.
}
\label{dndym-1500c1lcp-yield}
\end{figure}

 In order to remove trivial size dependences
 we present ratios to proton yields, loosely writing
e.g. d/p for the ratio of deuterons to protons.
We shall see that even these 'size-reduced' observables
 (and hence the extracted effective
thermal parameters) are definitely still system size dependent.
Furthermore, using the isospin pairs $^3$H/$^3$He and $\pi^-$/$\pi^+$, we are
able to separate size and isospin dependences: like the study of
ground state masses, also the variation of the reaction systems along the
valley of stability to change the system size necessarily leads to strongly
 correlated simultaneous change of the isospin of the system.
For $0.15A$, $0.4A$, $1.0A$ and $1.5A$ GeV we show the results of this kind
of analysis, see the next four figures,
Figs.~\ref{dndym-150c1lcp-yield} to \ref{dndym-1500c1lcp-yield},
 and their captions. To each of these
figures we add a panel showing the characteristics of the studied systems in
terms of $N/Z$ versus $A_{sys}$.
To better see the system size variations relative to the 
lightest system (Ni+Ni at the lowest energy, Ca+Ca at the higher energies),
we also use double ratios.
As a main conclusion from these rich data we can say that there is a
significant influence of both system size and system composition
 on the outcoming yields of light clusters and of pions.
When {\it averaging} over the isospin partners $^3$He/$^3$H, resp.
$\pi^-/\pi^+$ we see (left lower panels in 
Figs.~\ref{dndym-400c1lcp-yield} to \ref{dndym-1500c1lcp-yield})
a regular evolution of the system-size dependence with the mass of the
ejectile. 
In particular we confirm the weak dependence of total, or
isospin averaged, pion multiplicities \cite{reisdorf07}. 

In contrast, in the upper two panels of the same figures the data for
 two isospin
partners are seen to branch off as the $N/Z$ is varied along with the size.
 In an isospin asymmetric $(N>Z)$ medium like $^{197}$Au + $^{197}$Au
 (in contrast to    $^{40}$Ca + $^{40}$Ca, Fig.~\ref{auca/y-cac1-a3})
 more n-n than p-p collisions take place, hence more
   $\pi^-$ than $\pi^+$ are produced (see also our work ref.~\cite{reisdorf07}).
   For similar combinatorics reasons more $^3$H than $^3$He are synthesized.
   The more interesting question is whether
   this is also influenced by density-dependent long-range isovector potentials.
   This question, so far, is not settled. For instance, for pion ratios, see
   the contradictig references \cite{ferini06,zgxiao09,zqfeng10}.
Concerning $^3$He and $^3$H, to our knowledge no microscopic transport
code has reproduced so far the special 'anomaly' features discussed earlier
in connection with Figs.~\ref{dndym-au400c1p} and \ref{dndym-au400c1A3-a}
in section \ref{clusterization} and Fig.~\ref{ekinavc4-r} (right panel) 
in section \ref{radial}.
This makes also a convincing rendering of the experimental yield data
difficult. Obviously more work is necessary.

\begin{figure}
\begin{center}
\epsfig{file=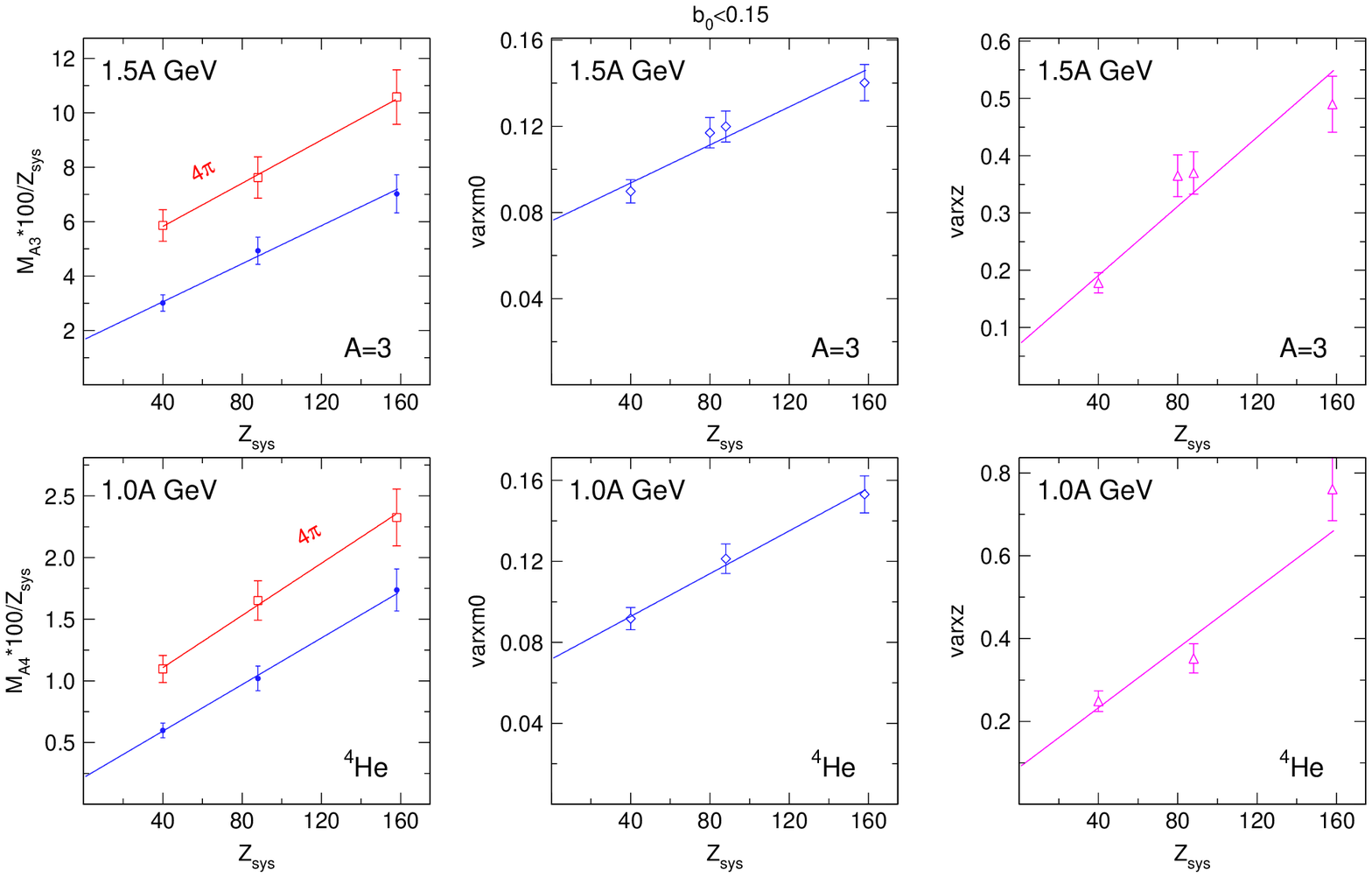,width=150mm}
\end{center}
\caption{
System size dependences of symmetric
central collisions at incident beam energies of $1.0A$ GeV
 (lower panels, $^4$He fragments)
and $1.5A$ GeV (upper panels, average of $^3$He and $^3$H fragments).
All straight lines are linear fits. 
Left panels:
 Multiplicities (reduced to 100 protons).
The data marked $4\pi$ with open square symbols are $4\pi$ integrated,
while the closed symbols represent mid-rapidity yields
(constrained to $|y_{z0}|<0.5$).
Middle panels:
Variance {\it varxm0} of the scaled constrained transverse rapidity
 distributions
 ($dn/dy_{xm0}$). 
Right panels: stopping {\it varxz}.
}
\label{y-A4}
\end{figure}

A final important point to make on the chemistry is the strong correlation
of the yields, the transverse rapidity variances and the fragment
specific stopping when varying the
system sizes $Z_{sys}$.
This effect is best illustrated when feeding from higher mass clusters is
small
so that one can envision the synthesis of new nuclei from a hot expanding,
originally {\it nucleonic}, soup.
In view of the steep decrease of yields with fragment size,
this condition could be safely assumed to be fulfilled at $0.4A$ GeV for Li
 \cite{reisdorf04b}.
As is shown in Fig.~\ref{y-A4}  the same can be assumed to be true at
 $1.0A$ GeV for $^4$He clusters  and at $1.5A$ GeV for mass three clusters.
To remove superimposed isospin effects we have added the mass three H and He
isotopes.
Increased stopping (right panels) indicates increased compression.
The increasing constrained transverse rapidity variances {\it varxm0} 
(middle panels) are interpreted
 as increasing radial
flow developed thereafter in the expansion phase which is coupled to
 increased cooling ('droplet formation').
(We remind the reader that not only the variances, but also the shapes of
 these distributions were well described by a thermal shape assuming an
equivalent temperature $T_{eq}$, see section \ref{LCP} and figs. 11-16.)
The interpretation is supported by the remarkable, close-to-quadratic
(the power is $1.7\pm0.1$) system-size dependences of the mid-rapidity cluster
 multiplicities already  stressed in ref. \cite{reisdorf04b}.
Kaon ($K^+$) production in the SIS energy range is also  known
 \cite{foerster07} to show  stronger than linear
size dependences  associated to varying achieved compressions.
The mechanisms are different however: in the kaon case there is
increased production in the high density case which thereafter, in
contrast to pion production \footnote{the insensitivity of pion yields to
assumptions on the EOS and the system size is due, in this interpretation,
 to a compensation of higher production at higher
achieved maximum density by a stronger absorption-cooling when expanding}, 
is partially 'memorized' due to small $K^+$ reaction cross
sections making the cooling phase 'inefficient'. 
For clusters we start 'from zero' at maximum achieved density,
 but increased compression leading to subsequent
increased cooling creates more clusters.
In this sense we suggest that this process, like $K^+$ production, is
sensitive to the EOS because it partially memorizes the initial compression
which, in turn, is largest for softest EOS.

Loosely speaking,
we are thus witnessing the creation of nuclei in a mini-'supernova'.
With increasing beam energy we catch earlier, hotter, stages forming
smaller nuclei.
For a theoretical treatment of the $0.4A$ GeV data in this connection see 
\cite{santini05}. 
This work supports our interpretation and shows that understanding
this 'fast' synthesis of nuclei is not outside our present theoretical
 capabilities, although there is room for improvement.

\section{Summary}\label{summary}

We summarize here  the results presented in sections \ref{rapidity} to
\ref{chemistry}.
In a study involving 25 system-energies we have been able to
establish a vast systematics of very central heavy ion collisions in the
SIS energy range.
We started characterizing the data for identified ejectiles by
rapidity distributions both in the longitudinal and the transverse
directions relative to the beam axis.
As a very general rule we found that the variance of the longitudinal
distribution was always broader than the variance of the transverse
direction. This effect was most pronounced at the highest incident energy
(Figs.~\ref{dndy-au150c1Z1A1th},\,\ref{dndy-au1500c1Z1A1}).

We also introduced a transverse distribution constrained by a cut
$|y_{z0}|<0.5$ on the scaled longitudinal rapidity (the definitions
were given in section \ref{centrality}) and found that its shape, plotted on
a linear scale, could always be very well described by a thermal distribution
defining thus an equivalent 'temperature' $T_{eq}$ that was characteristic
for each particle, incident energy and centrality.
This equivalent temperature was found in general to rise with the
particle's mass, Fig.~\ref{dndym-au400c1pdt}, an effect that could be
qualitatively reproduced by IQMD simulations, Fig.~\ref{dndym-au400c1smpdt-a},
although the shapes of the simulated distributions looked less perfectly thermal
and the $T_{eq}$ followed the experimental data only approximately.

At $1.5A$ GeV beam energy the rise of $T_{eq}$ with mass could also be
followed with the He isotopes and as a consequence of this behaviour it
could be shown that the spectra of heavier particles could not 
simply be derived by power law from the proton spectra, 
Fig.~\ref{dndym-ruru1500c1he-b} somewhat in contradiction to messages
published in ref.~\cite{swang95}.  

The $T_{eq}$ were shown to rise significantly with centrality, 
Fig.~\ref{dndym-au400c1he}.
At lower incident energies a remarkable deviation of the mass trend of
$T_{eq}$ was observed: the $^4$He value tended to be lower than the $^3$He
value. This 'He anomaly' \cite{neubert00} and large 
differences between $^3$H and $^3He$ spectra were tentatively associated
with clusterization phenomena,
 Figs.~\ref{dndym-au400c1p},\,\ref{dndym-au400c1A3-a},
illustrating the fact that a proper rendering of cluster probabilities was
essential if simulations were to be used for extracting basic information
 from the data.

The systematic comparison of transverse and longitudinal rapidity
distributions, Fig.~\ref{dndy-400c1Z1A2}, turned out to be a rich source of
information, as one observes remarkable variations with incident energy and
system size. 
Although the shapes of the distributions showed features that cannot be 
fully described by one parameter,
we proposed to summarize the differential information by a stopping
observable {\it varxz}, the ratio of variances of transverse to longitudinal
rapidity distributions. 
Stopping was always found to have values below one, a fact that we
interpreted as partial transparency or incomplete stopping on account
of the system size dependences.
The hydrodynamic rebound expected in the full stopping scenario could
also be definitely excluded by use of isospin tracer methods introduced
earlier \cite{rami00} by our Collaboration: an increase of transparency
at the high end of the studied energy range, Fig.~\ref{r4y-400a-r4stop},
could be unambiguously confirmed. 

We were able to confront the global stopping observable \cite{reisdorf04a}
with simulations varying the stiffness of the nuclear EOS.
The most striking feature of the excitation function for the Au+Au system,
namely the existence of a pronounced maximum for $E/A=0.2A$ to $0.8A$ GeV
could be qualitatively reproduced, but the strong descent at higher energies
to decreased stopping was underestimated, Fig.~\ref{vartl-indra}.
The predictions for {\it varxz} were found to be sufficiently sensitive to
the EOS to exceed the experimental uncertainty: this adds another constraint
from {\it central} collisions to efforts to fix the EOS from heavy ion data.

Moving beyond global observables to individual stopping data associated to
specific identified particles, we found stopping to be essentially a collective
phenomenon, Fig.~\ref{var-aupdt-h}, but a closer inspection also established
that there was a {\it hierarchy} of stopping suggesting that heavier
clusters were in part connected on the average with a less violent collision
 history. This hierarchy was weakest in the maximum of the stopping excitation
function, Fig.~\ref{varauc1pdt-4}, but otherwise followed as a function
of isotope mass, incident energy and system size a rather complex behaviour
reflecting the rich data presented earlier in Fig.~\ref{dndy-400c1Z1A2} and
representing a challenge to future simulations, Fig.~\ref{var400c1pdt}.
Surprisingly, we could not find a significant dependence of stopping on
isospin at $0.4A$ and $1.5A$ GeV, Fig.~\ref{varruzrc1pdt},
while the expected strong dependence on centrality, Fig.~\ref{varzx-au1500},
is a warning that centrality must be well matched to experiment when
simulations are done.

For the lower energies we have confirmed that the stopping hierarchy is
omnipresent also for all heavier clusters up to $Z=8$.
In an attempt to join up to  energies higher than were studied
here we showed the strong trend for increasing transparency of leading
protons as the energy is raised to AGS and SPS energies.  

Passing from the compression stage, characterized by the stopping observable,
we studied the expansion stage, characterized by radial flow, which is
generally inferred by mass dependences of kinetic energies or equivalent
temperatures.
We found that average kinetic energies of identified particles emitted at
$90^{\circ}$ follow extremely regular patterns as a function of
incident energy, Fig.~\ref{ekinavc4-r}, the details of which, like the
$^3$H-$^3$He difference, can only be understood if the simulation predicts
clusterization correctly, Figs.~\ref{ekinavc4p},\,\ref{ekinav-au120c4}.
If we combine the data for protons, deuterons and the average of the two
mass three isotopes ('A3 method'), 
we can establish a radial flow systematics for Au+Au
extending from $0.09A$ to $1.5A$ GeV which is in fair agreement in the
overlapping energy range with \cite{lisa95}.
We find a strong system size dependence of radial flow,
Fig.~\ref{ekinav-150c4}, which is however not reproduced quantitatively by
IQMD simulations, Fig.~\ref{ekinav-1000c4q}.
In agreement with \cite{lisa95} we find that the mass dependences of
kinetic energies are strongly dependent on the emission angles,
Fig.~\ref{ekinav-au90c1-6}, leading to a necessary reassessment of the origin of
the 'flow' deduced in our earlier work on heavy clusters \cite{reisdorf97}.
In general the observed partial transparencies and stopping hierarchies
call for a more qualified interpretation of the data, as illustrated in
Fig.~\ref{ekinav-au90c1-A3-2Z}.

Using the equivalent temperatures, based on constrained transverse
rapidity distributions, we observe the same structural characteristics if
we plot $T_{eq}$ versus mass for the five LCP, Fig.~\ref{dndym-150c1lcp-teq}.
The system size evolution of $T_{eq}$ is quite different for nucleonic
clusters than for pions, Fig.~\ref{dndym-1000c1lcp-teq-2} reflecting
the different mechanisms of creation/annihilation.
The presence of 'spectator' matter in non-central collisions influences
the resulting $T_{eq}$, Fig.~\ref{dndym-au1000c1pdt-Teq}.

A systematics of $T_{eq}$ for all five LCP versus impact parameter is
presented in Fig.~\ref{varau400} and allows to extrapolate to $b_0=0$.

The final state of the system after expansion-cooling is described by its
chemical composition which we have tabulated in the appendix for the
twenty-five system energies. 
For energies below $0.2A$ GeV we are consistent with INDRA-ALADIN data
\cite{andronic06,lefevre04,lavaud01}, while quite satisfactory agreement
with the pioneering PLASTIC BALL experiments \cite{doss88} could be
demonstrated, see Table~\ref{tab-PB}, for cases of overlap.
Throughout the SIS-BEVALAC energy range clusterization is an important
feature, Fig.~\ref{y-auc1-cluster-4}, that cannot be treated by
perturbative coalescence models and influences significantly the
spectra and flow of all particles: hence conclusions from the data with
the help of transport codes must handle correctly this aspect of the
reactions.
At the same time, particle creation (pions) in this energy range
 rises gradually from
a perturbative level to as much as $20\%$ in terms of the energy fraction at
$1.5A$ GeV (many more details, connected with pions and obtained with the FOPI
setup, were published in \cite{reisdorf07}).
At the lower energy end $\alpha$ particles play a special role: below $0.5A$
GeV more $\alpha$ particles are emitted than $^3$He ejectiles,
Fig.~\ref{y-auc1-he}, and  $\alpha$ particle condensation is probably needed to
explain in detail the very different $^3$He and $^3$H spectra as suggested
by our discussion of Figs.\ref{dndym-au400c1p} and \ref{dndym-au400c1A3-a}.
There is a notable system-size and isospin dependence of the chemical
'temperatures' that we evidenced using isospin pairs like $^3$He-$^3$H and
$\pi^+$-$\pi^-$ in various systems, 
Figs.~\ref{dndym-150c1lcp-yield}-\ref{dndym-1500c1lcp-yield}. 

Finally, our yield data suggest that there is at least a partial memory of the
initial degree of compression, and hence the stiffness of the EOS:
more compression leads to more efficient subsequent cooling enhancing the
degree of clusterization. This is suggested by
the close to quadratic dependence on system size and
 the conspicuous correlations
between cluster yields, radial flow and stopping evidenced in the last
figure of this paper and in \cite{reisdorf04b}.


\newpage

{\bf Appendix: Yield tables}
\begin{center}
Multiplicities and charge balance for Au+Au at $E/A=0.09$ GeV and $b_0 <
0.15$.
            
\begin{tabular}{cccccccc}
\hline
Z=1 & $40.1 \pm 2.1$ & p & $17.0\pm 0.9$ & Z=2 & $23.2\pm 1.7$ & 
                                                       $^3$He &              \\
    &                 & d & $13.3\pm 1.0$ &     &               & 
                                                       $^4$He &               \\
    &                 & t & $9.7\pm 1.0$ &     &               & & \\
\hline
\end{tabular}
\begin{tabular}{cccc}
Li & $4.1\pm 0.4$ &
Be & $1.77\pm 0.18$ \\
B  & $1.79\pm 0.18$ & 
C  & $1.43\pm 0.15$ \\
\hline
\end{tabular}
\end{center}
Charge balance  LCP: 86.5 or $54.8\%$ \\
Charge in HC Z3-Z6 (measured): $36.8\pm 3.7$ or $(23.3\pm 2.4)\%$ \\
Charge measured: 123.3 or $78.0\%$\\
Charge in HC:     158-86.5=71.5 or $45.3\%$ \\                            

\begin{center}
Multiplicities and charge balance for Au+Au at $E/A=0.12$ GeV and $b_0 <
0.15$.
            
\begin{tabular}{cccccccc}
\hline
Z=1 & $51.5 \pm 2.6$ & p & $21.5\pm 1.1$ & Z=2 & $24.2\pm 1.7$ & 
                                                       $^3$He & $8.0\pm 0.8$ \\
    &                 & d & $17.6\pm 1.2$ &     &               & 
                                                       $^4$He & $16.2\pm 1.6$ \\
    &                 & t & $12.2\pm 1.2$ &     &               & & \\
\hline
\end{tabular}
\begin{tabular}{cccc}
Li & $4.7\pm 0.5$ &
Be & $1.75\pm 0.18$ \\
B  & $1.62\pm 0.16$ & 
C  & $1.16\pm 0.12$ \\
N  & $0.71\pm 0.07$ &
O  & $0.37\pm 0.04$ \\
\hline
\end{tabular}
\end{center}
Charge balance  LCP: 99.7 or $63.1\%$ \\
Charge in HC Z3-Z8 (measured): $43.9\pm 4.4$ or $(27.8\pm 2.8)\%$ \\
Charge measured: 143.6 or $90.9\%$ \\
Charge in HC Z9-Z14 (estimated): $8.2\pm 3.0$ or $(5.2\pm 1.9)\%$ \\
Total balance: $96.1\%$ \\
Adopted IMF charge: $55.2\pm 5$ or $34.9\%$ \\

\begin{center}
Multiplicities and charge balance for Ni+Ni at $E/A=0.15$ GeV and $b_0 <
0.15$.
            
\begin{tabular}{cccccccc}
\hline
Z=1 & $27.3 \pm 1.4$ & p & $15.1\pm 0.8$ & Z=2 & $10.8\pm 0.8$ & 
                                                    $^3$He & $4.06\pm 0.41$ \\
    &                 & d & $8.4\pm 0.6$ &     &               & 
                                                    $^4$He & $6.71\pm 0.68$ \\
    &                 & t & $3.74\pm 0.4$ &     &               & & \\
\hline
\end{tabular}
\begin{tabular}{cccc}
Li & $1.24\pm 0.13$ &
Be & $0.42\pm 0.04$ \\
B  & $0.25\pm 0.03$ & C & $0.18\pm 0.02$\\
\hline
\end{tabular}
\end{center}
Charge balance  LCP: 48.84 or $87.2\%$ \\
Charge in HC: 7.69 or  $(13.7\pm 1.4)\%$ \\
Total balance: 56.53  or $100.9\%$ \\

\begin{center}
Multiplicities and charge balance for Xe+CsI at $E/A=0.15$ GeV and $b_0 <
0.15$.
            
\begin{tabular}{cccccccc}
\hline
Z=1 & $46.0 \pm 2.3$ & p & $20.1\pm 1.0$ & Z=2 & $18.8\pm 1.3$ & 
                                                       $^3$He & $5.9\pm 0.6$ \\
    &                 & d & $15.6\pm 1.1$ &     &               & 
                                                       $^4$He & $12.9\pm 1.3$ \\
    &                 & t & $10.3\pm 1.0$ &     &               & & \\
\hline
\end{tabular}
\begin{tabular}{cccc}
Li & $2.77\pm 0.28$ &
Be & $0.89\pm 0.09$ \\
B  & $0.67\pm 0.07$ & 
C  & $0.41\pm 0.04$ \\
N  & $0.19\pm 0.02$ &
O  & $0.10\pm 0.01$ \\
\hline
\end{tabular}
\end{center}
Charge balance  LCP: 83.53 or $77.3\%$ \\
Charge in HC: 19.83 or $(18.4\pm 1.8)\%$ \\
Total balance: 103.36 or $95.7\%$ \\


\begin{center}
Multiplicities and charge balance for Au+Au at $E/A=0.15$ GeV and $b_0 <
0.15$.
            
\begin{tabular}{cccccccc}
\hline
Z=1 & $59.9 \pm 3.0$ & p & $24.1\pm 1.2$ & Z=2 & $25.0\pm 1.8$ & 
                                                       $^3$He & $8.4\pm 0.9$ \\
    &                 & d & $20.4\pm 1.4$ &     &               & 
                                                       $^4$He & $16.6\pm 1.7$ \\
    &                 & t & $15.1\pm 1.5$ &     &               & & \\
\hline
\end{tabular}
\begin{tabular}{cccc}
Li & $5.0\pm 0.5$ &
Be & $1.69\pm 0.17$ \\
B  & $1.44\pm 0.15$ & 
C  & $0.90\pm 0.09$ \\
N  & $0.50\pm 0.05$ &
O  & $0.26\pm 0.03$ \\
\hline
\end{tabular}
\end{center}
Charge balance  LCP: 109.9 or $69.6\%$ \\
Charge in HC Z3-Z8 (measured): $40.0\pm 4$ or $(25.3\pm 2.5)\%$ \\
Charge measured: 149.9 or $94.9\%$ \\
Charge in HC Z9-Z14 (estimated): $4.4\pm 1.5$ or $(2.8\pm 1)\%$\\
Total balance: $97.7\%$\\


\begin{center}
Multiplicities and charge balance for Ni+Ni at $E/A=0.25$ GeV and $b_0 <
0.15$.
            
\begin{tabular}{cccccccc}
\hline
Z=1 & $34.9 \pm 1.8$ & p & $19.2\pm 1.0$ & Z=2 & $9.0\pm 0.7$ & 
                                                    $^3$He & $3.24\pm 0.33$ \\
    &                 & d & $10.5\pm 0.8$ &     &               & 
                                                    $^4$He & $5.79\pm 0.58$ \\
    &                 & t & $5.1\pm 0.5$ &     &               & & \\
\hline
\end{tabular}
\begin{tabular}{cccc}
Li & $0.91\pm 0.09$ &
Be & $0.26\pm 0.03$ \\
B  & $0.10\pm 0.01$ & & \\
\hline
\end{tabular}
\end{center}
Charge balance  LCP: 52.93 or $94.5\%$\\
Charge in HC: 4.28 or  $(7.6\pm 0.8)\%$ \\
Total balance: 57.21  or $102.2\%$


\begin{center}
Multiplicities and charge balance for Xe+CsI at $E/A=0.25$ GeV and $b_0 <
0.15$.
            
\begin{tabular}{cccccccc}
\hline
Z=1 & $59.6 \pm 3.0$ & p & $26.3\pm 1.4$ & Z=2 & $16.7\pm 1.2$ & 
                                                       $^3$He & $6.0\pm 0.6$ \\
    &                 & d & $19.3\pm 1.4$ &     &               & 
                                                       $^4$He & $10.8\pm 1.1$ \\
    &                 & t & $14.1\pm 1.5$ &     &               & & \\
\hline
\end{tabular}
\begin{tabular}{cccc}
Li & $2.33\pm 0.23$ &
Be & $0.60\pm 0.06$ \\
B  & $0.29\pm 0.03$ & 
C  & $0.14\pm 0.02$ \\
N  & $0.05\pm 0.01$ & &\\
\hline
\end{tabular}
\end{center}
Charge balance  LCP: 92.99 or $86.1\%$\\
Charge in HC: 12.65 or $(11.7\pm 1.2)\%$ \\
Total balance: 105.64 or $97.8\%$


\begin{center}
Multiplicities and charge balance for Au+Au at $E/A=0.25$ GeV and $b_0 <
0.15$.
            
\begin{tabular}{cccccccc}
\hline
Z=1 & $79.9 \pm 4.0$ & p & $33.7\pm 1.7$ & Z=2 & $24.8\pm 1.7$ & 
                                                       $^3$He & $8.8\pm 0.9$ \\
    &                 & d & $26.7\pm 1.9$ &     &               & 
                                                       $^4$He & $16.0\pm 1.6$ \\
    &                 & t & $19.5\pm 2.0$ &     &               & & \\
\hline
\end{tabular}
\begin{tabular}{cccc}
Li & $4.5\pm 0.4$ &
Be & $1.32\pm 0.13$ \\
B  & $0.67\pm 0.07$ & 
C  & $0.35\pm 0.04$ \\
N  & $0.13\pm 0.02$ & &\\
\hline
\end{tabular}
\end{center}
Charge balance  LCP: $82.1\%$ \\
Charge in HC: 26.0 or  $(16.45\pm 1.7)\%$ \\
   measured HC charge (Z3-Z7): 25.08, estim (Z8-Z10): 0.91 \\
Total balance: $98.5\%$\\


\begin{center}
Multiplicities and charge balance for Au+Au at $E/A=0.40$ GeV and $b_0 <
0.15$.
            
\begin{tabular}{cccccccc}
\hline
Z=1 & $106.0 \pm 4.0$ & p & $52.9\pm 2.7$ & Z=2 & $21.3\pm 1.9$ & 
                                                       $^3$He & $9.4\pm 0.9$ \\
    &                 & d & $34.2\pm 2.1$ &     &               & 
                                                       $^4$He & $11.9\pm 1.1$ \\
    &                 & t & $19.4\pm 1.8$ &     &               & & \\
$\pi^+$ & $0.95\pm 0.08$ & $\pi^-$ & $2.80\pm 0.14$ & & & & \\
\hline
\end{tabular}
\begin{tabular}{cccc}
Li & $3.5\pm 0.4$ &
Be & $0.84\pm 0.09$ \\
B  & $0.27\pm 0.03$ & 
C  & $0.096\pm 0.01$ \\
\hline
\end{tabular}
\end{center}
Charge balance  LCP: 148.6 or $94.1\%$ \\
Charge balance  LCP+pions: 146.8 or $92.9\%$ \\
Charge in HC (Z3-Z6): $15.77\pm 1.6$ or $(10.0\pm 1.0)\%$ \\
Total balance: $102.9\%$\\


\begin{center}
Multiplicities and charge balance for Au+Au at $E/A=0.6$ GeV and $b_0 <
0.15$.
\begin{tabular}{cccccccc}
\hline
Z=1 & $121.9 \pm 6.1$ & p & $68.5\pm 3.5$ & Z=2 & $16.4\pm 1.5$ & 
                                                       $^3$He & $9.1\pm 0.9$ \\
    &                 & d & $36.6\pm 2.2$ &     &               & 
                                                       $^4$He & $7.4\pm 0.7$ \\
    &                 & t & $16.8\pm 1.5$ &     &               & & \\
$\pi^+$ & $3.14\pm 0.25$ & $\pi^-$ & $7.08\pm 0.36$ & & & & \\
\hline
\end{tabular}
\end{center}
Charge balance  LCP+pions: $95.4\%$ \\
Estimate IMF: $(6\pm 2)\%$ \\
Total balance: $101.4\%$\\


\begin{center}
Multiplicities and charge balance for Au+Au at $E/A=0.8$ GeV and $b_0 <
0.15$.
\begin{tabular}{cccccccc}
\hline
Z=1 & $132.51\pm 6.6$ & p & $79.4\pm 4.0$ & Z=2 & $13.5\pm 1.2$ & 
                                                       $^3$He & $8.2\pm 0.8$ \\
    &                 & d & $37.7\pm 2.3$ &     &               & 
                                                       $^4$He & $5.2\pm 0.5$ \\
    &                 & t & $15.4\pm 1.6$ &     &               & & \\
$\pi^+$ & $6.26\pm 0.50$ & $\pi^-$ & $12.49\pm 0.63$ & & & & \\
\hline
\end{tabular}
\end{center}
Charge balance  LCP+pions: 153.3 or $97.0\%$ \\
Estimate IMF: $(4\pm 2)\%$ \\
Total balance: $101.0\%$\\


\begin{center}
Multiplicities and charge balance for Au+Au at $E/A=1.0$ GeV and $b_0 <
0.15$.
\begin{tabular}{cccccccc}
\hline
Z=1 & $140.9\pm 7.0$ & p & $91.5\pm 4.0$ & Z=2 & $11.0\pm 1.2$ & 
                                                       $^3$He & $7.3\pm 0.7$ \\
    &                 & d & $38.8\pm 2.3$ &     &               & 
                                                       $^4$He & $3.7\pm 0.4$ \\
    &                 & t & $13.9\pm 1.3$ &     &               & & \\
$\pi^+$ & $9.87\pm 0.79$ & $\pi^-$ & $18.26\pm 0.91$ & & & & \\
\hline
\end{tabular}
\end{center}
Charge balance  LCP+pions: 154.5 or $97.8\%$ \\
IMF  order $2\%$ ($< 4\%$)\\                    
Total balance: $99.8\%$\\


\begin{center}
Multiplicities and charge balance for Au+Au at $E/A=1.2$ GeV and $b_0 <
0.15$.
\begin{tabular}{cccccccc}
\hline
Z=1 & $148.2\pm 7.0$ & p & $99.3\pm 4.0$ & Z=2 & $8.74\pm 1.3$ & 
                                                       $^3$He & $6.1\pm 0.9$ \\
    &                 & d & $36.8\pm 2.3$ &     &               & 
                                                       $^4$He & $2.6\pm 0.4$ \\
    &                 & t & $12.1\pm 1.3$ &     &               & & \\
$\pi^+$ & $14.13\pm 1.13$ & $\pi^-$ & $24.37\pm 1.22$ & & & & \\
\hline
\end{tabular}
\end{center}
Charge balance  LCP+pions: 155.5 or $98.4\%$ \\
IMF  neglected \\                               
Total balance: $98.4\%$ (IMF neglected) \\


\begin{center}
Multiplicities and charge balance for Au+Au at $E/A=1.5$ GeV and $b_0 <
0.15$.
\begin{tabular}{cccccccc}
\hline
Z=1 & $158.4\pm 7.0$ & p & $111.3\pm 5.6$ & Z=2 & $6.76\pm 1.3$ & 
                                                       $^3$He & $5.1\pm 0.8$ \\
    &                 & d & $36.0\pm 2.2$ &     &               & 
                                                       $^4$He & $1.7\pm 0.3$ \\
    &                 & t & $11.0\pm 1.1$ &     &               & & \\
$\pi^+$ & $21.16\pm 1.70$ & $\pi^-$ & $33.34\pm 1.67$ & & & & \\
\hline
\end{tabular}
\end{center}
Charge balance  LCP+pions: 159.7 or $101.1\%$ \\
IMF  neglected \\                               
Total balance: $101.3\%$ (IMF neglected) \\


\begin{center}
Multiplicities and charge balance for Ru+Ru at $E/A=0.4$ GeV and $b_0 <
0.15$.
\begin{tabular}{cccccccc}
\hline
Z=1 & $59.8\pm 3.0$ & p & $33.2 \pm 1.7$ & Z=2 & $11.0\pm 1.0$ & 
                                                     $^3$He & $5.6\pm 0.5$ \\
    &                 & d & $18.3\pm 1.1$ &     &               & 
                                                     $^4$He & $5.4\pm 0.5$ \\
    &                 & t & $8.3\pm 0.8$ &     &               & & \\
$\pi^+$ & $0.70\pm 0.06$ & $\pi^-$ & $1.07\pm 0.06$ & & & & \\
\hline
\end{tabular}
\begin{tabular}{cccc}
Li & $1.28\pm 0.13$ &
Be & $0.28\pm 0.03$ \\
\hline
\end{tabular}
\end{center}
Charge balance  LCP+pions: 81.37 or $92.5\%$ \\
Total charge IMF : 5.5  or $6.2\%$  (estimate 0.5 for $Z>4$) \\
Total balance: $92.5+6.2=98.7\%$ \\

\begin{center}
Multiplicities and charge balance for Zr+Zr at $E/A=0.4$ GeV and $b_0 <
0.15$.
\begin{tabular}{cccccccc}
\hline
Z=1 & $56.0\pm 2.8$ & p & $28.8 \pm 1.5$ & Z=2 & $10.2\pm 1.0$ & 
                                                     $^3$He & $4.8\pm 0.4$ \\
    &                 & d & $18.3\pm 1.1$ &     &               & 
                                                     $^4$He & $5.4\pm 0.5$ \\
    &                 & t & $8.9\pm 0.8$ &     &               & & \\
$\pi^+$ & $0.56\pm 0.05$ & $\pi^-$ & $1.42\pm 0.07$ & & & & \\
\hline
\end{tabular}
\begin{tabular}{cccc}
Li & $1.22\pm 0.13$ &
Be & $0.29\pm 0.03$ \\
\hline
\end{tabular}
\end{center}
Charge balance  LCP+pions: 75.49 or $94.4\%$ \\
Total charge IMF : 5.3  or $6.6\%$  (estimate 0.5 for $Z>4$) \\
Total balance: $94.4+6.6=101.0\%$ \\


\begin{center}
Multiplicities and charge balance for Ru+Ru at $E/A=1.0$ GeV and $b_0 <
0.15$.
\begin{tabular}{cccccccc}
\hline
Z=1 & $79.1\pm 4.0$ & p & $53.9 \pm 2.7$ & Z=2 & $5.17\pm 0.47$ & 
                                                     $^3$He & $3.68\pm 0.33$ \\
    &                 & d & $19.7\pm 1.2$ &     &               & 
                                                     $^4$He & $1.49\pm 0.14$ \\
    &                 & t & $5.5\pm 0.5$ &     &               & & \\
$\pi^+$ & $6.05\pm 0.49$ & $\pi^-$ & $8.09\pm 0.41$ & & & & \\
\hline
\end{tabular}
\end{center}
Charge balance  LCP+pions: 87.38 or $99.3\%$ \\
IMF  neglected \\                               


\begin{center}
Multiplicities and charge balance for Ru+Ru at $E/A=1.5$ GeV and $b_0 <
0.15$.
\begin{tabular}{cccccccc}
\hline
Z=1 & $87.03\pm 4.4$ & p & $64.5 \pm 3.2$ & Z=2 & $3.32\pm 0.3$ & 
                                                     $^3$He & $2.65\pm 0.24$ \\
    &                 & d & $18.4\pm 1.1$ &     &               & 
                                                     $^4$He & $0.67\pm 0.06$ \\
    &                 & t & $4.1\pm 0.4$ &     &               & & \\
$\pi^+$ & $12.49\pm 1.00$ & $\pi^-$ & $15.63\pm 0.78$ & & & & \\
\hline
\end{tabular}
\end{center}
Charge balance  LCP+pions: 90.52 or $102.9\%$ \\
IMF  neglected \\                               

\begin{center}
Multiplicities and charge balance for Zr+Zr at $E/A=1.5$ GeV and $b_0 <
0.15$.
\begin{tabular}{cccccccc}
\hline
Z=1 & $84.3 \pm 4.2$ & p & $61.3 \pm 3.1$ & Z=2 & $3.26\pm 0.32$ & 
                                                     $^3$He & $2.61\pm 0.24$ \\
    &                 & d & $18.6\pm 1.1$ &     &               & 
                                                     $^4$He & $0.65\pm 0.06$ \\
    &                 & t & $4.3\pm 0.4$ &     &               & & \\
$\pi^+$ & $11.45\pm 0.92$ & $\pi^-$ & $17.19\pm 0.86$ & & & & \\
\hline
\end{tabular}
\end{center}
Charge balance  LCP+pions: 85.09 or $106.4\%$ \\
IMF  neglected \\                               

\begin{center}
Multiplicities and charge balance for Ca+Ca at $E/A=0.4$ GeV and $b_0 <
0.15$.
\begin{tabular}{cccccccc}
\hline
Z=1 & $30.5\pm 1.6$ & p & $19.1\pm 1.0$ & Z=2 & $4.9\pm 0.5$ & 
                                                       $^3$He & $2.8\pm 0.3$ \\
    &                 & d & $8.6\pm 0.5$ &     &               & 
                                                       $^4$He & $2.1\pm 0.2$ \\
    &                 & t & $2.7\pm 0.3$ &     &               & & \\
$\pi^+$ & $0.45\pm 0.04$ & $\pi^-$ & $0.47\pm 0.03$ & & & & \\
\hline
\end{tabular}
\end{center}
Charge balance  LCP+pions: 40.29 or $100.7\%$ \\
 Z3 $0.34 \pm 0.04$ \\                               
Charge balance: 40.29+1.02=41.31  or $103.3\%$ \\

\begin{center}
Multiplicities and charge balance for Ca+Ca at $E/A=0.6$ GeV and $b_0 <
0.15$.
\begin{tabular}{cccccccc}
\hline
Z=1 & $33.2\pm 1.7$ & p & $22.9\pm 1.2$ & Z=2 & $3.7\pm 0.4$ & 
                                                       $^3$He & $2.6\pm 0.3$ \\
    &                 & d & $8.1\pm 0.5$ &     &               & 
                                                       $^4$He & $1.1\pm 0.1$ \\
    &                 & t & $2.2\pm 0.2$ &     &               & & \\
$\pi^+$ & $1.13\pm 0.09$ & $\pi^-$ & $1.28\pm 0.07$ & & & & \\
\hline
\end{tabular}
\end{center}
Charge balance  LCP+pions: 40.54 or $101.3\%$ \\
IMF  neglected \\                               

\begin{center}
Multiplicities and charge balance for Ca+Ca at $E/A=0.8$ GeV and $b_0 <
0.15$.
\begin{tabular}{cccccccc}
\hline
Z=1 & $35.2\pm 1.8$ & p & $25.3\pm 1.3$ & Z=2 & $2.6\pm 0.2$ & 
                                                     $^3$He & $1.85\pm 0.23$ \\
    &                 & d & $8.0\pm 0.5$ &     &               & 
                                                     $^4$He & $0.75\pm 0.09$ \\
    &                 & t & $1.9\pm 0.2$ &     &               & & \\
$\pi^+$ & $2.11\pm 0.17$ & $\pi^-$ & $2.28\pm 0.11$ & & & & \\
\hline
\end{tabular}
\end{center}
Charge balance  LCP+pions: 40.24 or $100.6\%$ \\

\begin{center}
Multiplicities and charge balance for Ca+Ca at $E/A=1.0$ GeV and $b_0 <
0.15$.
\begin{tabular}{cccccccc}
\hline
Z=1 & $36.5\pm 1.8$ & p & $26.5\pm 1.4$ & Z=2 & $2.1\pm 0.2$ & 
                                                     $^3$He & $1.65\pm 0.15$ \\
    &                 & d & $7.8\pm 0.5$ &     &               & 
                                                     $^4$He & $0.45\pm 0.04$ \\
    &                 & t & $1.52\pm 0.14$ &     &               & & \\
$\pi^+$ & $3.16\pm 0.26$ & $\pi^-$ & $3.33\pm 0.17$ & & & & \\
\hline
\end{tabular}
\end{center}
Charge balance  LCP+pions: 40.57 or $101.4\%$ \\

\begin{center}
Multiplicities and charge balance for Ca+Ca at $E/A=1.5$ GeV and $b_0 <
0.15$.
\begin{tabular}{cccccccc}
\hline
Z=1 & $38.6\pm 1.8$ & p & $30.4\pm 1.6$ & Z=2 & $1.10\pm 0.16$ & 
                                                     $^3$He & $0.93\pm 0.16$ \\
    &                 & d & $7.2\pm 0.5$ &     &               & 
                                                     $^4$He & $0.17\pm 0.03$ \\
    &                 & t & $1.22\pm 0.15$ &     &               & & \\
$\pi^+$ & $5.62\pm 0.45$ & $\pi^-$ & $5.78\pm 0.30$ & & & & \\
\hline
\end{tabular}
\end{center}
Charge balance  LCP+pions: 40.64 or $101.6\%$ \\

\begin{center}
Multiplicities and charge balance for Ca+Ca at $E/A=1.93$ GeV and $b_0 <
0.15$.
\begin{tabular}{cccccccc}
\hline
Z=1 & $39.6\pm 1.8$ & p & $33.7\pm 1.7$ & Z=2 & $0.82\pm 0.20$ & 
                                                     $^3$He & $0.70\pm 0.14$ \\
    &                 & d & $6.8\pm 0.5$ &     &               & 
                                                     $^4$He & $0.13\pm 0.03$ \\
    &                 & t & $1.11\pm 0.15$ &     &               & & \\
$\pi^+$ & $7.82\pm 0.63$ & $\pi^-$ & $7.86\pm 0.40$ & & & & \\
\hline
\end{tabular}
\end{center}
Charge balance  LCP+pions: 41.22 or $103.1\%$ \\

\begin{ack}
This work has been supported by the German BMBF,
contract 06HD154 and within the framework of the WTZ
program (Project RU8 02/021), by the DFG (Project 446-KOR-113/76),
the DAAD (PPP D /03/44611) and the IN2P3/GSI agreement 97/29.
This work was also supported by the National Research Foundation of Korea
 grant (No. 2010-0015692)
\end{ack}
\newpage



\end{document}